\documentclass[preprint,tightenlines,aps,eqsecnum,floatfix,showpacs]{revtex4-1}
\usepackage{ulem} 
\usepackage{dcolumn}
\usepackage{bm}
\usepackage{color}
\usepackage{amssymb}
\usepackage{amsmath}
\usepackage{graphicx}
\usepackage{amsfonts}
\usepackage{slashed}
\usepackage{pstricks}
\usepackage{float}
\allowdisplaybreaks

\begin{document}
\title{Nuclear Axial Currents in Chiral Effective Field Theory}
\author{A. Baroni$^{\,{\rm a},\dagger}$,
L.\ Girlanda$^{\,{\rm b}}$,
S.\ Pastore$^{\,{\rm  c,d}}$,
R.\ Schiavilla$^{\,{\rm a,e}}$,
and M.\ Viviani$^{\,{\rm f}}$}
\affiliation{
$^{\rm a}$\mbox{Department of Physics, Old Dominion University, Norfolk, VA 23529} \\
$^{\rm b}$\mbox{Department of Mathematics and Physics, University of Salento and INFN-Lecce, 73100 Lecce, Italy}\\
$^{\rm c}$\mbox{Department of Physics and Astronomy, University of South Carolina, Columbia, SC 29208}\\
$^{\rm d}$\mbox{Theoretical Division, Los Alamos National Laboratory, Los Alamos, NM 87545}\\
$^{\rm e}$\mbox{Theory Center, Jefferson Lab, Newport News, VA 23606}\\
$^{\rm f}$\mbox{INFN-Pisa, 56127 Pisa, Italy}\\
$^{\dagger}$\mbox{Supported by a Jefferson Science Associates Theory Fellowship.}
}

\date{\today}

\begin{abstract}
Two-nucleon axial charge and current operators are derived in chiral effective field
theory up to one loop.  The derivation is based on time-ordered perturbation theory, and
accounts for cancellations between the contributions of irreducible diagrams and the
contributions due to non-static corrections from energy denominators of reducible diagrams.
Ultraviolet divergencies associated with the loop corrections are isolated in dimensional
regularization.  The resulting axial current is finite and conserved
in the chiral limit, while the axial charge requires
renormalization. A complete set of contact terms for the axial charge up to the
relevant order in the power counting is constructed.
\end{abstract}
\pacs{21.45.-v, 23.40-s}
\maketitle
\section{Introduction}

\label{sec:chieft}
Chiral symmetry is an approximate symmetry of Quantum Chromodynamics (QCD), the
fundamental theory that describes the interactions of quarks and gluons---the symmetry
becomes exact in the limit of vanishing quark masses.  Chiral effective field theory ($\chi$EFT)
is the theoretical framework that permits the derivation of nuclear potentials and electroweak
currents from the symmetries of QCD---the exact Lorentz, parity, and time-reversal symmetries, and the
approximate chiral symmetry.  Pions and nucleons (and low-energy excitations
of the nucleon, such as the $\Delta$ isobar), rather than quarks and gluons, are the degrees of
freedom of $\chi$EFT.  Chiral symmetry requires the pion to couple to these baryons, as well as
to other pions, by powers of its momentum $Q$ and, as a consequence, the Lagrangian describing
their interactions can be expanded in powers of $Q/\Lambda_\chi$, where $\Lambda_\chi\sim 1$
GeV is the chiral symmetry breaking scale.  Classes of Lagrangians emerge, each characterized
by a given power of $Q/\Lambda_\chi$, or equivalently a given order in the derivatives of the pion
field and/or pion mass factors, and each containing a certain number of unknown parameters,
the so called low-energy constants (LECs).  These LECs could in principle be calculated from
the underlying QCD theory of quarks and gluons, but the non-perturbative nature of this theory
at low energies makes this task extremely difficult.  Hence, in practice, the LECs are fixed by
comparison with experimental data, and therefore effectively encode short-range
physics and the effects of baryon resonances, such
as the $\Delta$ isobar, and heavy-meson exchanges, not explicitly retained in the
chiral Lagrangians.

Within $\chi$EFT a variety of studies have been carried out in the
strong-interaction sector dealing with the derivation of two- and three-nucleon
potentials~\cite{Ord95,Epelbaum98,Ent03,Machleidt11,Nav07,Epe02,Bira94,Bern11,Gir11}
and accompanying isospin-symmetry-breaking corrections~\cite{Fri99,Epe99,Fri04,Fri05},
and in the electroweak sector dealing with the derivation of parity-violating
two-nucleon potentials induced by hadronic weak interactions~\cite{Hax13,Zhu05,Gir08,Viv14}
and the construction of nuclear electroweak
currents~\cite{Park93,Park96,Pastore08,Pastore09,Pastore11,Piarulli13,Koelling09,Koelling11}.
Most of these studies have been based on a formulation of $\chi$EFT in which
nucleons and pions are the explicit degrees of freedom.  A few, however, have also
retained $\Delta$ isobars as explicit degrees of freedom. 

In this paper, the focus is on nuclear axial charge and current operators.
These were originally derived up to one loop in heavy-baryon covariant perturbation
theory (HBPT) in a pioneering work by Park {\it et al.}~\cite{Park93}.  Here we re-derive them by employing a formulation of time-ordered perturbation theory (TOPT), which accounts for cancellations
occurring at a given order in the power counting between the contributions of irreducible diagrams
and the contributions due to non-static corrections from energy denominators of reducible
diagrams~\cite{Pastore08}.  Because of the different treatment of reducible diagrams in the HBPT
and TOPT approaches, we find differences between the operators obtained in these two formalisms
as well as additional differences due to the omission of a number of
contributions in Ref.~\cite{Park93}, as discussed in Sec.~\ref{sec:disc}.

An accurate theory of nuclear electroweak structure and dynamics is relevant in several
areas of current interest.  One such area is that of low-energy tests of physics beyond the Standard
Model in $\beta$-decay experiments~\cite{Severijns06}.  Phenomenologically, the weak interactions
are known to couple only to left-handed neutrinos, and to violate parity maximally.  However, beyond
the Standard Model (BSM) theories have been constructed in which small deviations from these
properties are introduced.  These deviations affect the correlation coefficients entering $\beta$-decay
rates, and can in principle be detected.  For a proper interpretation of these measurements and, in
particular, to unravel possible signatures of BSM physics, it is crucial to have control of the nuclear
structure and weak interactions in nuclei.

Another area of interest is that of neutrino interactions with nuclei and neutron matter.
The low-energy inelastic neutrino scattering from nuclei is important in astrophysics
and for neutrino detectors.  The spallation of neutrons from nuclei by neutrino interactions
is relevant in setting the neutron to seed ratio in core-collapse supernovae.   Accurate
predictions for neutrino-nucleus scattering cross sections, specifically from the argon
nucleus, are key to the measurements of supernovae neutrino fluxes, a major
component of the Deep Underground Neutrino Experiment (DUNE).  At temperatures
of a few MeV, neutrino processes are also very important in core-collapse supernovae.
One significant issue is the decoupling of various flavors of neutrinos and antineutrinos
at the surface of the proto-neutron star.  This sets the initial temperatures (flux versus
energy) of $e$, $\mu$ and $\tau$ neutrinos and antineutrinos. Understanding this initial
flux is critical to interpreting the subsequent evolution of neutrinos and their role in the
r-process.  Neutrino and antineutrino interactions in neutron matter are also of importance
in understanding the evolution of the very neutron-rich matter formed in neutron-star
mergers, since they can potentially alter the neutron to proton ratio and significantly
impact the r-process in neutron star mergers, currently considered to be an important 
source for r-process nucleosynthesis.

The present paper is organized as follows.  In Sec.~\ref{sec:hl} pion-nucleon ($\pi N$) and
pion-pion ($\pi\pi$) interaction Hamiltonians are constructed from the chiral Lagrangian
formulation of Refs.~\cite{Fettes00,Sch12}---for convenience these Lagrangians are
listed in Appendix~\ref{app:ls}, where a number of details relative to the construction
of the Hamiltonians up to the relevant chiral order are also provided.  In Sec.~\ref{sec:road}
the power counting scheme and TOPT formulation adopted in the present work
are described.  These along with the interaction vertices obtained in
Appendix~\ref{app:vert} are utilized to derive two-nucleon axial charge and current
operators up to one loop in Secs.~\ref{sec:axc} and~\ref{sec:axj}, respectively.
Ultraviolet divergencies associated with the loop corrections are isolated
in dimensional regularization: the resulting axial current is then found to be finite, while
the axial charge requires renormalization.  All this along with the
renormalization of the one-pion-exchange (tree-level) axial charge is discussed-
in Sec.~\ref{sec:renor}.  A number of details are relegated
to Appendix~\ref{app:cta0}, where a complete
set of contact terms for the axial charge (up to the
relevant order) is constructed, to Appendix~\ref{app:jforms}, where
loop functions entering the axial current are defined,
and to Appendix~\ref{app:renora},
where a listing of counter-terms is given.
In Sec.~\ref{sec:disc} a summary and discussion of our results as well as 
a comparison between the expressions for the axial operators
obtained here and those of Park {\it et al.}~\cite{Park93} are provided.
Conclusions are  summarized in Sec.~\ref{sec:concl}.

\section{Interaction Hamiltonians from chiral Lagrangians}
\label{sec:hl}

The chiral Lagrangian describing the interactions of pions and
nucleons is given by
\begin{equation}
\label{eq:ltot}
{\cal L}={\cal L}_{\pi N}+{\cal L}_{\pi\pi}\ ,
\end{equation}
where
\begin{eqnarray}
{\cal L}_{\pi N}&=&{\cal L}_{\pi N}^{(1)}+{\cal L}_{\pi N}^{(2)}+{\cal L}_{\pi N}^{(3)}+\dots\ ,\\
{\cal L}_{\pi\pi}&=&{\cal L}_{\pi\pi}^{(2)}+{\cal L}_{\pi\pi}^{(4)}+\dots\ ,
\end{eqnarray}
and the superscript $n$ specifies the chiral order $Q^n$ ($Q$ denotes generically the low-momentum
scale), i.e., the number of derivatives of the pion field and/or insertions of the pion mass.
External fields are counted as being of order $Q$.
Since we are interested in deriving nuclear potentials and currents up
one loop, it suffices to retain in ${\cal L}$ up to ${\cal L}_{\pi N}^{(3)}$ and ${\cal L}_{\pi\pi}^{(4)}$.
The Lagrangians ${\cal L}_{\pi N}^{(n)}$ (in fact up to order $n=4$) and ${\cal L}_{\pi \pi}^{(n)}$ have been
given, for example, in Refs.~\cite{Fettes00} and~\cite{Sch12}, respectively, and are
listed in Appendix~\ref{app:ls} of the present paper for completeness.  The total
Lagrangian can be written as
\begin{eqnarray}
{\cal L} &=& \overline{N } \left( i\, \slashed{\partial} -m +\Gamma^0_a \, \partial_0\pi_a
 +\Lambda^i_a \, \partial_i\pi_a  + \Delta \right) N  \nonumber \\
&&+\frac{1}{2}\left(  \partial^0 \pi_a \, G_{ab} \, \partial_0 \pi_b
 +\partial^i \pi_a \, \widetilde{G}_{ab} \, \partial_i \pi_b
 -m_\pi^2\, \pi_a H_{ab}\, \pi_b \right) -f_\pi\,  A^\mu_a \, F_{ab}\,\left(\partial_\mu \pi_b\right) \ ,
\label{eq:2.3}
\end{eqnarray}
where $\pi_a$ is the pion field of isospin component $a$, $N$ is the iso-doublet of nucleon fields,
$A^\mu_a$ is the axial-vector field of isospin component $a$, $f_\pi$ is the pion decay
constant, and $m$ and $m_\pi$ are, respectively, the nucleon and pion masses.
The symbols $\Gamma^0_a$, $\Lambda_a^i$, and $\Delta$ denote
combinations of the pion and axial-vector fields (and their derivatives) and/or
of pion mass factors, having the following expansions
\begin{equation}
\label{eq:eg0}
\Gamma^0_a = \Gamma^0_a(0)+\Gamma^0_a(1)+\Gamma^0_a(2)\ ,
\end{equation}
and similarly for $\Lambda^i_a$, and
\begin{equation}
\Delta  = \Delta(1)+\Delta(2)+\Delta(3) \ ,
\label{eq:edelta}
\end{equation}
where the argument $n$ in $\Gamma^0_a(n)$, $\Lambda^i_a(n)$, and $\Delta(n)$
specifies the power counting $Q^n$.  The symbols $G_{ab}$, $\widetilde{G}_{ab}$,
$H_{ab}$, and $F_{ab}$ denote three-by-three matrices in isospin space,
containing powers of the pion field and/or pion mass.
A listing of all these quantities, limited to the terms
relevant for the construction of the currents at one loop, is provided in Appendix~\ref{app:ls}.
At this stage the various fields, masses, and coupling constants
are to be understood as bare (un-renormalized) quantities.

From the Lagrangian ${\cal L}$ in Eq.~(\ref{eq:2.3}) the conjugate momenta relative to the pion and
nucleon fields follow as
\begin{eqnarray}
\Pi^\dagger& =& \frac{\partial {\cal L}}{\partial(\partial_0{N})} = i\, \overline{N}\, \gamma^0 \ ,\\
\Pi_a&=& \frac{\partial {\cal L}}{\partial (\partial_0{\pi}_a)} = G_{ab} \, \partial^0{\pi}_b -f_\pi \, F_{ab} \, A^0_b
+\overline{N} \, \Gamma_a^0 \, N \ ,
\end{eqnarray}
and the Hamiltonian then reads
\begin{equation}
{\cal H}=\Pi^\dagger \, \partial_0 N+ \Pi_a\,  \partial_0 \pi_a -{\cal L}={\cal H}_0+{\cal H}_I \ ,
\end{equation}
where ${\cal H}_0$,
\begin{equation}
{\cal H}_0=\frac{1}{2}\left( \Pi_a\,\Pi_a - \partial^i \pi_a\, \partial_i \pi_a +m_\pi^2 \,\pi_a\,\pi_a \right)
+\overline{N} \left(-i\, \gamma^i \, \partial_i +m\right)N \ ,
\end{equation}
is the free pion and nucleon Hamiltonian, while ${\cal H}_I$ is
the Hamiltonian accounting for the interactions between pions and nucleons as well
as between these and the external field.  By only keeping terms linear in the latter,
the interaction Hamiltonian is given by
\begin{eqnarray}
\label{eq:hhii}
{\cal H}_I&=&\frac{1}{2}\, \Pi_a \left[\left(G^{-1}\right)_{ab}-\delta_{ab}\right]  \Pi_b 
-\frac{1}{2}\, \left[ \Pi_a \left( G^{-1} \right)_{ab} \left(\overline{N} \, \Gamma_b^0 \, N\right)+{\rm h.c.}\right]
\nonumber\\
&&+\frac{f_\pi}{2}\, \left[ \Pi_a\, \left( G^{-1} \right)_{ab} F_{bc} \, A_c^0 +{\rm h.c.}\right]
-\frac{f_\pi}{2} \, \left[  \left( \overline{N} \, \Gamma_a^0 \, N\right)
\left( G^{-1} \right)_{ab} F_{bc} \, A^0_c +{\rm h.c.}\right]\nonumber\\
&&+\frac{1}{2}\, \left( \overline{N} \, \Gamma_a^0 \, N\right)    \left( G^{-1} \right)_{ab}  
\left( \overline{N} \, \Gamma_b^0 \, N\right)
-\overline{N} \left( \Lambda_a^i \,\partial_i \pi_a  +\Delta\right)N \nonumber\\
&&-\frac{1}{2}\,
\partial^i \pi_a \left(\widetilde{G}_{ab}-\delta_{ab}\right) \partial_i \pi_b
+f_\pi \, A^i_a \, F_{ab}\, \partial_i \pi_b
+ \frac{m_\pi^2}{2} \pi_a \left(H_{ab}-\delta_{ab} \right)  \pi_b\ .
\end{eqnarray}
It admits the following expansion in powers of $Q$:
\begin{equation}
{\cal H}_I={\cal H}_I^{(1)}+{\cal H}_I^{(2)}+{\cal H}_I^{(3)}+\dots \ ,
\end{equation}
and the vertices corresponding to the various interaction terms are listed
in Appendix~\ref{app:vert}.
\section{From amplitudes to currents}
\label{sec:road}

The expansion of the transition amplitude for a given process is based
on TOPT.  Terms in this expansion are conveniently
represented by diagrams.  We distinguish between reducible diagrams
(diagrams which involve at least one pure nucleonic intermediate state) and irreducible
diagrams (diagrams which include pionic and nucleonic intermediate states).
The former are enhanced with respect to the latter by a factor of $Q$ for each
pure nucleonic intermediate state (see below).  In the static limit---in
the limit $m\rightarrow \infty$, i.e., neglecting nucleon kinetic
energies---reducible contributions are infrared-divergent.  The prescription
proposed by Weinberg~\cite{Weinberg90} to treat these is to define the
nuclear potential and currents as given by the irreducible contributions
only.  Reducible contributions, instead, are generated by solving the Lippmann-Schwinger
(or Schr\"odinger) equation iteratively with the nuclear potential (and currents) arising
from irreducible amplitudes.

The formalism developed by some of the present authors is based on this prescription~\cite{Pastore08}.
However, the omission of reducible contributions from the definition of nuclear operators
needs to be dealt with care when the irreducible amplitude is evaluated under an approximation.
It is usually the case that the irreducible amplitude is evaluated in the static limit approximation.
The iterative process will then generate only that part of the reducible
amplitude including the approximate static nuclear operators.  The reducible
part obtained beyond the static limit approximation needs to be incorporated order
by order---along with the irreducible amplitude---in the definition of nuclear operators.
This scheme in combination with TOPT, which is best suited to separate the
reducible content from the irreducible one, has been implemented in
Refs.~\cite{Pastore09,Pastore11,Piarulli13} and is briefly described below.  The method
leads to nuclear operators which are not uniquely defined due to the non-uniqueness of the
transition amplitude off-the-energy shell.  While non unique, the resulting operators are
nevertheless unitarily equivalent, and therefore the description of physical systems is not
affected by this ambiguity~\cite{Friar77,Pastore11}. 

We note that an alternative approach, implemented to face the difficulties posed by the reducible amplitudes,
has been introduced by Epelbaum and collaborators~\cite{Epelbaum00}.  The method, referred to as
the unitary transformation method, is based on TOPT
and exploits the Okubo (unitary) transformation~\cite{Okubo54}
to decouple the Fock space of pions and nucleons into two subspaces, one containing only pure
nucleonic states and the other involving states that retain at least one pion.  In this decoupled
space, the amplitude does not involve enhanced contributions associated with the reducible diagrams.
The subspaces are not-uniquely defined, since it is always possible to perform
additional unitary transformations onto them, with a consequent change in the formal definition
of the resulting nuclear operators.  This, of course, does not affect physical representations.

The two TOPT-based methods outlined above lead to formally equivalent operator structures for
the nuclear potential and electromagnetic charge and current up to loop corrections
included, which makes it plausible to conjecture that the two methods are closely related.
However, this topic has not been investigated further.  In what follows, we focus on the method
developed in Refs.~\cite{Pastore09,Pastore11,Piarulli13} and show how nuclear operators
are derived from transition amplitudes.  Here, we are especially interested in the construction
of the two-body weak axial charge and current operators.  We will not discuss the aforementioned
unitary equivalence between operators corresponding to different off-the-energy-shell extrapolations
of the transition amplitudes.  This issue has already been addressed in considerable detail in Ref.~\cite{Pastore11}
for the case of the two-body nuclear potential and electromagnetic charge and current operators.
Similar considerations apply to the present case.

The starting point is the conventional
perturbative expansion for the amplitude
\begin{equation}
 \langle f \! \mid T_5\mid\! i \rangle= 
 \langle f \!\mid H_I \sum_{n=1}^\infty \left( 
 \frac{1}{E_i -H_0 +i\, \eta } H_I \right)^{n-1} \mid\! i \rangle \ ,
\label{eq:pt}
\end{equation}
where $\mid\! i \rangle$ and $\mid\! f \rangle$ represent the initial and final
states, respectively $\mid\! N_1 N_2 A\rangle$ and
$\mid \! N_1^\prime N_2^\prime\rangle$ ($A$ denotes generically the external
axial field), of energies $E_i$ and $E_f$ with $E_i=E_f$, $H_0$ is the Hamiltonian
describing free pions and nucleons, and $H_I$ is the Hamiltonian
describing interactions among these particles ($H_0=\int{\rm d}{\bf x}\, {\cal H}_0({\bf x})$ and similarly
for $H_I$, with ${\cal H}_0$ and ${\cal H}_I$ as defined
in Sec.~\ref{sec:hl} with the various fields taken in the
Schr\"odinger picture).  The evaluation of this amplitude
is carried out in practice by inserting complete sets of $H_0$ eigenstates
between successive terms of $H_I$.   Power counting is then used
to organize the expansion in powers of $Q/\Lambda_\chi \ll 1$.

In the perturbative series, Eq.~(\ref{eq:pt}), a generic (reducible or
irreducible) contribution is characterized by a certain number, say
$N$, of vertices, each scaling as $Q^{\alpha_i}\times Q^{-\beta_i/2}$
($i$=$1,\dots,N$), where $\alpha_i$ is the power counting implied by the
specific term in the interaction Hamiltonian $H_I$ under consideration
and $\beta_i$ is the number of
pions in and/or out of the vertex, a corresponding $N$--1 number of energy
denominators, and $L$ loops.  Out of these
$N$--1 energy denominators, $N_K$ of them will involve only nucleon kinetic
energies and possibly, depending on the particular time
ordering under consideration, the energy $\omega_q$ associated with
the external field, both of which scale as $Q^2$, while the remaining $N-N_K-1$
energy denominators will involve,
in addition, pion energies, which are of order $Q$.  Loops, on the other hand,
contribute a factor $Q^3$ each, since they imply integrations
over intermediate three momenta.  Hence the power counting
associated with such a contribution is
\begin{equation}
\left(\prod_{i=1}^N  Q^{\alpha_i-\beta_i/2}
\right)\times \left[ Q^{-(N-N_K-1)}\, Q^{-2N_K} \right ]
\times Q^{3L} \ .
\label{eq:count}
\end{equation}
Clearly, each of the $N-N_K-1$ energy denominators can be further expanded as
\begin{equation}
\frac{1}{E_i-E_I-\omega_\pi}= -\frac{1}{\omega_\pi}
\bigg[ 1 + \frac{E_i-E_I}{\omega_\pi}
+\frac{(E_i-E_I)^2}{\omega^2_\pi} + \dots\bigg] \ ,
\label{eq:deno}
\end{equation}
where $E_I$ denotes the energy of the intermediate state (including the
kinetic energies of the two nucleons and, where appropriate, the energy of the external field), and
$\omega_\pi$ the pion energy (or energies, as the case may be)---the ratio $(E_i-E_I)/\omega_\pi$ is of order $Q$.
The leading order term $-1/\omega_\pi$ represents the static limit,
while the sub-leading terms involving powers of $(E_i-E_I)/\omega_\pi$ represent
non-static corrections of increasing order; elsewhere~\cite{Pastore08,Pastore09}, we have
referred to these as recoil corrections.

Interactions with the external axial field are treated in first order
in Eq.~(\ref{eq:pt}), and inspection of the $Q$ scaling of the various
terms shows that the associated transition amplitude admits the following expansion
\begin{equation}
T_5=T_5^{(-3)}+T_5^{(-2)}+T_5^{(-1)} +\dots \ ,
\end{equation}
where $T_5^{(n)}$ is of order $Q^n$.  Next, we denote the two-nucleon
strong-interaction potential with $v$ and the weak-interaction potential with
$v_5= A^0_a\, \rho_{5,a}-{\bf A}_a\cdot {\bf j}_{5,a}$, where
$\rho_{5,a}$ and ${\bf j}_{5,a}$ are, respectively, the nuclear weak axial
charge and current operators and $A_a^\mu=(A_a^0,{\bf A}_a)$ is the external
axial field.  We construct $v+v_5$ by requiring that
iterations of $v+v_5$ in the Lippmann-Schwinger equation~\cite{Pastore11}
\begin{equation}
(v+v_5)+(v+v_5)\, G_0 \, (v+v_5) + (v+v_5) \, G_0 \, (v+v_5) \, G_0\, (v+v_5) +\dots \ ,
\end{equation}
match the $T_5$ amplitude, on the energy shell $E_i=E_f$, order by order in the power counting;
here $G_0$ denotes the  propagator $G_0=1/(E_i-E_I+i\eta)$.  The
potentials $v$ and $v_5$ have the following expansions
\begin{eqnarray}
v&=&v^{(0)}+v^{(2)}+v^{(3)} +\dots \ , \\
v_5&=&v_5^{(-3)}+v_5^{(-2)}+v_5^{(-1)}+v_5^{(0)}+v_5^{(1)}+\dots \ ,
\end{eqnarray}
where the potentials $v^{(n)}$ have been derived in Refs.~\cite{Pastore09,Pastore11}, in particular
$v^{(1)}$ vanishes~\cite{Pastore11}, and
$v_5^{(n)}=A^0_a\, \rho_{5,a}^{(n)}-{\bf A}_a\cdot {\bf j}_{5,a}^{(n)}$.  
The superscript $(n)$ on $v_5$ and $T_5$ only refers to the power counting of $\rho_{5,a}^{(n)}$ and ${\bf j}_{5,a}^{(n)}$,
and does not include the power of $Q$ associated with the external field.
The matching between $T_5^{(n)}$ and $v_5^{(n)}$
leads to the following relations~\cite{Pastore11}
\begin{eqnarray}
v_5^{(-3)}= T_5^{(-3)}
&& \ , \label{eq:vg3m} \\
v_5^{(-2)}= T_5^{(-2)}
&&-\left[ v_5^{(-3)}\, G_0\, v^{(0)}+
v^{(0)}\, G_0\, v_5^{(-3)} \right] \ , \\
v_5^{(-1)}=T_5^{(-1)}
&&-\left[ v_5^{(-3)}\, G_0\, v^{(0)}\, G_0\, v^{(0)}
+{\rm permutations} \right] \nonumber \\
&&-
\left[v_5^{(-2)}\, G_0\, v^{(0)}+v^{(0)}\, G_0\, v_5^{(-2)}\right]  \ ,\\
\label{eq:vg0}
v_5^{(0)}= T_5^{(0)}
&&-\Big[v_5^{(-3)}\, G_0\, v^{(0)}\, G_0\, v^{(0)}\, G_0\, v^{(0)}\,+\,{\rm permutations}\Big] \nonumber \\
&&-\left[v_5^{(-2)} \, G_0\, v^{(0)} \, G_0\, v^{(0)}+ {\rm permutations}\right]\nonumber\\
&&-\left[ v_5^{(-1)}\, G_0\, v^{(0)} +
v^{(0)}\, G_0\, v_5^{(-1)}\right] \nonumber \\
&&-\left[ v_5^{(-3)}\, G_0\, v^{(2)}+ v^{(2)}\, G_0\, v_5^{(-3)}\right] \ ,
\end{eqnarray}
\begin{eqnarray}
v_5^{(1)}= T_5^{(1)}&&-
\,\Big[v_5^{(-3)}\, G_0\, v^{(0)}\, G_0\, v^{(0)}\, G_0\, v^{(0)}\, G_0\,
v^{(0)}+ \,{\rm permutations}\Big]\nonumber\\
&&-\,\Big[v_5^{(-2)}\, G_0\, v^{(0)}\, G_0\, v^{(0)}\, G_0\, v^{(0)}+ \,{\rm permutations}\Big] \nonumber \\
&&-\,\Big[v_5^{(-1)}\, G_0\, v^{(0)}\, G_0\, v^{(0)}+ \,{\rm permutations}\Big] \nonumber \\
&&-\left[v_5^{(0)}\, G_0\, v^{(0)}+v^{(0)}\, G_0\, v_5^{(0)}\right] \nonumber \\
&&-
 \left[v_5^{(-3)}\, G_0\, v^{(2)}\, G_0\, v^{(0)}+{\rm permutations}\right]\nonumber\\
 &&-\left[v_5^{(-2)}\, G_0\, v^{(2)}+ v^{(2)} \, G_0\, v_5^{(-2)}\right] \nonumber\\
&&-
\left[v_5^{(-3)}\, G_0\, v^{(3)}+ v^{(3)} \, G_0\, v_5^{(-3)}\right] \ ,
\label{eq:vg1}
\end{eqnarray}
and a similar set of relations is obtained between $T^{(n)}$ and $v^{(n)}$, i.e., the amplitudes
and potentials in the presence of strong interactions only~\cite{Pastore11}.  These
relations allow us to construct $v^{(n)}$ and $v_5^{(n)}$ from $T^{(n)}$ and
$T_5^{(n)}$.
\begin{figure}[bth]
\includegraphics[width=2.5in]{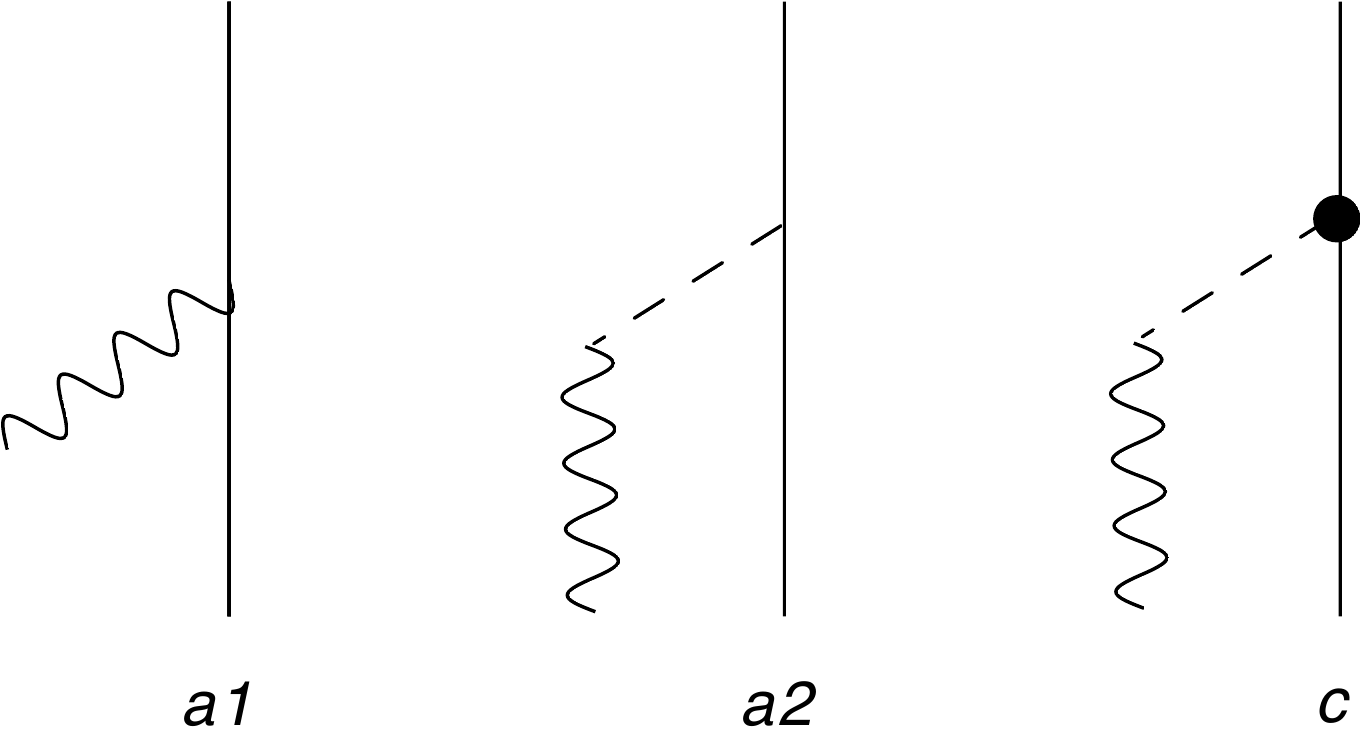}
\caption{Diagrams a1 and a2 contribute to the one-body axial current operator at order $Q^{(-3)}$.
Diagram c contributes to the one-body axial charge operator at order $Q^{(-2)}$.
Nucleons, pions, and axial fields are denoted by solid, dashed, and wavy lines,
respectively. Only a single time ordering
is shown for diagrams a2 and c.  The full dot in c is from the interaction
vertex $H_{\pi NN}^{(2)}$, see Appendix~\ref{app:vert}.  }
\label{fig:f1cc}
\end{figure}

The weak axial charge and current operators at leading order consist of the single-nucleon
contributions shown in Fig.~\ref{fig:f1cc} and are given by
\begin{eqnarray}
\rho_{5,a}^{(-2)}({\bf q})&=& -\frac{g_A}{4\,m} \, \tau_{1,a}\, {\bm\sigma}_1
\cdot\left( {\bf p}_1^\prime+{\bf p}_1\right) \,(2\pi)^3\delta({\bf p}_1+{\bf q}-{\bf p}_1^\prime) 
+ \left(1 \rightleftharpoons 2\right) \ ,
\label{eq:rr-2} \\
{\bf j}_{5,a}^{(-3)}({\bf q})&=& -\frac{g_A}{2} \, \tau_{1,a}
\left[ {\bm\sigma}_1-\frac{{\bf q}}{q^2+m_\pi^2}\,{\bm\sigma}_1\cdot{\bf q} \right] \,(2\pi)^3\delta({\bf p}_1+{\bf q}-{\bf p}_1^\prime)
+ \left( 1 \rightleftharpoons 2\right) \ ,
\label{eq:r-3}
\end{eqnarray}
where ${\bf q}$ is the momentum carried by the external field, and ${\bf p}_i$ and
${\bf p}_i^\prime$ are the initial and final momenta of nucleon $i$.
The counting $Q^{-3}$ of ${\bf j}_{5,a}$ (panel a1 in Fig.~\ref{fig:f1cc})
follows from the product of a factor $Q^0$ associated with the $ANN$
current vertex (recall that the $Q$ scaling of the external field is not counted),
and a factor $Q^{-3}$ due to the momentum-conserving $\delta$-function
$\delta\left({\bf p}_2^\prime-{\bf p}_2\right)$ implicit in disconnected terms of this type.
Evaluation of the pion-pole contribution (panel c), in which
the axial source couples directly to the pion which is then absorbed by
the nucleon, leads to the $\rho^{(-2)}_{5,a}$ expression in Eq.~(\ref{eq:rr-2}).
In this disconnected term, the counting $Q^{-2}$ accounts for the
$Q^{-3}$ factor due to $\delta\left({\bf p}_2^\prime-{\bf p}_2\right)$,
the factors $Q$ and $Q^2$ of the $\pi A$ and $\pi NN$ vertices, respectively, and
the factor $Q^{-2}$ from the pion field normalization and energy denominator
associated with the intermediate state.  A similar counting is applied to
panel a2 in Fig.~\ref{fig:f1cc} contributing to ${\bf j}_{5,a}$.

There is no direct coupling of the nucleon to $A^0_a$: the interaction
$- (g_A/2) \overline{N}\, {\bm \tau}\cdot {\bf A}_0\, \gamma^0\gamma^5  \,N$
in 
\[
-\overline{N}\, \Delta(2) \, N \ , 
\]
with $\Delta(2)$ as given by Eq.~(\ref{eq:d2}) occurs with the opposite
sign in 
\[
-\left(f_\pi/2\right) \, \left[  \overline{N} \, \Gamma_a^0(1) \, N\,
\left( G^{-1} \right)_{ab} F_{bc} \, A^0_c +{\rm h.c.}\right]\ ,
\]
with $\Gamma_a^0(1)$ as in the first term of Eq.~(\ref{eq:g1}) and
$\left(G\right)^{-1}_{ab}=F_{ab}=\delta_{ab}$ up to $\pi_a\pi_b$ or $m_\pi^2$
terms, and hence cancels out in Eq.~(\ref{eq:hhii}).  The single-nucleon
axial charge of the correct sign and strength follows from the sum of the two
time-ordered contributions of diagram c with the full dot representing
the interaction $\left(g_A/2f_\pi\right) \overline{N}\, {\bm\tau}\cdot{\bm \Pi}\,\gamma^0\gamma^5\, N$
from 
\[
-\left(1/2\right)\, \left[ \Pi_a \left( G^{-1} \right)_{ab} \, \overline{N} \, \Gamma_b^0(1) \, N+{\rm h.c.}\right]\ .
\]

Because of the different power counting of the leading order terms in the current
and charge operators, the strong interaction potentials needed to construct these
operators up to order $n=1$ include corrections up to $n=3$, i.e., $v^{(3)}$, in the
case of the current and up to $n=2$, i.e., $v^{(2)}$, in the case of the charge.
The leading order (LO) term $v^{(0)}$ consists of (static) one-pion-exchange (OPE)
and contact interactions, while the next-to-leading order (NLO) term $v^{(1)}$ (as already noted)
vanishes (see Ref.~\cite{Pastore11}).  The next-to-next-to leading order (N2LO) term
$v^{(2)}$ contains two-pion-exchange (TPE) and contact interactions, the latter
involving two gradients of the nucleon fields.  The $v^{(2)}$ term was originally derived
in Ref.~\cite{Ord95}, and is well known.  However,  at N2LO there is also a recoil correction
to the OPE, which we write as~\cite{Friar77}
\begin{equation}
\label{eq:3.15}
v^{(2)}_\pi(\nu)=v^{(0)}_\pi({\bf k})\, \frac{(1-\nu)\left[(E_1^\prime-E_1)^2+(E_2^\prime-E_2)^2\right]
-2\,\nu\, (E_1^\prime-E_1)(E_2^\prime-E_2)}{2\,\omega_k^2} \ ,
\end{equation}
where $v^{(0)}_\pi({\bf k})$ is the leading order OPE potential, defined as
\begin{equation}
v^{(0)}_\pi({\bf k})=-\frac{g^2_A}{4\, f_\pi^2}\, {\bm \tau}_1\cdot{\bm \tau}_2
\,\, {\bm \sigma}_1\cdot{\bf k} \,\,{\bm \sigma}_2 \cdot{\bf k}\,\frac{1}{\omega_k^2} \ ,
\end{equation}
$E_i$ (${\bf p}_i$) and $E_i^\prime$ (${\bf p}^\prime_i$) are
the initial and final energies (momenta) of nucleon $i$, and
${\bf k}={\bf p}_1-{\bf p}^\prime_1$.
There is an infinite class of corrections $v^{(2)}_\pi(\nu)$, labeled by
the parameter $\nu$, which, while equivalent on the energy shell ($E_1^\prime+E_2^\prime=E_1+E_2$)
and hence independent of $\nu$, are different off the energy shell.
Friar~\cite{Friar77} has in fact shown that these different off-the-energy-shell
extrapolations $v^{(2)}_\pi(\nu)$ are unitarily equivalent. 

The next-to-next-to-next-to-leading order (N3LO) term $v^{(3)}$ includes
interactions generated by vertices from the sub-leading Lagrangian
${\cal L}^{(2)}_{\pi N}$---these are of no interest for the present discussion---as well as
non-static corrections to the N2LO potentials $v^{(2)}$.  Among these,
the TPE correction $v^{(3)}_{2\pi}(\nu)$ (from direct and crossed box diagrams)  depends
on the specific choice made for $v^{(2)}_\pi(\nu)$.  However, as shown in Ref.~\cite{Pastore11},
the unitary equivalence remains valid also for $v^{(3)}_{2\pi}(\nu)$. In the derivation
of the axial current ${\bf j}^{(n)}_{5,a}$ at $n=1$ below, the choice
$\nu=0$ is made for $v^{(2)}_\pi(\nu)$ and $v^{(3)}_{2\pi}(\nu)$, specifically
Eq.~(\ref{eq:3.15}) above and Eq.~(19) of Ref.~\cite{Pastore11}.
The remaining non-static corrections in the potential $v^{(3)}$ are as given
in Eqs.~(B8), (B10), and (B12) of that work.  Clearly, different
choices in the off-the-energy-shell extrapolations of these potentials
will lead to different forms for (some of) the ${\bf j}^{(1)}_{5.a}(\nu)$ corrections
to the axial current.  As shown in the case of the electromagnetic
charge operator~\cite{Pastore11}, one would expect these different
forms to be unitarily equivalent.  However, this has not been verified
explicitly in the present case.
\section{Axial Charge}
\label{sec:axc}
The nuclear weak axial charge two-body operator can be written as
\begin{eqnarray}
\rho_{5,a}&=&\rho_{5,a}^{\mbox{OPE}}+\rho_{5,a}^{\mbox{TPE}}+\rho_{5,a}^{\mbox{CT}}\ , 
\end{eqnarray}
namely as a sum of terms due to one-pion exchange (OPE), two-pion
exchange (TPE), and contact contributions (CT).  We defer the
discussion of loop corrections to the OPE axial charge (and current) and of their
renormalization to a later section.  In the following, and in Sec.~\ref{sec:axj} as well,
contributions to the OPE and TPE (or MPE in Sec.~\ref{sec:axj}) operators are
labeled by the power counting superscript $(n)$.  While each individual
contribution is not explicitly identified as being OPE and TPE (or MPE),
this is obvious from the context.
 
Here and throughout this paper, we  adopt the following conventions.
The momenta ${\bf k}_i$ and ${\bf K}_i$ are defined as
\begin{equation}
{\bf K}_i =\left({\bf p}_i^\prime+{\bf p}_i\right)/2\ , \qquad {\bf k}_i={\bf p}_i^\prime-{\bf p}_i \ ,
\end{equation}
where ${\bf p}_i$ (${\bf p}_i^\prime$) is the initial (final) momentum of nucleon $i$.
A symmetrization $(1\rightleftharpoons 2)$
and an overall momentum-conserving $\delta$-function
$(2\pi)^3\delta({\bf k}_1+{\bf k}_2-{\bf q})$ are understood in
all terms listed below unless otherwise noted.

\subsection{Leading one-pion and two-pion exchange contributions}
\label{sec:oper}
\begin{figure}[bth]
\includegraphics[width=5in]{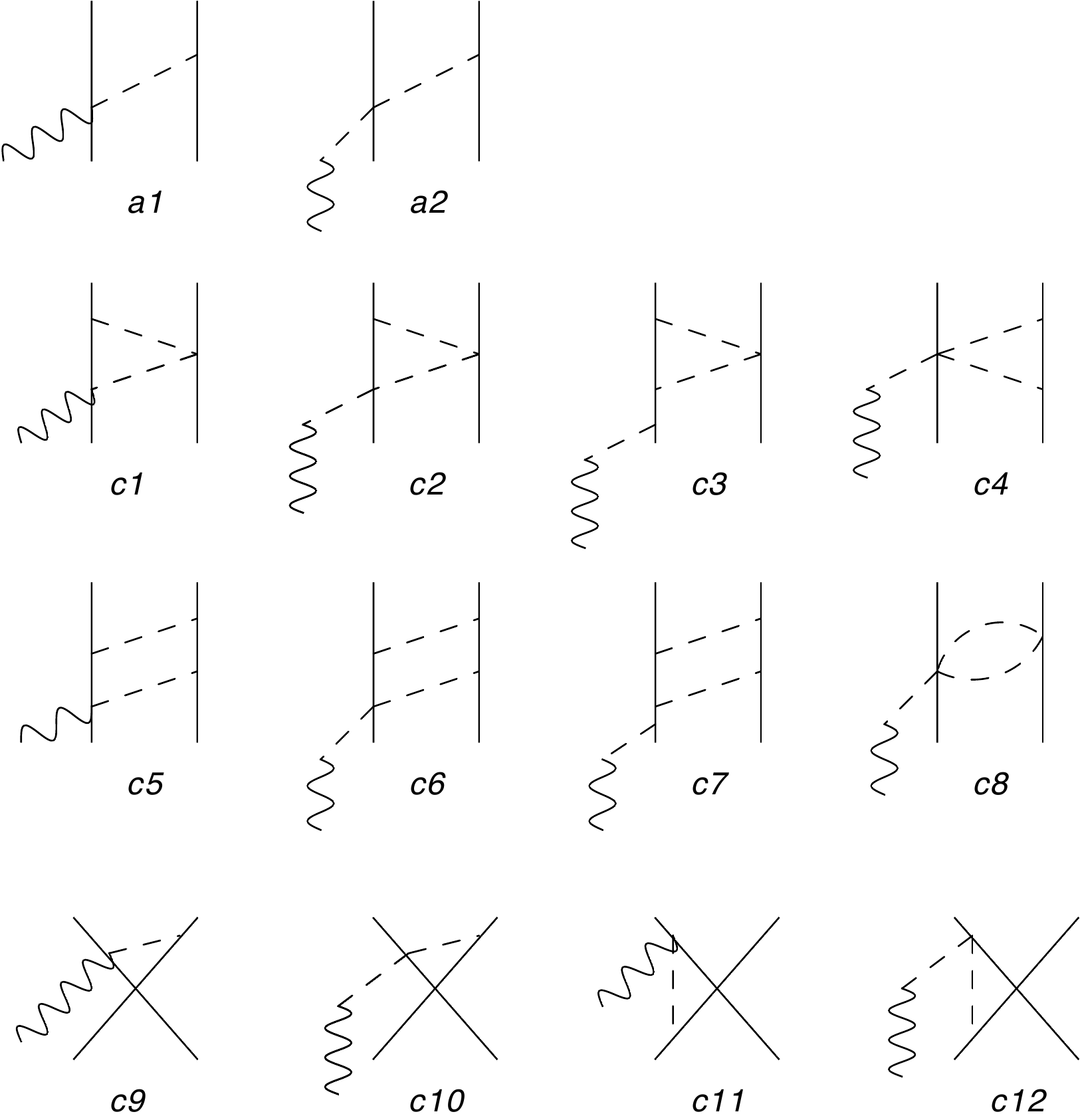}\\
\caption{Diagrams contributing to the OPE axial charge at leading order $Q^{-1}$ (panels a1 and a2),
and to the TPE axial charge operator at order $Q$.
Nucleons, pions, and axial fields are denoted by solid, dashed, and wavy lines,
respectively.  Only a single time ordering is shown for each topology. }
\label{fig:f5}
\end{figure}

Diagrams contributing to $\rho_{5,a}^{\mbox{OPE}}$ at leading order and to
$\rho_{5,a}^{\mbox{TPE}}$ are shown, respectively, in panels a1 and a2, and
panels c1-c12 of Fig.~\ref{fig:f5}.  The contributions of a1-a2, and c1-c2 and c5-c6
are given by
\begin{eqnarray}
\label{eq:rhoa1}
\rho_{5,a}^{(-1)}({\rm a1})&=&i\,\frac{g_A}{8f_\pi^2}\left({\bm \tau}_1\times{\bm \tau}_2\right)_a
{\bm \sigma}_2\cdot{\bf k}_2 \, \frac{1}{\omega_2^2}  \ , \\
\label{eq:rhoa2}
\rho_{5,a}^{(-1)}({\rm a2})&=&\rho_{5,a}^{(-1)}({\rm a1}) \ , \\
\rho^{(1)}_{5,a}({\rm c1+c2})&=&i\,\frac{ g_A}{16\, f_\pi^4}({\bm \tau}_1\times{\bm \tau}_2)_a\, {\bm \sigma}_1\cdot{\bf k}_2 \, I^{(0)}(k_2) \ ,\\
\rho^{(1)}_{5,a}({\rm c5}+{\rm c6})&=&i\,
\frac{\,g_A^3}{16\, f_\pi^4}\bigg[ 4\,{\bf \tau}_{1,a}\,  \sigma_{1i} \,
\left({\bm \sigma}_2\times{\bf k}_2\right)_j \, J^{(2)}_{ij}({\bf k}_2)\nonumber\\
&&+({\bm \tau}_1\times{\bm \tau}_2)_a\, \left[k_2^2\, J^{(0)}(k_2) -J^{(2)}(k_2)\right] 
{\bm \sigma}_1\cdot {\bf k}_2\bigg]\ , 
\end{eqnarray}
while those of c3-c4, c7-c8, and c9-c12 vanish. Corrections proportional to $1/m$ to topologies a1 and a2, due to non-static corrections to the energy denominators, that enter at order $Q$, vanish after summing over all time orderings.
Contributions, coming from ${\cal H}_{\pi NN}^{(2)}$ and ${\cal H}_{2\pi NN}^{(2)}$, to topologies a1 and a2, that enter at order $Q$, turn out to vanish.
The freedom
in the choice of pion field, parametrized by the parameter
$\alpha$ in Appendix~\ref{app:ls}, introduces an $\alpha$-dependence
in the interaction vertices with three or four pions, see Appendix~\ref{app:vert}.   
The contributions of diagrams c4 and c8, which include a $3\pi$ vertex, turn out to vanish
identically.   But in general this $\alpha$ dependence  must cancel out
exactly, as is indeed the case for the two-nucleon axial charge and
current operators obtained in this work.  The loop functions have
been defined as
\begin{eqnarray}
\label{eq:i0l}
I^{(0)}(k) &=& \int\frac{d{\bf p}}{(2\pi)^3} \, f(\omega_-,\omega_+) \ ,\\
J^{(0)}(k) &=&\int \frac{d{\bf p}}{(2\pi)^3}\,g(\omega_+,\omega_-) \ ,\\
J^{(2)}(k) &=&\int \frac{d{\bf p}}{(2\pi)^3}\, p^2 g(\omega_+,\omega_-)\ ,\\
J^{(2)}_{ij}({\bf k})&=&\int\frac{d{\bf p}}{(2\pi)^3} \, p_i p_j\, g(\omega_+,\omega_-)\ ,
\end{eqnarray}
with
\begin{eqnarray}
f(\omega_-,\omega_+)&=&\frac{1}{\omega_+\,\omega_-\, (\omega_+ + \omega_-)} \ , \\
g(\omega_-,\omega_+)&=& \frac{\omega_+^2+\omega_+\,\omega_-+\omega_-^2}
{\omega_+^3\,\omega_-^3(\omega_++\omega_-)} \ ,
\end{eqnarray}
and
\begin{equation}
\omega_\pm = \sqrt{({\bf p}\pm{\bf k})^2+ 4\, m_{\pi}^2} \ .
\end{equation}
They are evaluated in dimensional regularization~\cite{Pastore09}.  Insertion of
the finite parts of these loop functions leads to
\begin{eqnarray}
\rho^{(1)}_{5,a}({\rm c1+c2})&=&-i\frac{g_A}{128\,\pi^2\, f_\pi^4}({\bm \tau}_1\times{\bm \tau}_2)_a\, {\bm \sigma}_1
\cdot {\bf k}_2\, \frac{s_2}{k_2}\ln\left(\frac{s_2+k_2}{s_2-k_2}\right)\ , \\
\rho^{(1)}_{5,a}({\rm c5+c6})&=&-i\frac{g_A^3}{128\,\pi^2\, f_\pi^4}\bigg[4\, 
\tau_{1,a}\left({\bm \sigma}_1\times{\bm \sigma}_2\right)\cdot{\bf k}_2\,\frac{s_2}{k_2}\ln\frac{s_2+k_2}{s_2-k_2}\nonumber\\
&&-({\bm \tau}_1\times{\bm \tau}_2)_a\, {\bm \sigma}_1\cdot {\bf k}_2
\, \frac{k_2^2+2\, s_2^2}{k_2\, s_2}\ln\frac{s_2+k_2}{s_2-k_2}\bigg]\ ,
\end{eqnarray}
where
\begin{equation}
\label{eq:sdef}
s_j=\sqrt{4\,m_\pi^2+{\bf k}_j^2} \ .
\end{equation}
The divergent parts read
\begin{eqnarray}
\rho^{(1)}_{5,a}({\rm c1+c2})|_\infty&=&-i\,\frac{g_A}{128\pi^2\, f_\pi^4}({\bm \tau}_1\times{\bm \tau}_2)_a\, 
{\bm \sigma}_1\cdot{\bf k}_2\left(d_\epsilon-1\right)\, ,\\
\rho^{(1)}_{5,a}({\rm c5+c6})|_\infty&=&-i\,\frac{g_A^3}{32\pi^2 f_\pi^4}\bigg[\tau_{1,a}
\left({\bm \sigma}_1\times{\bm \sigma}_2\right)\cdot{\bf k}_2\left(d_\epsilon-\frac{1}{3}\right)\nonumber \\
&&- \frac{3}{4} \, ({\bm \tau}_1\times{\bm \tau}_2)_a\, {\bm \sigma}_1\cdot{\bf k}_2 \left(d_\epsilon+\frac{1}{3}\right)\bigg]\ ,\end{eqnarray}
with the constant $d_\epsilon$ defined as
\begin{equation}
\label{eq:deee}
d_\epsilon=-\frac{2}{\epsilon}+\gamma-\ln 4\pi+\ln \frac{m_\pi^2}{\mu^2} -1 \ ,
\end{equation}
where $\epsilon=3-d$ ($d$ is the number of dimensions), $\gamma$ is Euler's constant, and $\mu$ is a renormalization scale.
\subsection{Contact contributions}
\label{sec:rhocctt}
At order $Q^0$ there are no contact terms contributing to $\rho_{5,a}^{\mbox{CT}}$.  Those at
order $Q$ are given by (see Appendix~\ref{app:cta0})
\begin{equation}
\label{eq:rctct}
\rho_{5,a}^{\rm CT}=\sum_{i=1}^4 z_i \, O_i \ ,
\end{equation}
where the $z_i$ are (unknown) LECs and the operators $O_i$ with $i=1,\dots,4$, symmetrized
with respect to the exchange $1 \rightleftharpoons 2$, have been
defined as
\begin{eqnarray}
O_1&=&i\left({\bm\tau}_1\times{\bm\tau}_2\right)_a\,\left({\bm \sigma}_1\cdot{\bf k}_2-{\bm \sigma}_2\cdot{\bf k}_1\right) \ ,\\
\label{eq:4.22}
O_2&=&i\left({\bm\tau}_1\times{\bm\tau}_2\right)_a\,\left({\bm \sigma}_1\cdot{\bf k}_1-{\bm \sigma}_2\cdot{\bf k}_2\right) \ ,\\
O_3&=&i\,\left({\bm \sigma}_1\times{\bm \sigma}_2\right)\cdot\left(\tau_{1,a}\, {\bf k}_2-\tau_{2,a}\, {\bf k}_1\right)\ , \\
O_4&=&\left(\tau_{1,a}-\tau_{2,a}\right)\,\left({\bm \sigma}_1-{\bm \sigma}_2\right)\cdot\left({\bf K}_1+{\bf K}_2\right)\ .
\end{eqnarray}
We observe that the loop divergencies from c1-c2 and c5-c6 can be reabsorbed in the LECs $z_1$ and
$z_3$. 
\section{Axial Current}
\label{sec:axj}
\begin{figure}[bth]
\includegraphics[width=1.5in]{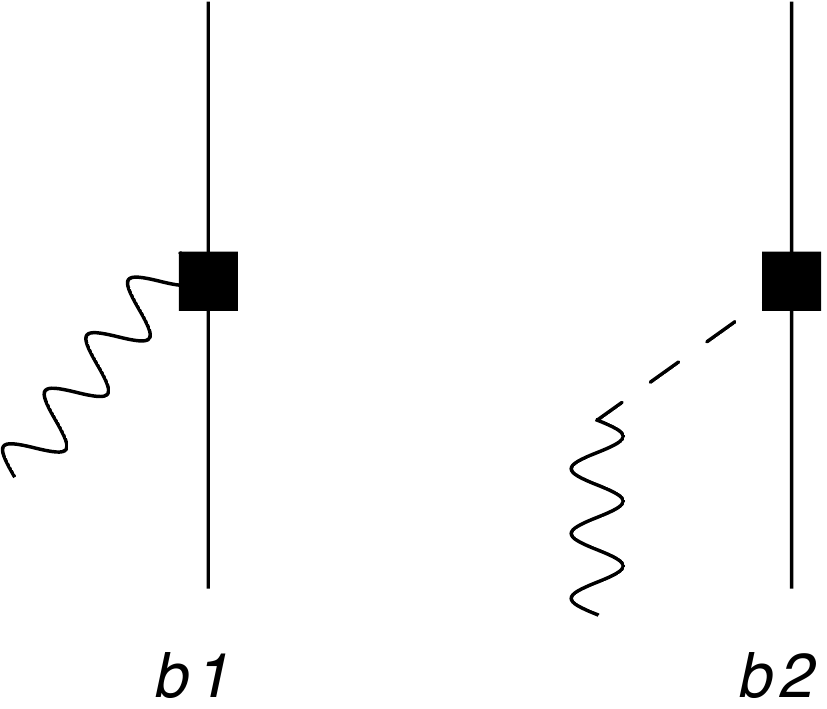}
\caption{Diagrams illustrating the relativistic corrections to the one-body axial current.
Nucleons, pions, and axial fields are denoted by solid, dashed, and wavy lines,
respectively.  Only a single time ordering is shown for diagram b2.  See text for further
explanations.}
\label{fig:f1rel}
\end{figure}

Before considering the two-body contributions, we note that
at order $Q^{-1}$ there are relativistic corrections to the
one-body current represented in diagrams b1
and b2 of Fig.~\ref{fig:f1rel}, given by
\begin{eqnarray}
\label{eq:ed0}
{\bf j}_{5,a}^{(-1)}({\rm b1})&=&\frac{g_A}{4\, m^2}\,\tau_{1,a} \bigg[ K_1^2\, {\bm \sigma}_1
+\frac{i}{2}\,{\bf k}_1 \times{\bf K}_1 - {\bm \sigma}_1\cdot{\bf K}_1\,\, {\bf K}_1
+\frac{1}{4}\,{\bm \sigma}_1\cdot{\bf k}_1 \, {\bf k}_1\bigg] \ ,\\
{\bf j}_{5,a}^{(-1)}({\rm b2})&=&-\frac{\bf q}{q^2+m_\pi^2} \left[\, {\bf q}\cdot{\bf j}_{5,a}^{(-1)}({\rm b1})
+\frac{g_A}{2\, m^2}\, \tau_{1,a}\,
{\bm\sigma}_1 \cdot{\bf K}_1\,\, {\bf k}_1\cdot{\bf K}_1\right]\ ,
\label{eq:ed1}
\end{eqnarray}
where b2 contains two contributions at
order $Q^{-1}$: one is from the $1/m^2$ terms originating from the
non-relativistic expansion of the $\pi NN$ interaction
$H^{(1)}_{\pi NN}$; the other is due to the $1/m$ terms in $H^{(2)}_{\pi NN}$
and the (leading) non-static corrections (proportional to $1/m$) to energy
denominators.  The b1 current has been found to give a significant
contribution to the cross section for proton weak capture on $^3$He of interest in
solar physics~\cite{Park03}.

As for the charge, the two-body current
is written as a sum of one-pion exchange (OPE), multi-pion exchange (MPE),
and contact (CT) terms (notation and conventions are as in Sec.~\ref{sec:axc}), 
\begin{eqnarray}
{\bf j}_{5,a}&=&{\bf j}_{5,a}^{\mbox{OPE}}+{\bf j}_{5,a}^{\mbox{MPE}}+{\bf j}_{5,a}^{\mbox{CT}}\ .
\end{eqnarray}
We discuss ${\bf j}^{\rm CT}_{5,a}$ here.
It is well known~\cite{Park03} that a single contact term occurs at order $Q^0$,
which we choose as
\begin{equation}
\label{eq:jctct}
{\bf j}_{5,a}^{\rm CT}=z_0 \big[\left({\bm \tau}_1\times{\bm \tau}_2\right)_a\, {\bm \sigma}_1\times{\bm \sigma}_2-\frac{{\bf q}}{q^2+m_\pi^2}\left({\bm \tau}_1\times{\bm \tau}_2\right)_a\, {\bf q}\cdot\left({\bm \sigma}_1\times{\bm \sigma}_2\right) \big]\ ,
\end{equation}
(where the second term of Eq.(~\ref{eq:jctct}) is the pion-pole contribution)
and none at order $Q$ (see Appendix~\ref{app:cta0}). This term is due to the interaction
$\left(\overline{N}\gamma^\mu\gamma_5\, u_\mu\, N\right) \overline{N} N$ and,
as first pointed by the authors of Ref.~\cite{Gar06}, the LEC $z_0$ is related to
the LEC $c_D$ (in standard notation) entering the three-nucleon potential at
leading order.  The two LECs $c_D$ and $c_E$ which fully characterize this
potential have been recently constrained by reproducing the empirical value
of the Gamow-Teller matrix element in
tritium $\beta$ decay and the binding energies of the trinucleons~\cite{Gazit09,Marcucci12}.
\subsection{Leading one-pion and multi-pion exchange and short-range contributions}
\begin{figure}[bth]
\includegraphics[width=6in]{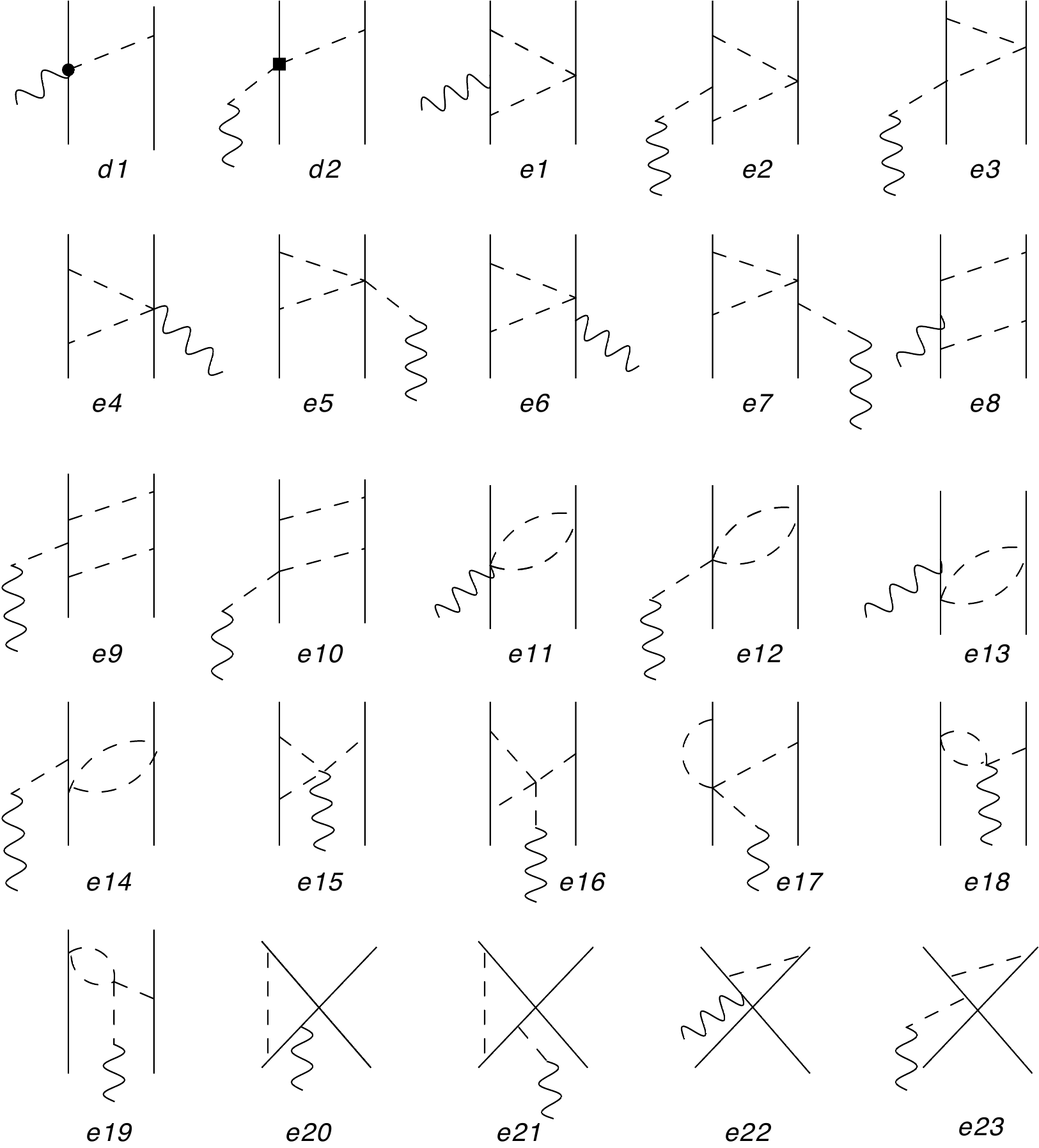}
\caption{Diagrams contributing to the OPE
axial current operator at order $Q^0$ and to the MPE axial current
at order $Q$.  Nucleons, pions, and axial
fields are denoted by solid, dashed, and wavy lines, respectively. 
Only a single time ordering is shown for each topology. }
\label{fig:f7}
\end{figure}

Leading contributions to ${\bf j}_{5,a}^{\mbox{OPE}}$ and
${\bf j}_{5,a}^{\mbox{MPE}}$ are shown, respectively, in panels d1-d2, and
panels e1-e23 of Fig.~\ref{fig:f7}.
There are no contributions at order $Q^{-1}$ from diagrams d1 and d2:
in d1 the interaction $H^{(1)}_{\pi NNA}$ contains no coupling to the
field ${\bf A}_a$, while in d2 the sum over the 6 time orderings, when leading order
vertices from $H^{(2)}_{\pi A}$, $H^{(1)}_{2\pi NN}$, and $H^{(1)}_{\pi NN}$
are considered, vanishes.  The first non-vanishing contributions enter at order $Q^{0}$, and read
\begin{eqnarray}
{\bf j}_{5,a}^{(0)}({\rm d1})\!&=&\!\frac{g_A}{2\, f_\pi^2} 
\left({\bm \tau}_1\times{\bm \tau}_2\right)_a
\left[i\frac{{\bf K}_1}{2\, m}-\frac{c_6+1}{4\, m}\, {\bm \sigma}_1\times {\bf q}
+\left(c_4+\frac{1}{4\,m}\right){\bm \sigma_1}\times{\bf k}_2\right]
{\bm\sigma}_2\cdot{\bf k}_2\, \frac{1}{\omega_2^2} \nonumber\\
&&+\,\frac{g_A}{f_\pi^2} \, c_3\, \tau_{2,a}\, \, {\bf k}_2\,\,{\bm\sigma}_2\cdot{\bf k}_2\,
\frac{1}{\omega_2^2}\, , \label{eq:ai}\\
{\bf j}_{5,a}^{(0)}({\rm d2})\!&=&\!-\frac{g_A}{2\, f_\pi^2}\frac{{\bf q}}{q^2+m_\pi^2}\left[\tau_{2,a}\left(4\, c_1\, m_\pi^2+2\, c_3\, {\bf q}\cdot {\bf k}_2 \right)-c_4\, \left({\bm \tau}_1\times{\bm \tau}_2\right)_a\, 
{\bm \sigma}_1\cdot\left({\bf q}\times{\bf k}_2\right) \right]
{\bm\sigma}_2\cdot{\bf k}_2 \frac{1}{\omega_2^2} \nonumber\\\,
&& -i\,\frac{g_A}{16 \, m\, f_\pi^2}\,\frac{\bf q}{q^2+m_\pi^2} \left({\bm \tau}_1\times{\bm \tau}_2\right)_a
\left(2\, {\bf K}_1 +i\,{\bm \sigma}_1\times{\bf k}_1\right)\cdot\left( {\bf q}+{\bf k}_2\right)\,
{\bm \sigma}_2 \cdot {\bf k}_2\, \frac{1}{\omega_2^2}\nonumber\\
&&+i\, \frac{g_A}{8\, m\, f_\pi^2}\,\frac{\bf q}{q^2+m_\pi^2} \left({\bm \tau}_1\times{\bm \tau}_2\right)_a 
\left({\bf K}_1\cdot{\bf k}_1+2\, {\bf K}_2\cdot{\bf k}_2\right){\bm\sigma}_2\cdot{\bf k}_2\, \frac{1}{\omega_2^2} \ .
\label{eq:aj}
\end{eqnarray}

For the diagrams contributing to ${\bf j}_{5,a}^{\mbox{MPE}}$
only a single time ordering is displayed for each topology.
It is understood that denominators involving pion energies in the
reducible topologies of diagrams e1-e2, e6-e7, e8-e10, e13-e14, e20-e21, and e22-e23
are expanded as in Eq.~(\ref{eq:deno}).  The resulting contributions
depend on the off-the-energy-shell prescription adopted
for the non-static corrections to the OPE, TPE, and OPE-contact potentials~\cite{Pastore11}.
Different prescriptions lead to different formal expressions for these
corrections as well as the accompanying weak axial current operators, which, however,
are expected to be related to each other
by unitary transformations.  This unitary equivalence was discussed in considerable
detail in Ref.~\cite{Pastore11}, where it was explicitly verified to hold in
the case of the electromagnetic charge operator.  Here we reiterate that the axial current
operators derived below are obtained by adopting the $\nu=0$ prescription for the non-static corrections
to the OPE, TPE, and OPE-contact potentials, as given in Eq.~(\ref{eq:3.15}) of the present
work and in Eqs.~(19), (B8), (B10), and (B12) of Ref.~\cite{Pastore11}.
We find that the contributions of diagrams e3, e6-e7, e11-e14, e18-e19, e22-e23 vanish,
while those of the remaining diagrams are given by
\begin{eqnarray}
{\bf j}_{5,a}^{(1)}({\rm e1})&=&-\frac{g_A^3}{16\, f_\pi^4}\,\tau_{2,a}\Big[R^{(2)}_{ij}({\bf k}_2)\, \sigma_{1j}
-{\bf k}_2 \, R^{(0)}(k_2) \, {\bm \sigma}_1\cdot{\bf k}_2\Big]\ ,\\
{\bf j}_{5,a}^{(1)}({\rm e2})&=&-\frac{{\bf q}}{q^2+m_\pi^2}\, {\bf q}\cdot{\bf j}_{5,a}^{(1)}({\rm e1}) \ ,\\
{\bf j}_{5,a}^{(1)}({\rm e4})&=&-\frac{g_A^3}{16\, f_\pi^4}\, \tau_{2,a}\, 
\Big[ k_1^2\, R^{(0)}(k_1)-R^{(2)}(k_1)\Big]{\bm \sigma}_2\ , \\
{\bf j}_{5,a}^{(1)}({\rm e5})&=&\frac{g_A^3}{32\, f_\pi^4}\frac{{\bf q}}{q^2+m_\pi^2}\bigg[\tau_{2,a}
\Big[k_1^2\,R^{(0)}(k_1)-R^{(2)}(k_1)\Big] 
\left[ \left(10\, \alpha -1\right){\bm \sigma}_2\cdot{\bf k}_2+{\bm \sigma}_2\cdot{\bf k}_1\right]\nonumber\\
&&-\left({\bm \tau}_1\times{\bm \tau}_2\right)_a\, R^{(2)}_{ij}({\bf k}_1)
\left({\bm \sigma}_1\times{\bf k}_1\right)_i\sigma_{2,j}
\bigg]\ ,\\
{\bf j}_{5,a}^{(1)}({\rm e8})&=&-\frac{g_A^5}{16\, f_\pi^4}\Bigg[\tau_{2,a} 
\bigg[ \left(  {\bm \sigma}_1\times {\bf k}_2\right)\times {\bf k}_2\Big[k_2^2 \, S^{(0)}(k_2) -S^{(2)}(k_2) \Big] 
\nonumber\\
&&+\Big[\,k_2^2 \, S^{(2)}(k_2) -S^{(4)}(k_2) \Big] {\bm \sigma}_1
-\Big[ k_2^2\, S^{(2)}_{ij}({\bf k}_2)-S^{(4)}_{ij}({\bf k}_2)\Big] \sigma_{1j}\bigg]\nonumber\\
&&+\,\frac{4}{3}\, \tau_{1,a}  \left(  {\bm \sigma}_2\times {\bf k}_2\right)\times {\bf k}_2\,S^{(2)}(k_2) \Bigg] \ ,\\
{\bf j}_{5,a}^{(1)}({\rm e9})&=&-\frac{{\bf q}}{q^2+m_\pi^2}\, {\bf q}\cdot{\bf j}_{5,a}^{(1)}({\rm e8})\ ,\\
{\bf j}_{5,a}^{(1)}({\rm e10})&=&\frac{g_A^3}{32\, f_\pi^4}\frac{{\bf q}}{q^2+m_\pi^2}\bigg[\left(2\, \tau_{2,a}-\tau_{1,a}\right)\Big
[k_2^2R^{(0)}(k_2)-R^{(2)}(k_2)\Big]{\bm \sigma}_1\cdot{\bf k}_2\nonumber\\
&&+\left({\bm \tau}_1\times{\bm\tau}_2\right)_aR^{(2)}_{ij}({\bf k}_1)\left({\bm \sigma}_2\times{\bf k}_2\right)_i\sigma_{1j}\bigg] \ ,\\
{\bf j}_{5,a}^{(1)}({\rm e15})&=&\frac{g_A^3}{32\, f_\pi^4}\bigg[\tau_{2,a}\left(10\, \alpha\, {\bf q}-3\, {\bf k}_1+{\bf k}_2\right)\Big[k_1^2R^{(0)}(k_1)-R^{(2)}(k_1)\Big]
\nonumber\\
&&-4\left({\bm \tau}_1\times{\bm \tau}_2\right)_a\, R_{ij}^{(2)}({\bf k}_1)
\left({\bm \sigma}_1\times{\bf k}_1\right)_j \bigg] \frac{{\bm \sigma}_2\cdot{\bf k}_2}{\omega_2^2}\ ,\\
{\bf j}_{5,a}^{(1)}({\rm e16})
&=&\frac{g_A^3}{64\, f_\pi^4}\, \tau_{2,a}\,\frac{{\bf q}}{q^2+m_\pi^2}\,
\bigg[2 \left(5\, m_\pi^2+2\, k_1^2+ k_2^2+ q^2\right)\left[k_1^2\,R^{(0)}(k_1)-R^{(2)}(k_1)\right]\nonumber\\
&&+\left[ k_1^4\, R^{(0)}(k_1)-R^{(4)}(k_1)\right] -20\,\alpha\left(q^2+k_2^2+2\,m_\pi^2\right)
\Big[k_1^2 \, R^{(0)}(k_1) -R^{(2)}(k_1) \Big] \nonumber\\
&&+ \,80\, \alpha\, J_{12}\bigg] 
\frac{{\bm \sigma}_2\cdot{\bf k}_2}{\omega_2^2} \nonumber\\
&&+\frac{g_A^3}{16\, f_\pi^4}
\left({\bm \tau}_1\times{\bm \tau}_2\right)_a\frac{{\bf q}}{q^2+m_\pi^2}
\, R^{(2)}_{ij}({\bf k}_1)\left({\bm \sigma}_1\times{\bf k}_1\right)_i
\left({\bf k}_2+{\bf q}\right)_j 
 \frac{{\bm \sigma}_2\cdot{\bf k}_2}{\omega_2^2}\ ,\\
{\bf j}_{5,a}^{(1)}({\rm e17})&=&\frac{g_A^3}{8\, f_\pi^4}\tau_{2,a}\frac{{\bf q}}{q^2+m_\pi^2}\left(1-10\,\alpha\right)J_{12}\frac{{\bm \sigma}_2\cdot{\bf k}_2}{\omega_2^2} \ , \\
{\bf j}_{5,a}^{(1)}({\rm e20})&=&\frac{\ g_A^3}{3\, f_\pi^2}\, C_T\, \tau_{1,a} \, J_{14}\, {\bm \sigma}_2\ ,\\
{\bf j}_{5,a}^{(1)}({\rm e21})&=&-\frac{\bf q}{q^2+m_\pi^2}\, {\bf q}\cdot{\bf j}_{5,a}^{(1)}({\rm e20})
 \ ,
\end{eqnarray}
where the constants $J_{mn}$ are as in Eq.~(\ref{eq:jmn}), and the loop functions
$R_{ij}^{(n)}$ have been defined as
\begin{eqnarray}
\!\!\!\!\!\!R^{(0)}(k)&=&\int\frac{d{\bf p}}{(2\pi)^3}\, \tilde{f}(\omega_+,\omega_-)\ ,\\
\!\!\!\!\!\!R^{(2)}(k)&=&\int\frac{d{\bf p}}{(2\pi)^3}\, p^2\,\tilde{f}(\omega_+,\omega_-)\ ,\\
\!\!\!\!\!\!R^{(2)}_{ij}({\bf k})&=&\int\frac{d{\bf p}}{(2\pi)^3}\, p_ip_j\,\tilde{f}(\omega_+,\omega_-)\ ,\\
\!\!\!\!\!\!R^{(4)}(k)&=&\int\frac{d{\bf p}}{(2\pi)^3}\, p^4\,\tilde{f}(\omega_+,\omega_-)\ ,\\
\!\!\!\!\!\!R^{(4)}_{ij}({\bf k})&=&\int\frac{d{\bf p}}{(2\pi)^3}\,p_ip_j\, p^2\,\tilde{f}(\omega_+,\omega_-)\ ,\\
\end{eqnarray}
with
\begin{equation}
\tilde{f}(\omega_+,\omega_-)=\frac{1}{\omega_+^2\, \omega_-^2}\ .\\
\end{equation}
The loop functions  $S_{ij}^{(n)}$ are defined similarly with $\tilde{f}(\omega_+,\omega_-)$ replaced by
\begin{equation}
\tilde{g}(\omega_+,\omega_-)=\frac{\omega_+^2+\omega_-^2}{\omega_+^4 \, \omega_-^4}=-\frac{1}{4}\frac{d}{d\, m_\pi^2}\tilde{f}(\omega_+,\omega_-)\ .
\label{eq:ggg}
\end{equation}
After dimensional regularization, we obtain
\begin{eqnarray}
\!\!\!\!\!\!R^{(0)}(k)&=&\frac{1}{16\pi}\int_0^1dz \,\frac{1}{M(k,z)}\ ,\\
\!\!\!\!\!\!R^{(2)}(k)&=&-\frac{3}{4\, \pi}  \int_0^1dz\bigg[M(k,z)-\frac{1}{12}\,\frac{(z-\overline{z})^2}{M(k,z)}\, k^2\bigg]\ ,\\
\!\!\!\!\!\!R^{(2)}_{ij}({\bf k})&=&-\frac{1}{4\, \pi} \int_0^1 dz\bigg[\delta_{ij} \,M(k,z)-\frac{1}{4}\,\frac{(z-\overline{z})^2}{M(k,z)}\, k_ik_j\bigg]\ ,\\
\!\!\!\!\!\!R^{(4)}(k)&=&\frac{5}{\pi}\int_0^1dz\bigg[M(k,z)^3-\frac{1}{2}\, (z-\overline{z})^2\, M(k,z)\,k^2 +\frac{1}{80}\frac{(z-\overline{z})^4}{M(k,z)}\, k^4\bigg]\ ,\\
\!\!\!\!\!\!R^{(4)}_{ij}({\bf k})&=&\frac{5}{3\, \pi} \int_0^1dz\bigg[\delta_{ij} \Big[M(k,z)^3-\frac{3 }{20}\,(z-\overline{z})^2\, M(k,z) \, k^2 \Big]\nonumber\\
&&- \frac{21}{20}\, \Big[ (z-\overline{z})^2 M(k,z)-\frac{1}{28}\,\frac{(z-\overline{z})^4}{M(k,z)}\, k^2\Big] k_ik_j\bigg]\ ,
\end{eqnarray}
where
\begin{equation}
\label{eq:mmfnt}
M(k,z)=\sqrt{z\overline{z}\,k^2+m_\pi^2}\ ,
\end{equation}
and
\begin{equation}
\label{eq:zzzz}
\overline{z}=1-z\ .
\end{equation}
The regularized $S^{(n)}_{ij}(k)$ loop functions easily follow from Eq.~(\ref{eq:ggg}).
Inserting these relations into the equations above, and noting that
the $\alpha$ dependence cancels out upon summing the contributions of
diagrams e5, e15, e16, and e17, we obtain the expressions reported in
Appendix~\ref{app:jforms}.  No divergencies occur in these loop corrections
at order $Q$, consistently with the fact that there are no contact terms in
the axial current at this order. 
Contributions coming from ${\cal L}_{\pi N}^{(3)}$, proportional to $d_i$'s, that enter through topologies d1 and d2 turn out to vanish.
\section{Renormalization of the one-pion exchange axial charge}
\label{sec:renor}
We now proceed to renormalize the order $Q$ loop corrections to the
OPE axial charge operator (as shown below, no renormalization at this
order is needed for the loop corrections to the OPE axial current).
We first construct the set of relevant counter-terms, and then carry out
the renormalization of the nucleon and pion masses, field rescaling factors
$Z_\pi$ and $Z_N$, pion decay constant $f_\pi$, nucleon axial
coupling constant $g_A$, and, lastly, loop corrections to the OPE axial charge.
We define
	\begin{eqnarray}
 \pi_a=\sqrt{Z_\pi}\,\pi_a^r\ ,\qquad
 N=\sqrt{Z_N}\, N^r\ ,
 \label{eq:zz}
	\end{eqnarray}
where ${\pi}_a^r$ and $N^r$ denote, respectively, the renormalized pion and nucleon fields,
and $Z_\pi$ and $Z_N$ are the corresponding field rescaling constants, assumed to have
the following expansions
\begin{eqnarray}
Z_{\pi}&=&1+\delta Z_\pi\ , \qquad \delta Z_\pi\sim Q^2\ ,\\
Z_N&=&1+\delta Z_N\ , \qquad \delta Z_N\sim Q^2\ .
\label{eq:zze}
\end{eqnarray}
We also define the physical pion mass $m^r_\pi$ and nucleon mass $m^r$ as
\begin{eqnarray}
m_\pi^{r\,2} &=& m_\pi^2+\delta m_\pi^2\ , \qquad \delta m_\pi^2\sim Q^4\ ,\\
m^r&=&m+\delta m\ , \qquad \delta m\sim Q^2\ .
\label{eq:mm}
\end{eqnarray}
As illustrated in Appendix~\ref{app:renora}, the total Lagrangian can be written as
\begin{eqnarray}
{\cal L} &=& \overline{N }^{\,r} \left( i\, \slashed{\partial} -m^r +\Gamma^{0\, \prime}_a \, \partial_0\pi^r_a
+\Lambda^{i\,\prime}_a \, \partial_i\pi^r_a  + \Delta^\prime \right) N^r\nonumber \\
&&+\frac{1}{2}\left(  \partial^0 \pi^r_a \, G^\prime_{ab} \, \partial_0 \pi^r_b
+\partial^i \pi^r_a \, \widetilde{G}^{\, \prime}_{ab} \, \partial_i \pi^r_b
-m_\pi^{r\,2}\, \pi^r_a\, H^\prime_{ab}\, \pi^r_b \right) -f_\pi\, 
A^\mu_a \, F^\prime_{ab}\,\partial_\mu \pi^r_b \nonumber\\
&&+\delta m\, \overline{N}^{\,r} N^r+\delta Z_N \,\overline{N}^{\,r}\left(i\gamma^\mu\partial_\mu-m^r\right)N^r
+\frac{\delta m_\pi^2}{2}\,\pi_a^r \pi_a^r \ ,
\label{eq:renL}
\end{eqnarray}
which is then expressed in terms of renormalized fields and masses, but
bare coupling constants $g_A$ and $f_\pi$ and LECs.  This Lagrangian
has essentially the same form as the bare one in Eq.~(\ref{eq:2.3})
(the primed quantities are defined in Appendix~\ref{app:renora}),
and leads to a similar interaction Hamiltonian as in Eq.~(\ref{eq:hhii}),
\begin{eqnarray}
\label{eq:hhiir}
{\cal H}_I&=& {\cal H}_I\big[{\rm Eq.~(\ref{eq:hhii})\,\,with\,\, primed\,\,quantities\,\,
and\,\, renormalized\,\, fields\,\, and \,\, masses}\, \big] \nonumber\\
&&-\delta m \, \overline{N}^{\,r}N^r-\delta Z_N\, \overline{N}^{\,r}\left(i\gamma^i\partial_i-m^r\right)N^r 
-\frac{\delta m_\pi^2}{2}\, \pi_a^r \pi_a^r\ .
\end{eqnarray} 
In addition to the vertices listed in Appendix~\ref{app:vert},
this Hamiltonian generates vertices corresponding to the set of counter-terms in
Eqs.~(\ref{eq:counter1})--(\ref{eq:counter6}), explicit expressions for which
follow from those in Appendix~\ref{app:vert}.
\subsection{Field and mass renormalization}
The determination of the scaling factors $Z_\pi=1+\delta Z_\pi$ and $Z_N=1+\delta Z_N$ for the pion and nucleon fields,
and the renormalization of the pion and nucleon masses have been discussed recently
and in considerable detail in Ref.~\cite{Viv14}.  We only quote the results here:
\begin{eqnarray}
\label{eq:deltam}
\delta m_\pi^2&=&2\, l_3\frac{m_\pi^{r\,4}}{f_\pi^2}+\frac{m_\pi^{r\,2}}{4f_\pi^2}J_{01}\ , \qquad
\delta Z_\pi= -2\frac{m_\pi^{r\,2}}{f_\pi^2}l_4+\frac{10\, \alpha-1}{2f_\pi^2}J_{01}\ , \\
\delta m&=&-4\, m_\pi^{r\,2}\, c_1-\frac{3\,g_A^2}{8\,f_\pi^2}\, J_{12} \ ,\qquad \delta Z_N=-\frac{3\,g_A^2}{8\,f_\pi^2}\, J_{13} \ ,
\end{eqnarray}
where the constants $J_{mn}$ are defined in Eq.~(\ref{eq:jmn}).  Only leading $Q^2$ corrections
are provided above, but for $\delta m$ which also includes the sub-leading term of order
$Q^3$ proportional to $J_{12}$.  The sign for $\delta m$ differs from that in Ref.~\cite{Viv14},
since there $m^r=m-\delta m$.
\subsection{Renormalization of the pion decay constant $f_\pi$}
\begin{figure}[bth]
\includegraphics[width=1.5in]{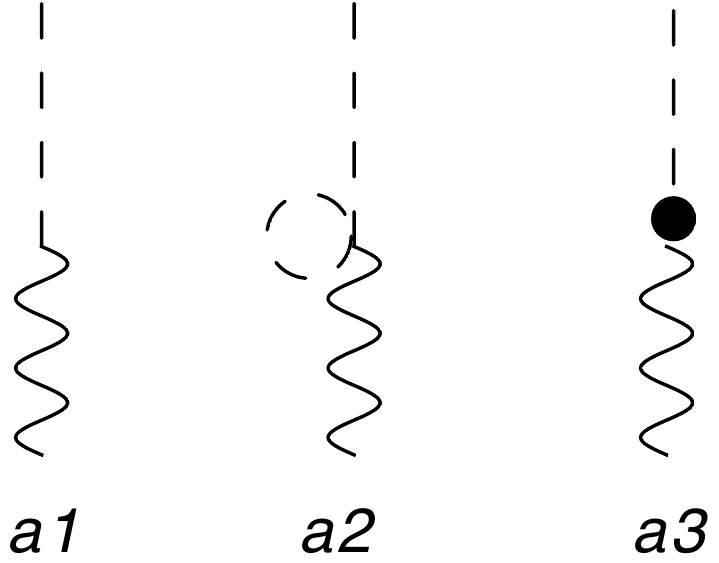}
\caption{Diagrams relevant for the renormalization of $f_\pi$.}
\label{fig:f}
\end{figure}

The relevant interaction Hamiltonians are
\begin{eqnarray}
\!\!\!\!H^{(2)\,\prime}_{\pi A}\!&=&\!f_\pi\int {\rm d}{\bf x} \left(  {\bf A}^i\cdot\partial_i{\bm \pi}^r+ {\bf A}^0\cdot{\bf \Pi}^r\right) \ ,\\
\!\!\!\! H_{3\pi A}^{(2)\,\prime}&=&\frac{1}{2f_\pi}\int {\rm d}{\bf x}\Big[2\,(1-2\,\alpha) {\bf A}^i\cdot{\bm \pi}^r
\,\, {\bm \pi}^r\cdot \partial_i{\bm \pi}^r -(2\, \alpha+1) {\bf A}^i\cdot\partial_i{\bm \pi}^r\,\, {\bm \pi}^r\cdot{\bm \pi}^r\nonumber\\
&&+ 2\left(\alpha-1/2\right)A^0_a \, \pi^r_b\, \Pi^r_a\,\pi^r_b
+ 2 \,\alpha \, A^0_a\,\left( \pi^r_a\, {\bm \pi}^r\cdot {\bm \Pi}^r
+{\bm \Pi}^r\cdot {\bm \pi}^r\, \pi^r_a \, \right) \Big]\ ,\\
\!\!\!\! H^{(4)\,\prime}_{\pi A}\!&=&\!\!\!\int {\rm d}{\bf x} \left[\frac{2\, m_\pi^{r\, 2}\, l_4}{f_\pi}\,
{\bf A}^i\cdot\partial_i{\bm \pi}^r\!
-\frac{\delta Z_\pi}{2} f_\pi \left(-{\bf A}^i\cdot\partial_i{\bm \pi}^r\!+\!{\bf A}^0\cdot{\bm \Pi}^r\right)\right]\ ,
\end{eqnarray}
where $H^{(2)\,\prime}_{\pi A}$ and $H_{3\pi A}^{(2)\,\prime}$ are the same as
in Eqs.~(\ref{eq:b40}) and~(\ref{eq:b44}) but in terms of renormalized pion field and mass,
while $H^{(4)\,\prime}_{\pi A}$ relative to Eq.~(\ref{eq:b41}) includes counter-terms.
The contributions illustrated in Fig.~\ref{fig:f} read
\begin{eqnarray}
{\rm a1}&=&-if_\pi\left({\bf k}\cdot{\bf A}_a-\omega A^0_a\right)\, ,\\
{\rm a2}&=&-\frac{i}{2f_\pi}\, J_{01} \Big[- \left(5\, \alpha+1/2\right) {\bf A}_a\cdot{\bf k}  
-(5\, \alpha-3/2) A^0_a\, \omega  \Big] \ , \\
{\rm a3}&=&-2\,i \frac{ m_\pi^{r\,2}\, l_4}{f_\pi}\,
{\bf k}\cdot {\bf A}_a+i\frac{\delta Z_\pi}{2}f_\pi\left(-{\bf k}\cdot{\bf A}_a-\omega A^0_a\right)\ .
\end{eqnarray}
We now require that the renormalized (physical) pion decay constant is equal to
\begin{eqnarray}
-if^r_\pi\left({\bf k}\cdot{\bf A}-\omega A^0_a\right)&=&{\rm a1+a2+a3}\ ,
\end{eqnarray}
implying
\begin{equation}
f^r_\pi=f_\pi\left(1+\frac{ m_\pi^{r\,2} \,l_4}{f_\pi^2}-\frac{J_{01}}{2\,f_\pi^2}\right) \ ,
\end{equation}
which to the order $Q^2$ of interest also gives
\begin{equation}
f_\pi=f^r_\pi \left(1-\frac{ m_\pi^{r\,2}\, l_4}{f^{r\, 2}_\pi}+\frac{J_{01}}{2\, f^{r\,2}_\pi}\right) \ .
\label{eq:f-fp}
\end{equation}
This result is in accord with that obtained in Ref.~\cite{Gasser84}.
\subsection{Renormalization of the $\pi N$ coupling constant $g_A/f_\pi$}
\begin{figure}[bth]
\includegraphics[width=2.5in]{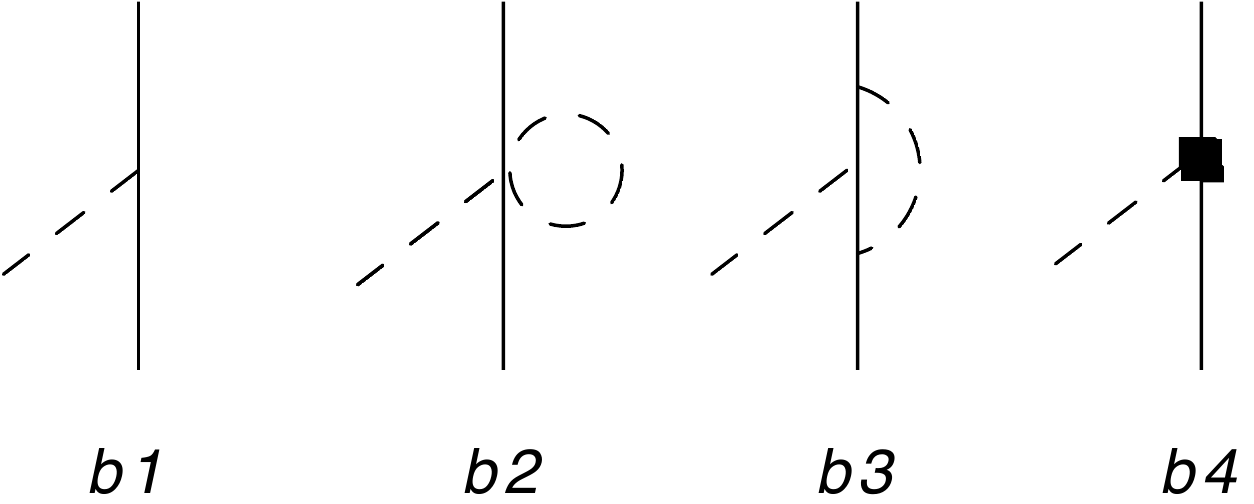}
\caption{Diagrams relevant for the renormalization of $g_A/f_\pi$.}
\label{fig:gf}
\end{figure}

Apart from $H^{(1)\,\prime}_{\pi NN}$ and $H_{3\pi NN}^{(1)\,\prime}$,
which are similar to those in Eqs.~(\ref{eq:b3}) and~(\ref{eq:b15}) (but again
expressed in terms of renormalized nucleon and pion fields, and pion mass),
the other interaction Hamiltonian needed is
\begin{equation}
\label{eq:6.19}
H^{(3)\, \prime}_{\pi NN}=
\int{\rm d}{\bf x}\,\left[\frac{m_\pi^{r\,2}}{f_\pi}(2\, d_{16}-d_{18})+\frac{g_A}{2f_\pi}
\left(\delta Z_N+\frac{\delta Z_\pi}{2}\right)\right]\overline{N}^r{\bm \tau}\cdot{\partial_i{\bm \pi}^r}\gamma^i\gamma^5 N^r \ .
\end{equation}
We find that the contributions of the diagrams in Fig.~\ref{fig:gf} are given by
\begin{eqnarray}
{\rm b1}&=&i\frac{g_A}{2f_\pi}\,{\bm \sigma}\cdot{\bf k}\,\tau_a\ ,\\
{\rm b2}&=&-i \frac{g_A}{8f_\pi^3}\left(10\,\alpha-1\right)J_{01}\,{\bm \sigma}\cdot{\bf k}\, \tau_a\, ,\\
{\rm b3}&=&i\frac{g_A^3}{48f_\pi^3}\, J_{13}\, {\bm \sigma}\cdot{\bf k}\, \tau_a\ , \\
{\rm b4}&=&i\bigg[\frac{m_\pi^{r\,2} }{f_\pi}\left(2\,d_{16}-d_{18}\right)-\frac{3\, g_A^3}{16 f_\pi^3}\,J_{13} \nonumber\\
&&+\frac{g_A}{4f_\pi^3}\left(-2\, m_\pi^{r\,2}\,l_4+\frac{10\,\alpha-1}{2}\,J_{01}\right)\bigg]
{\bm \sigma}\cdot{\bf k}\, \tau_a\ ,
\end{eqnarray}
and in terms of renormalized $g_A^r$ and $f_\pi^r$ it must be
\begin{eqnarray}
i \frac{g^r_A}{2\, f^r_\pi} \,{\bm \sigma}\cdot{\bf k}\, \tau_a={\rm b1+b2+b3+b4}\ ,
\end{eqnarray}
which leads to the following relation valid to order $Q^2$
\begin{eqnarray}
\frac{g^r_A}{f^r_\pi}&=&\frac{g_A}{f_\pi}\left[1+\frac{2\, m_\pi^{r\,2}}{g_A}\left(2\, d_{16}- d_{18}\right)
-\frac{g_A^2}{3f_\pi^2}J_{13}-\frac{m_\pi^{r\,2}\,l_4}{f_\pi^2}\right] \nonumber\\
&=&\frac{g_A}{f_\pi} \left(1+\frac{4\,m_\pi^{r\,2}}{g^r_A}\, d_{16}
-\frac{g^{r\,2}_A}{3f^{r\,2}_\pi}\, J _{13}-\frac{m_\pi^{r\,2}\, l_4}{f_\pi^{r\,2}}\right) 
\left(1-\frac{2\,m_\pi^{r\,2}}{g^r_A}\, d_{18}\right) \ ,
\label{eq:ratio}
\end{eqnarray}
where in the second line, in the terms of order $Q^2$, we have replaced $g_A$ and $f_\pi$
by their renormalized values $g_A^r$ and $f_\pi^r$,
which is correct at this order, and have isolated the Goldberger-Treiman discrepancy.
The above relations are in agreement with Eqs.~(102) and (103) of Ref.~\cite{Viv14}.

Since $f_\pi$ has already been renormalized, we can use Eq.~(\ref{eq:ratio}) to independently renormalize
$g_A$.  We find up to order $Q^2$
\begin{eqnarray}
g^r_A=g_A\left[1-\frac{1}{2\, f^{r\,2}_\pi}J_{01}-\frac{g_A^{r\,2}}{3\,f_\pi^{r\,2}}
J_{13}+\frac{4\,m_\pi^{r\,2}}{g^r_A}d_{16}\right]\left(1-\frac{2\,m_\pi^{r\,2}}{g^r_A}\, d_{18}\right)\ .
\label{eq:gfin}
\end{eqnarray}
As a check of this result, in the next subsection we provide a direct renormalization of $g_A$ by
considering the coupling of the axial field ${\bf A}_a$ to the nucleon. 
\subsection{Renormalization of the axial coupling constant $g_A$}
The relevant interaction Hamiltonians are $H_{ANN}^{(1)\,\prime}$ and $H_{2\pi NNA}^{(1)\,\prime}$
in Eqs.~(\ref{eq:b20}) and~(\ref{eq:b28}), and 
\begin{equation}
H_{ANN}^{(3)\,\prime}\!=\!-\!\!\int {\rm d}{\bf x}\overline{N}^r\bigg(2\,m_\pi^{r\,2} d_{16}\, {\bm \tau}\cdot{\bf A}_i\gamma^i\gamma_5+\delta Z_N\frac{g_A}{2}{\bm \tau}\cdot{\bf A}_i\gamma^i\gamma_5
+\frac{d_{22}}{2}{\bm \tau}\cdot\partial^j{\bf F}_{ij}\gamma^i\gamma_5\bigg)N^r\ .
\end{equation}
We consider a similar set of diagrams as in Fig.~\ref{fig:gf}, but for the
incoming pion line replaced by the external field.  Their contributions
are given by
\begin{eqnarray}
{\rm b1}&=&\frac{g_A}{2}\, \tau_a\, {\bm \sigma}\cdot {\bf A}_a\ ,\\
{\rm b2}&=&-\frac{g_A}{4f_\pi^2}\, J_{01}\,\tau_a\,{\bm \sigma}\cdot {\bf A}_a\ ,  \\
{\rm b3}&=&\frac{g_A^3}{48 f_\pi^2}\,J_{13}\,\tau_a\,{\bm \sigma}\cdot {\bf A}_a\ ,  \\
{\rm b4}&=& \left(\frac{g_A}{2}\,\delta Z_N+2\,m_\pi^{r\,2} \, d_{16}\right)\tau_a\, {\bm \sigma} \cdot {\bf A}_a
+\frac{d_{22}}{2}\tau_a\left({\bf q}\, {\bf q}\cdot{\bm \sigma}-q^2{\bm \sigma}\right)\cdot {\bf A}_a\ , 
\end{eqnarray}
and sum up to $\overline{g}_A^{\,r}\, {\bm \sigma}\, \tau_a/2$, with the renormalized
axial coupling constant (to order $Q^2$) obtained as
\begin{equation}
\overline{g}^{\,r}_A=g_A\left[1-\frac{1}{2\,f_\pi^{r\,2}}\, J_{01}-\frac{g_A^{r\,2}}{3\,f_\pi^{r\,2}}\,
J_{13}+\frac{4\, m_\pi^{r\,2}}{g^r_A}\, d_{16}\right]\ ,
\end{equation}
and $\overline{g}^{\,r}_A$, apart from the Goldberger-Treiman discrepancy, is in agreement
with Eq.~(\ref{eq:gfin}).  It is also in agreement with the results, to order $Q^2$, reported by
Schindler {\it et al.} in Ref.~\cite{Schindler07}.  The term proportional to $d_{22}$ quadratic in
${\bf q}$ contributes to the nucleon axial radius~\cite{Schindler07}.

\begin{figure}[bth]
\includegraphics[width=4in]{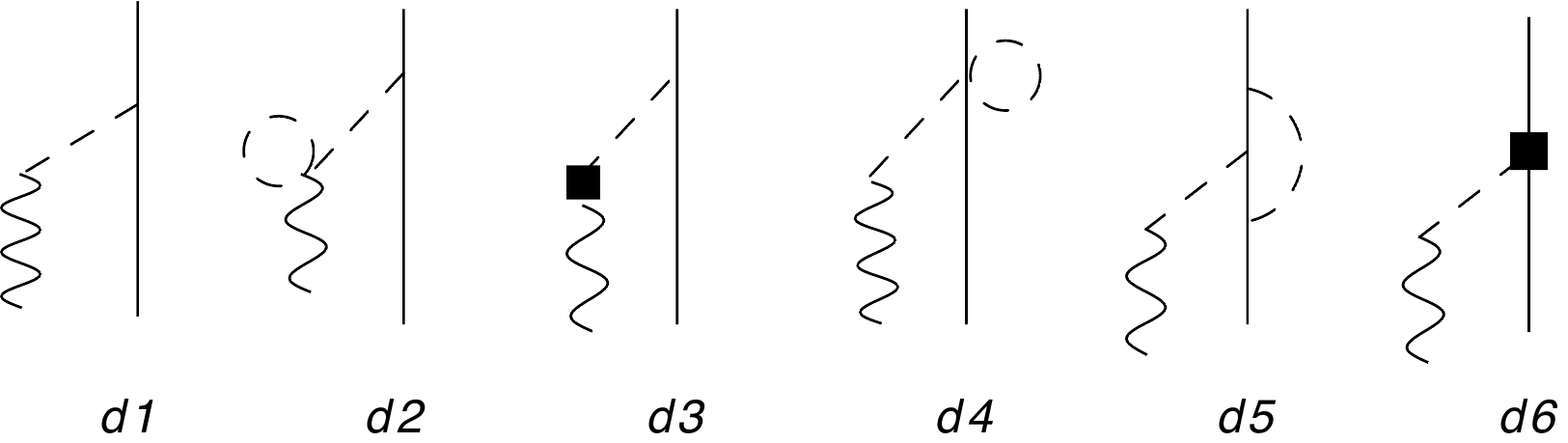}
\caption{Pion-pole diagrams.}
\label{fig:g}
\end{figure}
\subsection{Renormalization of pion-pole contributions}
We examine the pion-pole contributions illustrated in Fig.~\ref{fig:g}. We obtain
\begin{eqnarray}
{\rm d1}&=&-\frac{g_A}{2}\, {\bf A}_a\cdot {\bf q}\, \frac{{\bf q}\cdot {\bm \sigma}}{q^2+m_\pi^{r\,2}}\,  \tau_a\ ,\\
{\rm d2+d3}&=&\frac{g_A}{2f_\pi^2}\left(-m_\pi^{r\,2} \,l_4+\frac{J_{01}}{2}\right)
{\bf A}_a\cdot {\bf q}\, \frac{{\bf q}\cdot{\bm \sigma}}{q^2+m_\pi^{r\,2}}\tau_a\, ,\\
{\rm d4}&=& \frac{g_A}{8f_\pi^2}\left(10\, \alpha-1\right)J_{01}\,{\bf A}_a\cdot {\bf q}\, \frac{ {\bf q}\cdot{\bm \sigma}}{q^2+m_\pi^{r\,2}}\,\tau_a\ ,\\
{\rm d5}&=&-\frac{g_A^3}{48f_\pi^2}\,J_{13}\,{\bf A}_a\cdot {\bf q}\, \frac{{\bf q}\cdot{\bm{\sigma}}}{q^2+m_\pi^{r\,2}}\,\tau_a\ ,\\
{\rm d6}&=&\bigg[-m_\pi^{r\,2}\left(2\,d_{16}-d_{18}\right)+\frac{3\, g_A^3}{16 f_\pi^2}\,J_{13} \nonumber\\
&&-\frac{g_A}{4f_\pi^2}\left(-2\, m_\pi^{r\,2}\,l_4+\frac{10\,\alpha-1}{2}\,J_{01}\right)\bigg]
{\bf A}_a\cdot {\bf q}\, \frac{{\bf q}\cdot{\bm \sigma}}{q^2+m_\pi^{r\,2}}\, \tau_a\ .
\end{eqnarray}
Their sum reads
\begin{eqnarray}
{\rm d1+\dots+d6}&=&
-\frac{g_A}{2}\left[1-\frac{1}{2\, f^{r\,2}_\pi}J_{01}-\frac{g_A^{r\,2}}{3\,f_\pi^{r\,2}}
J_{13}+\frac{4\,m_\pi^{r\,2}}{g^r_A}d_{16}\right]\left(1-\frac{2\,m_\pi^{r\,2}}{g^r_A}\, d_{18}\right) \nonumber\\
&&\times\, {\bf A}_a\cdot {\bf q}\, \frac{{\bf q}\cdot{\bm \sigma}}{q^2+m_\pi^{r\,2}}\, \tau_a \ ,
\end{eqnarray}
and therefore the renormalized $g_A^r$ follows exactly as in Eq.~(\ref{eq:gfin}), including the
Goldberger-Treiman discrepancy.  The renormalized (single-nucleon) current is then given by
\begin{equation}
\label{eq:jjrre}
{\bf j}_{5,a}=-\frac{\overline{g}_A^{\, r}}{2} \, {\bm \sigma}\, \tau_a
+ \frac{g_A^{\, r}}{2}\, {\bf q}\, \frac{{\bf q}\cdot{\bm \sigma}}{q^2+m_\pi^{r\,2}}\, \tau_a \ ,
\end{equation}
and this current is conserved in the chiral limit ($m_\pi \rightarrow 0$), since in that limit
$g_A^r= \overline{g}^{\, r}_A$.
\subsection{Renormalization of OPE axial charge}
\begin{figure}[bth]
\includegraphics[width=6.0in]{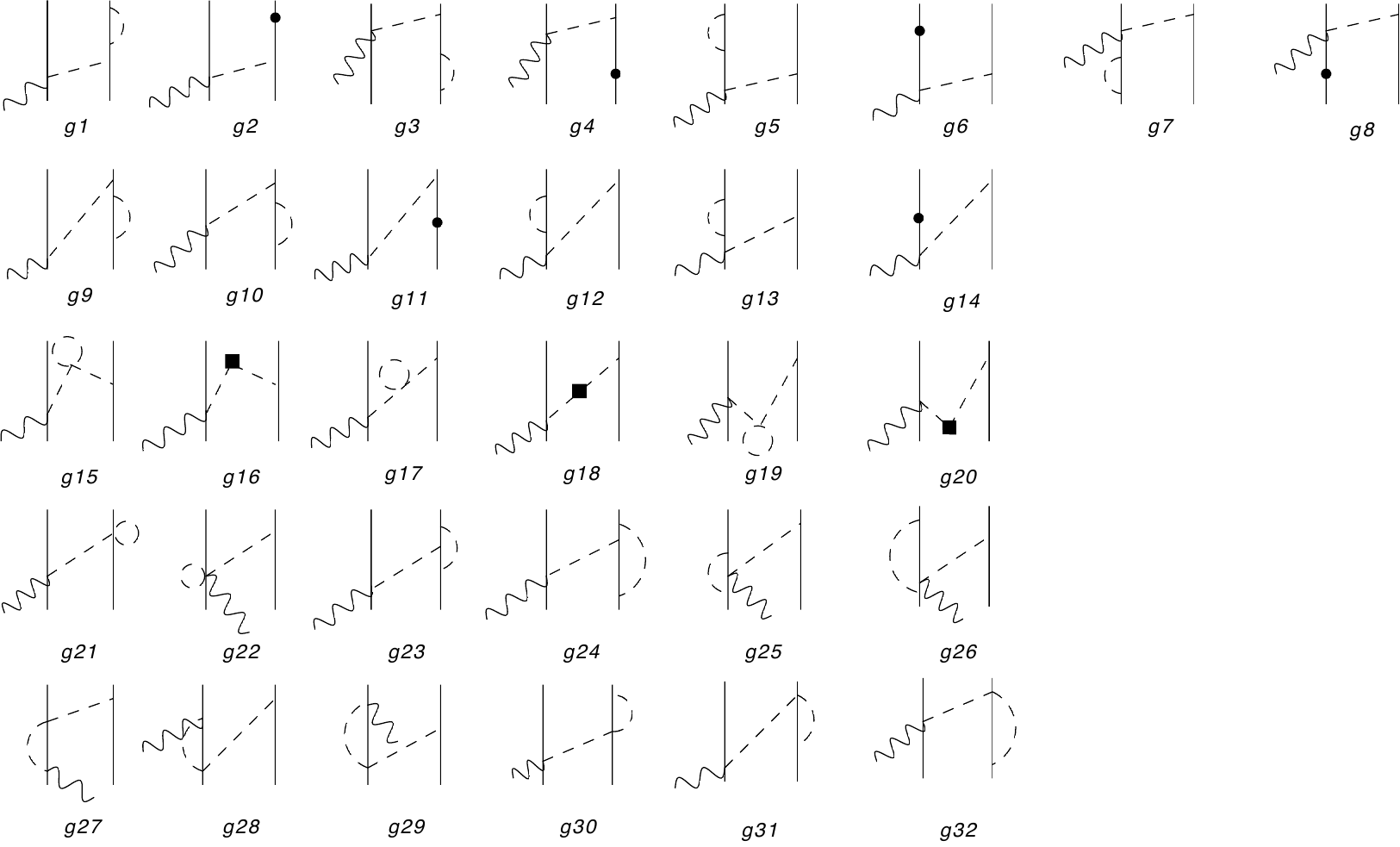}
\caption{Half of the possible time-ordered non-pole corrections to the OPE
axial charge at order $Q$.  Nucleons, pions, and axial
fields are denoted by solid, dashed, and wavy lines, respectively.  See text for further explanations.}
\label{fig:h}
\end{figure}

We begin by discussing the non-pion-pole contributions illustrated in Fig.~\ref{fig:h}.
In diagrams g2, g4, g6, g8, g11, and g14, the solid dot represents the interaction
$-\delta m-4\, m_\pi^{r\, 2}\, c_1$, where $\delta m$ is the nucleon mass counter-term.
The contributions associated with diagrams
g1-g2, g3-g4, g5-g6, g7-g8, g9-g11, and g12-g14 represent the renormalization
of nucleon external lines and, with the choice of $\delta m$ in Eq.~(\ref{eq:deltam}),
they are seen to vanish.

Next, the solid square in diagrams g16, g18, and g20 represents the interaction
\begin{eqnarray}
H_{2\pi}^{(4)\,\prime}&=&-\int{\rm d}{\bf x}\, \left(\frac{m_\pi^{r\,2}\, l_4}{f_\pi^2}+\frac{\delta Z_\pi}{2}\right)
\left( {\bm \Pi}^r\cdot{\bm \Pi}^r + \partial^i{\bm \pi}^r\cdot\partial_i{\bm \pi}^r \right) \nonumber\\
&&+\int{\rm d}{\bf x}\, \left[\frac{m_\pi^{r\,4}\left(l_3+l_4\right)}{f_\pi^2}+ \frac{m_\pi^{r\, 2}}{2}
\,\delta Z_\pi-\frac{\delta m_\pi^2}{2}\right]{\bm \pi}^r\cdot{\bm \pi}^r \ ,
\end{eqnarray}
with vertex (in the convention of Appendix~\ref{app:vert})
\begin{eqnarray}
\label{eq:b22r}
\langle  0\!\mid H_{2\pi}^{(4)\,\prime}\mid\!{\bf k}_1,a_1;{\bf k}_2,a_2\rangle&=&\delta_{a_1, a_2}
\bigg[ \left(\frac{2\, m_\pi^{r\,2}\,l_4}{f_\pi^2}+\delta Z_\pi\right) \left( \omega_1\omega_2-{\bf k}_1\cdot{\bf k}_2\right) \nonumber\\
&&+ \frac{2 \,m_\pi^{r\,4}\, (l_3+ l_4)}{f_\pi^2}+m_\pi^{r\,2}\, \delta Z_\pi -\delta m_\pi^2\bigg] \ ,
\end{eqnarray}
With $\delta Z_\pi$ and $\delta m_\pi^2$ as given in Eq.~(\ref{eq:deltam}),
the contributions of diagrams g15-g20 cancel out.

The remaining loop contributions in diagrams g21-g29 are given by
\begin{eqnarray}
\rho_{5,a}^{(1)}({\rm g21})&=&\rho_{5,a}^{(-1)}({\rm a1})\frac{1}{4\, f_\pi^{2}}(1-10\,\alpha)J_{01}\ ,\\
\rho_{5,a}^{(1)}({\rm g22})&=&\rho_{5,a}^{(-1)}({\rm a1})\frac{5}{8\, f_\pi^{2}}(1-4\alpha)J_{01}\ ,\\
\rho_{5,a}^{(1)}({\rm g23+g24})&=&\rho_{5,a}^{(-1)}({\rm a1})\frac{g_A^{2}}{24\, f_\pi^{2}}J_{13}\ ,\\
\rho_{5,a}^{(1)}({\rm g25+g26})&=&-\rho_{5,a}^{(-1)}({\rm a1})\frac{g_A^{2}}{8\, f_\pi^{2}}J_{13}\ ,\\
\rho_{5,a}^{(1)}({\rm g27+g28+g29})&=&\rho_{5,a}^{(-1)}({\rm a1})\frac{1}{4\, f_\pi^{2}}J_{01}\ ,
\end{eqnarray}
while those in diagrams g30-g32 vanish identically.  Here $\rho_{5,a}^{(-1)}(a1)$ is
defined as in Eq.~(\ref{eq:rhoa1}).

Finally, one needs to include the contributions due to the interactions
$H_{\pi NN}^{(3)\, \prime}$ in Eq.~(\ref{eq:6.19}) and 
\begin{equation}
H_{\pi NN A}^{(3)\, \prime}=-\left(\delta Z_N+\delta Z_\pi/2\right)\frac{1}{4\,f_\pi}\int d{\bf x}\, \overline{N}^{\, r}{\bf A}_0
\cdot\left({\bm \tau}\times{\bm \pi}^r\right)\gamma_0 N^r
\label{eq:pnnaren}\ ,
\end{equation}
in the OPE axial charge, which simply lead to the correction of order $Q$
\begin{equation}
\left [2\, \delta Z_N+\delta Z_\pi +\frac{2\, m_\pi^{r\, 2}}{g_A} \left( 2\, d_{16}-d_{18}\right) \right] \rho_{5,a}^{(-1)}({\rm a1}) \ .
\end{equation} 
Thus, the sum of the order $Q$ corrections to the axial charge from non-pole contributions, denoted as 
$\rho^{(1)}_{5,a}({\rm npp})$, reads
\begin{eqnarray}
\rho^{(1)}_{5,a}({\rm npp})&=&\rho_{5,a}^{(-1)}({\rm a1})\bigg[\frac{1}{f^2_\pi}\left(\frac{9}{8}-5\,\alpha\right) J_{01}
-\frac{g_A^2}{12\, f_\pi^2}J_{13} +2\, \delta Z_N \nonumber\\
&&+\,\delta Z_\pi +\frac{2\, m_\pi^{r\, 2}}{g_A} \left( 2\, d_{16}-d_{18}\right)\bigg] \ ,
\end{eqnarray}
which, which after insertion of $\delta Z_N$ and $\delta Z_\pi$,
is expressed as
\begin{eqnarray}
\rho_{5,a}^{(1)}({\rm npp})&=&i\frac{g^r_A}{8\,f_\pi^{r\,2}}\left({\bm \tau}_1\times{\bm \tau}_2\right)_a
{\bm \sigma}_2\cdot{\bf k}_2\frac{1}{\omega_2^2}
\bigg[ \frac{5}{8\, f_\pi^{r\, 2}}J_{01}-\frac{5\,g_A^{r\, 2}}{6\, f_\pi^{r\,2}}J_{13} \nonumber \\
&&-\frac{2\, m_\pi^{r\, 2}}{f_\pi^{r\, 2}}\, l_4 +\frac{2\, m_\pi^{r\, 2}}{g^r_A}\, \left(2\, d_{16}-d_{18}\right) \bigg]\ ,
\label{eq:4}
\end{eqnarray}
where the bare $g_A$ and $f_\pi$ have been replaced by their respective
renormalized values---this replacement is correct to the order of interest here.
The complete non-pole axial charge, denoted as $\rho_{5,a}^{\rm OPE}({\rm npp})$ below,
results from the sum of the leading-order contribution in Eq.~(\ref{eq:rhoa1})
with the ratio $g_A/f_\pi^2$
replaced by its renormalized value
\begin{equation}
\frac{g_A}{f_\pi^2}=\frac{g^r_A}{f_\pi^{r\,2}} \bigg[ 1 -\frac{1}{2\,f_\pi^{r\, 2}}J_{01}+\frac{g_A^{r\,2}}{3\,f_\pi^{r\,2}}
J_{13}+\frac{ 2\, m_\pi^{r\, 2} }{f_\pi^{r2}}\, l_4-\frac{2\,m_\pi^{r\, 2}}{g^r_A}\left(2\,d_{16}-d_{18}\right) \bigg] \ ,
\end{equation}
and the contribution $\rho_{5,a}^{(1)}({\rm npp})$.  We obtain
\begin{equation}
\rho_{5,a}^{\rm OPE}({\rm npp})=i\frac{g^r_A}{8\,f_\pi^{r\,2}}\left({\bm \tau}_1\times{\bm \tau}_2\right)_a
{\bm \sigma}_2\cdot{\bf k}_2\frac{1}{\omega_2^2}
\bigg(1+ \frac{1}{8\, f_\pi^{r\, 2}}\,J_{01}-\frac{g_A^{r\, 2}}{2\, f_\pi^{r\,2}}\, J_{13}\bigg) \ .
\label{eq:6.52}
\end{equation}
\begin{figure}[bth]
\includegraphics[width=4.0in]{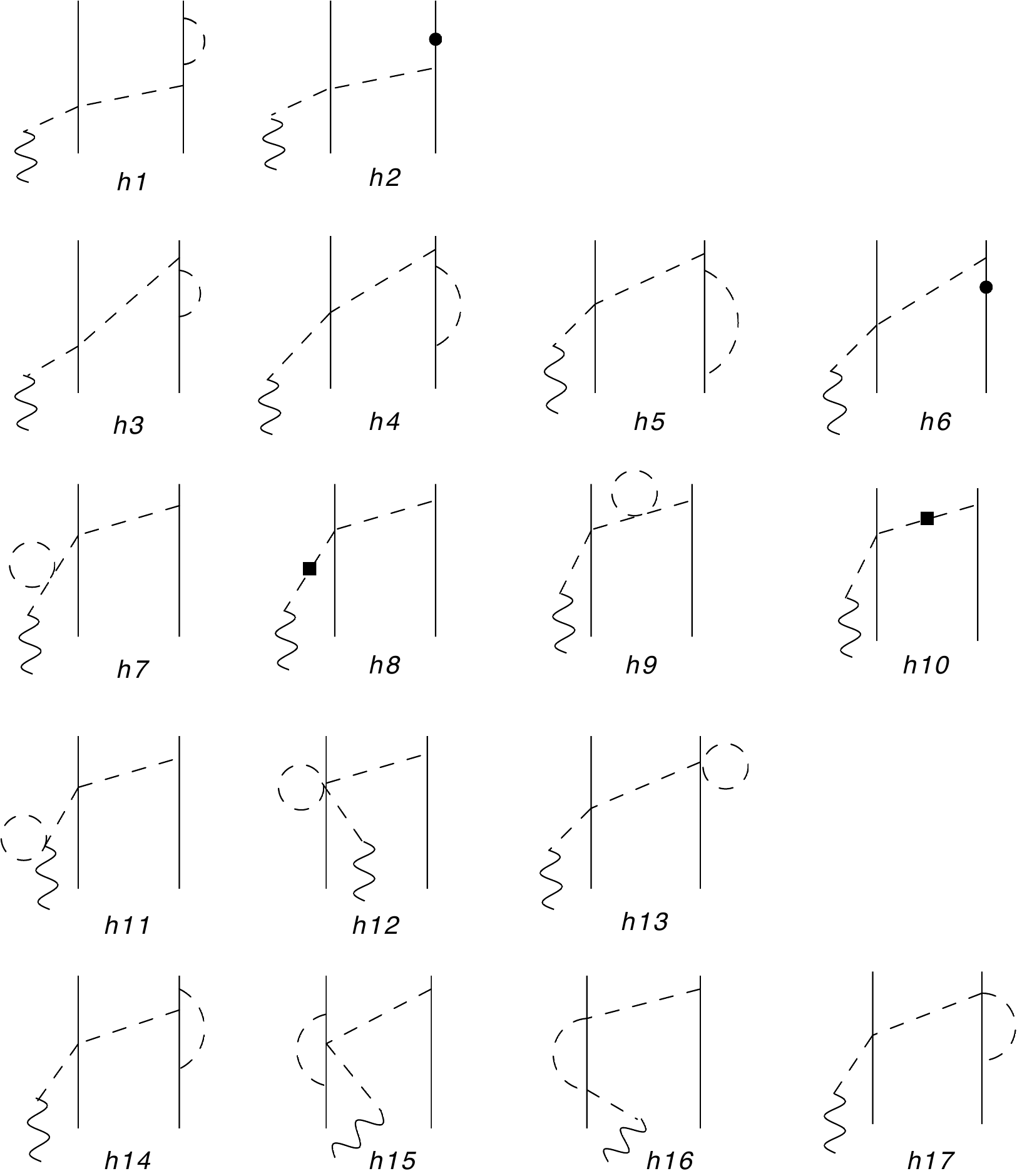}
\caption{Representative diagrams for each of the relevant classes contributing
to pole corrections to the OPE axial charge at order $Q$. 
Nucleons, pions, and axial fields are denoted by solid, dashed, and wavy lines, respectively.
More than a single time ordering is shown for some of the diagrams.}
\label{fig:hpp}
\end{figure}

The diagrams describing the pole corrections are illustrated in
Fig.~\ref{fig:hpp} (only representative diagrams for each of the relevant classes are drawn for
brevity), and are similar to those in Fig.~\ref{fig:h}.  A slightly more
complicated analysis along the lines illustrated above
leads to a pole OPE axial charge, denoted $\rho_{5,a}^{(1)}({\rm pp})$, given by
\begin{equation}
\rho_{5,a}^{\rm OPE}({\rm pp})=i\frac{g^r_A}{8\,f_\pi^{r\,2}}\left({\bm \tau}_1\times{\bm \tau}_2\right)_a
{\bm \sigma}_2\cdot{\bf k}_2\frac{1}{\omega_2^2}
\bigg(1- \frac{1}{8\, f_\pi^{r\, 2}}\,J_{01}-\frac{g_A^{r\, 2}}{2\, f_\pi^{r\,2}}\, J_{13}\bigg)\ .
\label{eq:6.53}
\end{equation}
The sum of the npp and pp contributions evaluated in dimensional regularization is
\begin{eqnarray}
\rho_{5,a}^{\rm OPE}({\rm npp}+{\rm pp})&=&i\frac{g^r_A}{8\,f_\pi^{r\,2}}\left({\bm \tau}_1\times{\bm \tau}_2\right)_a
{\bm \sigma}_2\cdot{\bf k}_2\frac{1}{\omega_2^2}
\bigg(1-\frac{g_A^{r\, 2}}{\, f_\pi^{r\,2}}\, J_{13}\bigg) \nonumber\\	
&=&\rho_{5,a}^{(-1)}({\rm a1})	\bigg[1-\frac{3\, m_\pi^{r\, 2}}{8\,\pi^2\,f_\pi^{r\, 2}} g_A^{r\, 2}\, \left(d_{\epsilon}-\frac{1}{3}\right)\bigg]\, .
\end{eqnarray}
\begin{figure}[bth]
\includegraphics[width=6in]{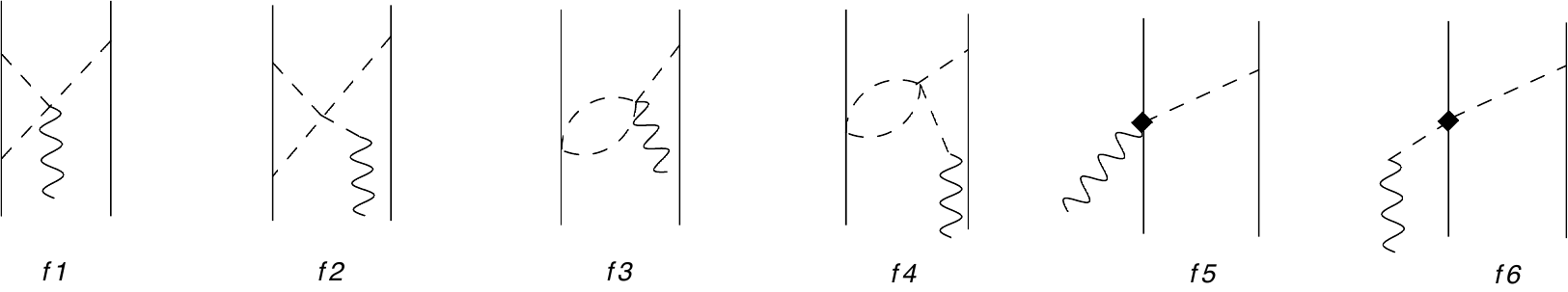}
\caption{Additional loop and tree-level corrections of order $Q$ to the OPE axial charge.
Nucleons, pions, and axial fields are denoted by solid, dashed, and wavy lines, respectively.  Only a single
time ordering is shown for each topology.  See text for further explanations.}
\label{fig:j}
\end{figure}
There are additional loop corrections to the OPE axial charge, see Fig.~\ref{fig:j}.  Their contributions
are obtained as
\begin{eqnarray}
\rho^{(1)}_{5,a}({\rm f1+f2})&=&-\frac{g_A^{r\,2}}{2\,f_\pi^{r\,2}}\, \rho_{5,a}^{(-1)}({\rm a1})
\left[ k_1^2\, I^{(0)}(k_1)-I^{(2)}(k_1)\right] \ ,\\
\rho^{(1)}_{5,a}({\rm f3+f4})&=&-\frac{1}{8\, f_\pi^{r\,2}}\, \rho_{5,a}^{(-1)}(a1)\, L(k_1)\ ,
\end{eqnarray}
where $\rho^{(-1)}_{5,a}({\rm a1})$ is again defined as in Eq.~(\ref{eq:rhoa1}),
except that $g_A$ and $f_\pi$ are replaced by their renormalized
values $g^r_A$ and $f^r_\pi$.  The loop function $I^{(0)}(k)$ has been defined in Eq.~(\ref{eq:i0l}), while $I^{(2)}(k)$ and
$L(k)$ read
\begin{eqnarray}
I^{(2)}(k)&=&  \int \frac{d{\bf p}}{(2\pi)^3}\, \, p^2\, f(\omega_-,\omega_+)\ ,\\
L(k) &=&\int\frac{d{\bf p}}{(2\pi)^3} (\omega_+-\omega_-)^2\, f(\omega_-,\omega_+)\ .
\end{eqnarray}
Evaluation in dimensional regularization leads to
\begin{eqnarray}
\label{eq:6.58}
\rho^{(1)}_{5,a}({\rm f1+f2})&=&\rho_{5,a}^{(-1)}({\rm a1})\frac{g_A^{r\,2}}{48\pi^2\, f_\pi^{r\,2}}\bigg[\frac{s_1}{k_1}\ln\left(\frac{s_1+k_1}{s_1-k_1}\right)\left(5\, k_1^2+8\, m_\pi^{r\,2}\right)\nonumber\\
&&\,+\, k_1^2\left(5\,
d_\epsilon-\frac{13}{3}\right)+18\, m_\pi^{r\,2}\left(d_\epsilon-\frac{2}{9}\right)\bigg]\ ,\\
\label{eq:6.59}
\rho^{(1)}_{5,a}({\rm f3+f4})&=&\rho_{5,a}^{(-1)}({\rm a1})\frac{1}{48\pi^2\,f_\pi^{r\,2}}\bigg[\frac{s_1^3}{k_1}\ln\frac{s_1+k_1}{s_1-k_1}-8\, m_\pi^{r\,2}+
k_1^2\left(d_\epsilon-\frac{5}{3}\right)\bigg]\ ,
\end{eqnarray}
with $d_\epsilon$ as given in Eq.~(\ref{eq:deee}).
We also need to account for tree-level contributions of order $Q$ originating
from the vertices $2\pi NN$ and $NN\pi A^0$ in Eqs.~(\ref{eq:b14}) and~(\ref{eq:b24}),
denoted by the solid diamonds in Fig.~\ref{fig:j}.  They can be written as
\begin{eqnarray}
\label{eq:6.70}
\rho^{(1)}_{5,a}({\rm f5}+{\rm f6})&=&2\,\rho_{5,a}^{(-1)}({\rm a1})\left(
\widetilde{d}_1\, k_1^2+\widetilde{d}_2\,k_2^2
+\widetilde{d}_3\, q^2 + \widetilde{d}_4 \,m_\pi^{r\,2} \right) \nonumber\\
&&+\,i\frac{g^r_A}{2\,f_\pi^{r\,2}}\, \widetilde{d}_5\,\tau_{2,a}\,
{\bm \sigma}_1 \cdot ({\bf q}\times{\bf k}_2)\,\, {\bm \sigma}_2\cdot{\bf k}_2\,\frac{1}{\omega_2^2} \ ,
\end{eqnarray}
where we have introduced the following combinations of LECs
\begin{eqnarray}
\widetilde{d}_1&=& 2\,d_2+d_6\ ,\\
\widetilde{d}_2&=&4\, d_1+2\, d_2+4\, d_3-d_6\ ,\\
\widetilde{d}_3&=& -2\,d_2+d_6\, ,\\
\widetilde{d}_4&=&4\,d_1+4\,d_2+4\,d_3+8\,d_5  \ ,\\
\widetilde{d}_5&=&d_{15}+2\,d_{23}\, .
\end{eqnarray}
The divergent parts of the $d_i$'s
(and hence $\widetilde{d}_i$'s) have been identified in the heavy-baryon
formalism, without considering any specific process, with the background-field
and heat-kernel methods, see Ref.~\cite{Gasser02} and references therein.
We report below the expressions for these divergent parts from Table 4 of that work:
\begin{eqnarray}
d_i&=&\frac{\beta_i}{f_\pi^2}\lambda+d_i^r(\mu)\  ,
\end{eqnarray}
where, in the conventions adopted in the present work,
\begin{eqnarray}
\lambda &=&\frac{1}{32\,\pi^2}\left(d_{\epsilon}+\ln\frac{\mu^2}{m_\pi^2}\right)\ , \\
d_i^r(\mu)&=&\frac{\beta_i}{32\, \pi^2\,f_\pi^2}\ln\frac{m_\pi^2}{\mu^2}+d^r_i(m_\pi)\ .
\end{eqnarray}
The $\beta_i$ functions of interest here are
\begin{eqnarray}
&&\beta_1=-\frac{g_A^4}{6}\ ,\qquad 
\beta_2=-\frac{1}{12}-\frac{5\,g_A^2}{12} \ ,\qquad
\beta_3=\frac{1}{2}+\frac{g_A^4}{6} \ ,\\
&&\beta_5=\frac{1}{24}+\frac{5\,g_A^2}{24}\ , \qquad
\beta_6=-\frac{1}{6}-\frac{5\,g_A^2}{6}\ , \qquad
\beta_{15}=\beta_{23}=0 \ ,
\end{eqnarray}
and $\beta_5$ is from Eq.~(B13) of Ref.~\cite{Gasser02} which
corresponds to our choice of operator basis in ${\cal L}^{(4)}_{\pi\pi}$.  For the
combinations $\widetilde{d}_i$ above we obtain
\begin{eqnarray}
\widetilde{d}_1&=&-\frac{1}{96\, \pi^2\, f_\pi^2}(1+5\,g_A^2)\,d_{\epsilon}+ \widetilde{d}_1^{\, r}\ ,\\
\widetilde{d}_2&=& \frac{1}{16\, \pi^2\,f_\pi^2}\, d_{\epsilon}+\widetilde{d}_2^{\, r}\ ,\\
\widetilde{d}_4&=&\frac{1}{16\, \pi^2\, f_\pi^2}\,d_{\epsilon}+\widetilde{d}_4^{\, r}\ ,
\end{eqnarray}
and $\widetilde{d}_3= \widetilde{d}_3^{\, r}$ and $\widetilde{d}_5=\widetilde{d}_{5}^{\, r}$.
We observe that the divergence proportional to $m_\pi^2$ from loop corrections
in $\rho_{5,a}^{\mbox{OPE}}(\mbox{npp}+\mbox{pp})$ cancels exactly that
present in $\mbox{f1}+\mbox{f2}$.  Next, the divergent part of $\widetilde{d}_1$
cancels exactly the term proportional to $k_1^2 \, d_\epsilon$ present in
$\mbox{f1}+\mbox{f2}$ and $\mbox{f3}+\mbox{f4}$.  The divergent parts of
$\widetilde{d}_2$ and $\widetilde{d}_4$ are the same, and therefore
can be reabsorbed in the LEC $z_2$ multiplying the contact term $O_2$.  Those of
$\widetilde{d}_3$  and $\widetilde{d}_5$ vanish, which is consistent with the fact that
there are no divergencies proportional to $q^2$ or in the operator multiplying $\widetilde{d}_5$.

Combining Eqs.~(\ref{eq:6.52}), (\ref{eq:6.53}), (\ref{eq:6.58}), (\ref{eq:6.59}),
and~(\ref{eq:6.70}), we then find that the renormalized OPE contributions
up to order $Q$ included read as
\begin{eqnarray}
\label{eq:6.66}
\rho_{5,a}^{\mbox{OPE}}&=&
i\frac{g^r_A}{4\,f_\pi^{r\,2}}\left({\bm \tau}_1\times{\bm \tau}_2\right)_a
{\bm \sigma}_2\cdot{\bf k}_2\frac{1}{\omega_2^2}
\bigg[1+\frac{g_A^{r\,2}}{96\pi^2\, f_\pi^{r\,2}}\Big[\Big(5\, k_1^2+8\, m_\pi^{r\,2}\Big)\frac{s_1}{k_1}
\ln\frac{s_1+k_1}{s_1-k_1}\nonumber\\
&&-\frac{13}{3}k_1^2+2\,m_\pi^2\Big]
+\frac{1}{96\pi^2\,f_\pi^{r\,2}}\left( \frac{s_1^3}{k_1}\ln\frac{s_1+k_1}{s_1-k_1}-\frac{5}{3}k_1^2-8\, m_\pi^{r\, 2}\right)
+\Big(\widetilde{d}^{\, r}_1\, k_1^2+\widetilde{d}^{\, r}_2\, k_2^2\nonumber\\
&&+\widetilde{d}^{\, r}_3\, q^2 + \widetilde{d}^{\, r}_4 \,m_\pi^{r\,2} \Big)\bigg]
+\,i\frac{g^r_A}{2\,f_\pi^{r\,2}}\, \widetilde{d}_5^{\,r}\,\tau_{2,a}\,
{\bm \sigma}_1 \cdot ({\bf q}\times{\bf k}_2)\,\, {\bm \sigma}_2\cdot{\bf k}_2\,\frac{1}{\omega_2^2}\ .
\end{eqnarray}
\subsection{OPE axial current}
\begin{figure}[bth]
\includegraphics[width=2.0in]{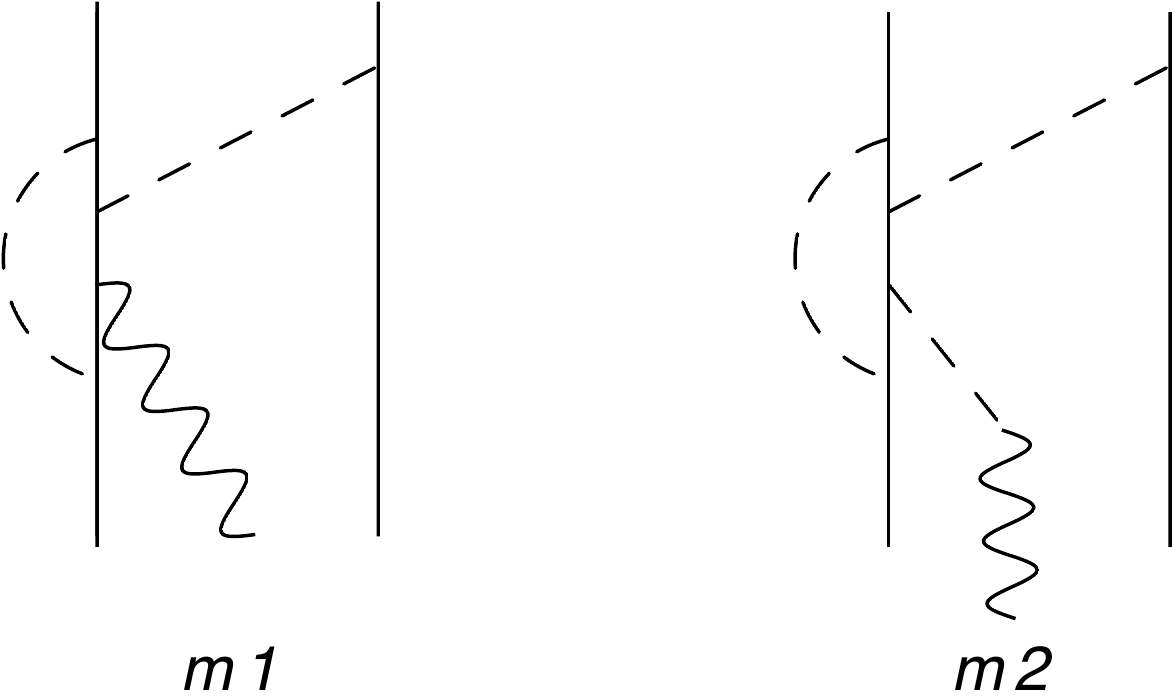}
\caption{The only non-vanishing loop corrections to the OPE axial current. 
Nucleons, pions, and axial fields are denoted by solid, dashed, and wavy lines,
respectively.  Only a single time ordering is shown for each topology. }
\label{fig:i}
\end{figure}

The loop corrections to the OPE axial current are shown in Figs.~\ref{fig:hpp}
and~\ref{fig:i}.  Those associated with panels h1-h17 are easily seen to vanish,
while the contributions of diagrams m1-m2 are obtained as
\begin{eqnarray}
{\bf j}_{5,a}^{(1)}({\rm m1})&=&-\frac{g_A^{r\,5}}{96\,f_\pi^{r\, 4}}\,J_{14}
\left[\,9\, \tau_{2,a} \, {\bf k}_2 - \left({\bm \tau}_1\times{\bm \tau}_2\right)_a 
\left({\bm \sigma}_1 \times {\bf k}_2\right)\right]
{\bm \sigma}_2\cdot{\bf k}_2 \,\frac{1}{\omega_2^2}\  , \\
{\bf j}_{5,a}^{(1)}({\rm m2})&=&-\frac{\bf q}{q^2+m_\pi^2}  \, {\bf q}\cdot{\bf j}_{5,a}^{(1)}({\rm m1}) \ ,
\end{eqnarray}
In dimensional regularization we find the finite result
 \begin{eqnarray}
 {\bf j}_{5,a}^{(1)}({\rm m1})&=&\frac{g_A^{r\, 5}\, m^r_\pi}{256\,\pi f_\pi^{r\, 4}}
 \left[\,9\, \tau_{2,a} \, {\bf k}_2 - \left({\bm \tau}_1\times{\bm \tau}_2\right)_a 
 \left({\bm \sigma}_1 \times {\bf k}_2\right)\right]
 {\bm \sigma}_2\cdot{\bf k}_2 \,\frac{1}{\omega_2^2} \ .
 \end{eqnarray}
No renormalization is necessary in this case, since
loop corrections to diagrams d1-d2 of Fig.~\ref{fig:f7} enter at order $Q^2$,
and are beyond the scope of the present work. 
\section{Discussion}
\label{sec:disc}
In this section we report the complete (and renormalized)
expressions for the weak axial charge and current operators,
compare these expressions to those obtained by the authors
of Ref.~\cite{Park93}, and discuss current conservation in the
chiral limit.  For simplicity, the superscript $r$ has been removed
from the pion and nucleon masses $m_\pi$ and $m$, the
nucleon axial coupling constant $g_A$, and pion decay constant
$f_\pi$.  However, all these quantities are understood to have
been renormalized.
 
The one-body operators and two-body contact operators are those
listed, respectively, in Eqs.~(\ref{eq:rr-2})--(\ref{eq:r-3}) and
Eqs.~(\ref{eq:ed0})--(\ref{eq:ed1}), and in Eqs.~(\ref{eq:rctct}) and
(\ref{eq:jctct}), while the two-body operators involving OPE, TPE or
MPE, and short-range terms follow in the next subsection.
Relativistic corrections (proportional to $1/m^3$) in the one-body axial
charge are neglected, those in the one-body axial current (proportional to $1/m^2$)
are retained in Eqs.~(\ref{eq:ed0})--(\ref{eq:ed1}), since they are known to be
important in weak transitions such as the proton weak capture on $^3$He at
low energies~\cite{Park03}. 

\subsection{Two-body axial charge and current operators up to one loop: summary}
\label{sec:summ}
The (renormalized) OPE contributions to the axial charge
are given in Eq.~(\ref{eq:6.66}), while those corresponding to the
axial current read
\begin{eqnarray}
\label{eq:opejfin}
\widetilde{{\bf j}}_{5,a}^{\,\,\mbox{OPE}}&=&{\bf j}_{5,a}^{\,\, \mbox{OPE}}-\frac{\bf q}{q^2+m_\pi^2}\,
{\bf q}\cdot {\bf j}_{5,a}^{\,\, \mbox{OPE}}-\frac{g_A}{2\, f_\pi^2}\,\frac{\bf q}{q^2+m_\pi^2} \bigg[
4\, m_\pi^2 \, c_1\, \tau_{2,a}\nonumber\\
&&-\frac{i}{2\, m }\left({\bm \tau}_1\times{\bm \tau}_2\right)_a\left({\bf K}_1\cdot{\bf k}_1+{\bf K}_2
\cdot{\bf k}_2\right)\bigg] {\bm \sigma}_2\cdot{\bf k}_2\frac{1}{\omega_2^2} \ ,
\end{eqnarray}
where 
\begin{eqnarray}
\label{eq:opej1fin}
{\bf j}_{5,a}^{\,\, \mbox{OPE}}&=& \frac{g_A}{2\,f_\pi^2} \Bigg[
\left(2\, c_3 -\frac{9}{128\, \pi} \frac{ g_A^4\, m_\pi}{f_\pi^2}\right) \tau_{2,a}\, {\bf k}_2
+\left({\bm \tau}_1\times{\bm \tau}_2\right)_a 
\bigg[ \frac{i}{2\, m} {\bf K}_1 - \frac{c_6+1}{4\, m}  {\bm \sigma}_1\times{\bf q} \nonumber\\
&&+\left( c_4+\frac{1}{4\, m}+ \frac{1}{128\, \pi} \frac{ g_A^4\, m_\pi}{ f_\pi^2} \right) {\bm \sigma}_1\times{\bf k}_2 \bigg] \Bigg] {\bm\sigma}_2\cdot{\bf k}_2\, \frac{1}{\omega_2^2}\ .
\end{eqnarray}
The TPE axial charge, and MPE and short-range axial current can be written, respectively, as
\begin{eqnarray}
\label{eq:tperfin}
\rho_{5,a}^{\mbox{TPE}}&=& i\frac{g^3_A}{128\pi^2f_\pi^4}
\bigg[\left({\bm \tau}_1\times{\bm \tau}_2\right)_a{\bm \sigma}_1\cdot{\bf k}_2\left(3-\frac{1}{g_A^2}-
\frac{4\, m_\pi^2}{k_2^2+4\,m_\pi^2 }\right)
-4\, \tau_{1,a}\left({\bm \sigma}_1\times{\bm \sigma}_2\right)\cdot{\bf k}_2\bigg] \nonumber\\
&&\times \frac{s_2}{k_2}\ln\left(\frac{s_2+k_2}{s_2-k_2}\right)	\ ,	
\end{eqnarray}
with with $s_j$ defined as in Eq.~(\ref{eq:sdef}), and
\begin{eqnarray}
\label{eq:mpejfin}
\widetilde{{\bf j}}_{5,a}^{\,\,\mbox{MPE}}&=&{\bf j}_{5,a}^{\, \mbox{MPE}}
-\frac{\bf q}{q^2+m_\pi^2}{\bf q}\cdot {\bf j}_{5,a}^{\, \mbox{MPE}} \nonumber\\
&&
+\frac{g_A^3}{128 \, \pi f_\pi^4}\,\frac{\bf q}{q^2+m_\pi^2}\Bigg[ \,
\tau_{2,a}\, \bigg[\, Z_1(k_1)\, {\bm \sigma}_2\cdot\left({\bf k}_1-{\bf k}_2\right)
+ Z_2({\bf k}_1)\,{\bm \sigma}_2\cdot{\bf k}_2\, \frac{1}{\omega_2^2}\, \bigg]\nonumber\\
&&+\left(2\,\tau_{2,a}-\tau_{1,a}\right)
 Z_1(k_2)\,{\bm \sigma}_1\cdot{\bf k}_2+\left({\bm \tau}_1\times{\bm \tau}_2\right)_a \,\bigg[ Z_3(k_1)\,\bigg[
\left({\bm \sigma}_1\times{\bm \sigma}_2\right)\cdot{\bf k}_1 \nonumber\\
&&-2\left({\bm \sigma}_1\times{\bf k}_1\right)\cdot\left({\bf k}_2+{\bf q}\right)
{\bm \sigma}_2\cdot{\bf k}_2 \, \frac{1}{\omega_2^2}\,\bigg] 
+Z_3(k_2)\left({\bm \sigma}_1\times{\bm \sigma}_2\right)\cdot{\bf k}_2\bigg]\Bigg]
\nonumber\\
&&+\frac{g_A^3}{128 \, \pi f_\pi^4}\, \tau_{2,a}\, Z_1(k_1)\bigg[\left({\bf k}_2-3\, {\bf k}_1\right)
{\bm \sigma}_2\cdot{\bf k}_2\, \frac{1}{\omega_2^2}
 -2\, {\bm \sigma}_2\bigg]\nonumber\\
 &&+\frac{g_A^3}{32 \, \pi f_\pi^4}\left({\bm \tau}_1\times{\bm \tau}_2\right)_a
Z_3(k_1)\, {\bm \sigma}_1\times{\bf k}_1 \,\, {\bm \sigma}_2\cdot{\bf k}_2\, \frac{1}{\omega_2^2} \ ,
\end{eqnarray}
where
\begin{eqnarray}
\label{eq:mpej1fin}
{\bf j}_{5,a}^{\, \mbox{MPE}}&=&\frac{g_A^3}{64 \, \pi f_\pi^4}\, \tau_{2,a}\,
\left[\, W_1(k_2)\, {\bm \sigma}_1 +W_2(k_2)\, {\bf k}_2\,\,{\bm \sigma}_1\cdot{\bf k}_2 \,\right] \nonumber\\
&&+\frac{g_A^5}{64\, \pi f_\pi^4}\, \tau_{1,a}\, W_3(k_2)\, \left({\bm \sigma}_2\times{\bf k}_2\right)\times{\bf k}_2 
-\frac{g_A^3\, m_\pi}{8\,\pi\, f_\pi^2} \,C_T \, \tau_{1,a}\, {\bm \sigma}_2 \ ,
\end{eqnarray}
and the loop functions $Z_i$ and $W_i$ are listed in Appendix~\ref{app:jforms}.
\subsection{Current conservation in the chiral limit}
\label{sec:chicons}
\begin{figure}[bth]
\includegraphics[width=5in]{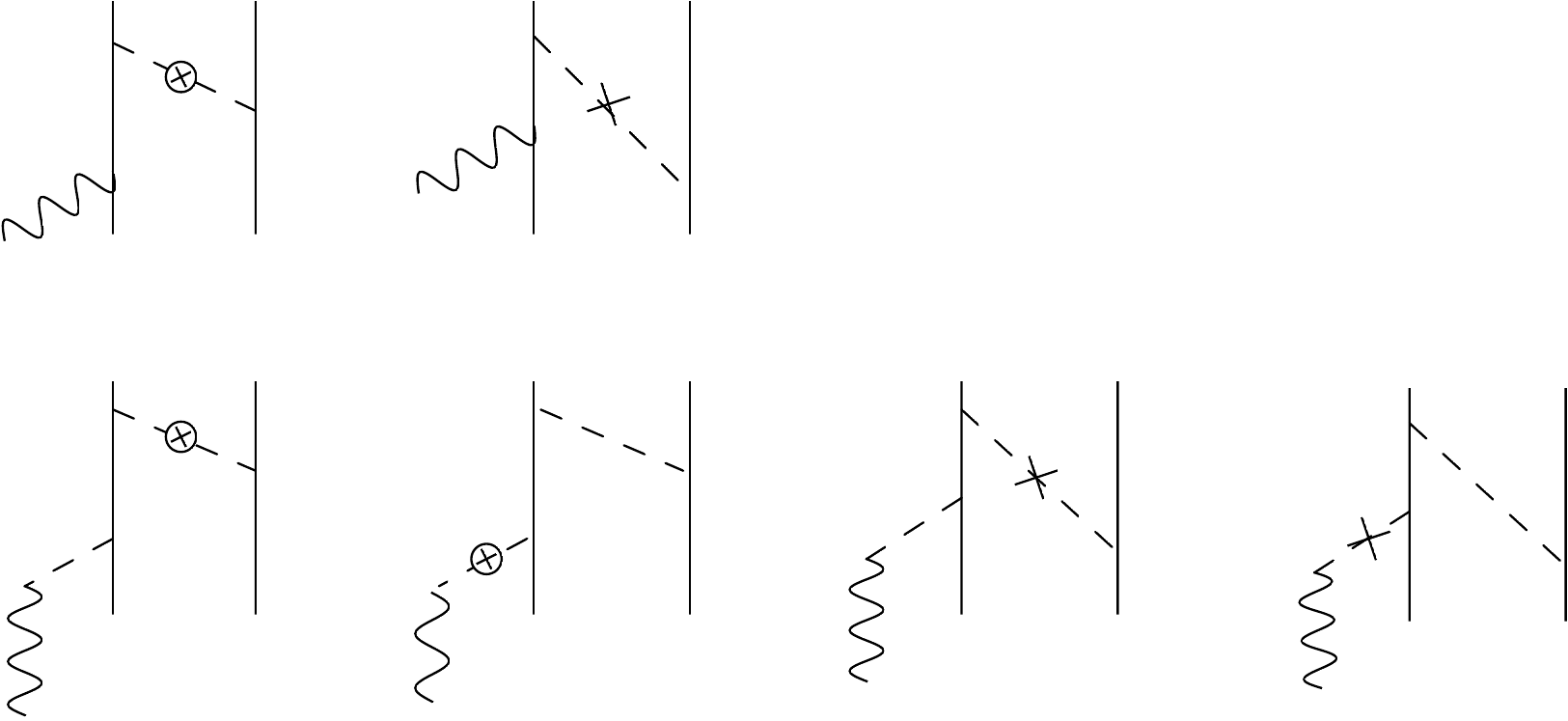}
\caption{Illustration of some of the non-static corrections to the axial current ignored in this
work.  Nucleons, pions, and axial fields are denoted by solid, dashed, and wavy lines,
respectively. See text for further explanations. }
\label{fig:fig12}
\end{figure}

In the chiral limit ($m_\pi \rightarrow 0$) the axial current is conserved and
\begin{eqnarray}
{\bf q}\cdot{\bf j}_{5,a}&=&\left[\, H\, ,\, \rho_{5,a}\,\right]\ ,
\end{eqnarray}
with the two-nucleon Hamiltonian given by
\begin{equation}
H=T^{(-1)}+v^{(0)}+v^{(2)}+\dots \ ,
\end{equation}
where the superscripts denote the power counting, the $v^{(n)}$ are the two-nucleon potentials
defined in Sec.~\ref{sec:road}, and the kinetic energy $T^{(-1)}$ (in momentum space) is
\begin{equation}
T^{(-1)}=\frac{{\bf p}^2_1}{2\,m}\, (2\pi)^3\delta({\bf p}_2^\prime-{\bf p}_2) +(1\rightleftharpoons 2) \ .
\end{equation}
Here, the potentials and axial charge and current operators (including the axial coupling
and pion decay constants and LECs entering them) are to be understood in the
chiral limit.  Order by order in the power counting, current conservation implies the following
set of relations
\begin{eqnarray}
{\bf q}\cdot{\bf j}_{5,a}^{(-3)}&=&0 \ ,\label{cc:LO}\\
{\bf q}\cdot{\bf j}_{5,a}^{(-1)}&=& \left[T^{(-1)} ,\, \rho_{5,a}^{(-2)}\right]\ , \label{cc:NLO}\\
{\bf q}\cdot{\bf j}_{5,a}^{(0)}&=&\left[ T^{(-1)} ,\, \rho_{5,a}^{(-1)}\right] +
 \left[ v^{(0)} ,\, \rho_{5,a}^{(-2)}\right] \ ,\label{cc:NNLO}\\
{\bf q}\cdot{\bf j}_{5,a}^{(1)}&=&\left[ T^{(-1)} ,\, \rho_{5,a}^{(0)}\right]
+ \left[ v^{(0)} ,\, \rho_{5,a}^{(-1)}\right]\ ,\label{cc:NNNLO}
\end{eqnarray}
where we have only kept up to terms of order $Q^2$.  Note that the
commutators implicitly bring in factors of $Q^3$.  The first of these
relations is obviously satisfied, see Eqs.~(\ref{eq:r-3}) or~(\ref{eq:jjrre}).
The second relation has
\begin{equation}
{\bf q}\cdot{\bf j}_{5,a}^{(-1)}=-\frac{g_A}{2\, m^2}\, \tau_{1,a}\, {\bf k}_1\cdot{\bf K}_1\, {\bm \sigma}_1\cdot {\bf K}_1 +(1\rightleftharpoons 2)\ ,
\end{equation}
where ${\bf j}_{5,a}^{(-1)}$ is given by the sum of the contributions
in Eqs.~(\ref{eq:ed0}) and~(\ref{eq:ed1}), and is also satisfied.  The
left-hand-side of the third relation has
\begin{eqnarray}
{\bf q}\cdot{\bf j}_{5,a}^{(0)}&=&i\frac{g_A}{4\,m f_\pi^2}\left({\bm \tau}_1\times{\bm \tau}_2\right)_a{\bm \sigma}_2\cdot{\bf k}_2\frac{1}{\omega_2^2}\left({\bf k}_1\cdot {\bf K}_1+ {\bf k}_2\cdot {\bf K}_2\right)+(1\rightleftharpoons 2)\ ,
\end{eqnarray}
and this matches the first commutator on the right-hand side, $\left[ T^{(-1)} ,\, \rho_{5,a}^{(-1)}\right]$
with $\rho_{5,a}^{(-1)}$ given by
\begin{equation}
\rho^{(-1)}_{5,a}=i\frac{g_A}{4\,f_\pi^{2}}\left({\bm \tau}_1\times{\bm \tau}_2\right)_a
{\bm \sigma}_2\cdot{\bf k}_2\frac{1}{\omega_2^2} +(1\rightleftharpoons2) \ ,
\end{equation}
i.e., the sum of terms a1 and a2 in Eqs.~(\ref{eq:rhoa1}) and~(\ref{eq:rhoa2}).
There are additional contributions to ${\bf j}_{5,a}^{(0)}$, which arise from non-static corrections
to the denominators involving pion energies in the diagrams illustrated in Fig.~\ref{fig:fig12},
where the crossed circle (cross) means that the these denominators are expanded as indicated in
Eq.~(\ref{eq:deno}) to order $Q$ ($Q^2$) {\it beyond} the leading-order static term.  These contributions are
needed in order to satisfy the commutator $\left[ v^{(0)} ,\, \rho_{5,a}^{(-2)}\right]$, but have been
neglected in the present work.

Lastly, we consider the fourth relation, Eq.~(\ref{cc:NNNLO}).  The axial current ${\bf j}_{5,a}^{(1)}$
obtained here is in the static limit, and one expects ${\bf q}\cdot {\bf j}_{5,a}^{(1)}$ to satisfy the
commutator
\begin{eqnarray}
\left[\, v^{(0)}\, ,\, \rho_{5,a}^{(-1)}\,\right]
&=&-\frac{g_A^3}{16\, f_\pi^4} \left(\tau_{1,a}-\tau_{2,a}\right)
\bigg[ \Big[ k_2^2\, R^{(0)}(k_2)-R^{(2)}(k_2)\Big] {\bm \sigma}_1\cdot{\bf k}_2 \nonumber\\
&&\qquad\qquad\qquad\qquad -\Big[ k_1^2\, R^{(0)}(k_1)-R^{(2)}(k_1)\Big] {\bm \sigma}_2\cdot{\bf k}_1\bigg] \nonumber\\
&&+\frac{g_A^3}{16\, f_\pi^4} \left({\bm \tau}_1\times{\bm \tau}_2\right)_a
\Big[ R^{(2)}_{ij}(k_2) \,\sigma_{1,i}\, \left({\bm \sigma}_2\times{\bf k}_2\right)_j \nonumber\\
&&\qquad\qquad\qquad\qquad-R^{(2)}_{ij}(k_1) \,\sigma_{2,i}\, \left({\bm \sigma}_1\times{\bf k}_1\right)_j\Big] \ ,\label{eq:comm}
\end{eqnarray} 
where the loop functions $R^{(n)}(k)$ and $R_{ij}^{(2)}(k)$ in the chiral limit read
\begin{eqnarray}
R^{(0)}(k) & \rightarrow & \frac{1}{16}\, \frac{1}{k} \ , \\
R^{(2)}(k) & \rightarrow & -\frac{1}{16}\, k \ , \\
R^{(2)}_{ij}(k) & \rightarrow & -\frac{1}{32}\, k\, \delta_{ij} +\dots \ ,
\end{eqnarray}
and the $\dots$ indicate a term proportional to $k_i\, k_j$, which
vanishes when inserted in Eq.~(\ref{eq:comm}).
The current-conservation constraint is seen to be satisfied by
noting the only non-vanishing contributions to
${\bf q}\cdot {\bf j}_{5,a}^{(1)}$ are those due
to diagrams e4, e5, e10, e15, e16, and e17 in Fig.~\ref{fig:f7},
proportional to the combination of coupling constants
$g_A^3/f_\pi^4$.  In particular the contributions of the purely
irreducible diagrams e4, e5, e15, e16, and e17 combine
to give
\begin{eqnarray}
&&{\bf q}\cdot {\bf j}_{5,a}^{(1)}({\rm e4+e5+e15+e16+e17})=-\frac{g_A^3}{32\, f_\pi^4}
\bigg[  \tau_{1,a}\, \Big[ k_2^2\, R^{(0)}(k_2)-R^{(2)}(k_2)\Big] {\bm \sigma}_1\cdot{\bf k}_2 \nonumber\\
&&\qquad\,\,\,\,\,\,+\,\tau_{2,a}\,\Big[ k_1^2\, R^{(0)}(k_1)-R^{(2)}(k_1)\Big] {\bm \sigma}_2\cdot{\bf k}_1\bigg] 
+\frac{g_A^3}{32\, f_\pi^4} \left({\bm \tau}_1\times{\bm \tau}_2\right)_a
\Big[ R^{(2)}_{ij}(k_2) \,\sigma_{1,i}\, \left({\bm \sigma}_2\times{\bf k}_2\right)_j \nonumber\\
&&\qquad\qquad\qquad\qquad-R^{(2)}_{ij}(k_1) \,\sigma_{2,i}\, \left({\bm \sigma}_1
\times{\bf k}_1\right)_j\Big] \ ,
\end{eqnarray}
with the remaining ``missing'' term being provided
by ${\bf q}\cdot {\bf j}_{5,a}^{(1)}({\rm e10})$.  The remaining commutator
$\left[ T^{(-1)} ,\, \rho_{5,a}^{(0)}\right]$
has a factor $1/m$, and therefore non-static
corrections need to be included in ${\bf j}_{5,a}^{(1)}$,
if the latter is to satisfy the complete Eq.~(\ref{cc:NNNLO}).
These corrections have been ignored in the present work.
\subsection{Comparison}
\label{sec:comp}
We compare the one- and two-body axial charge
and current operators derived here with those obtained by
Park {\it et al.}~in Refs.~\cite{Park93} and~\cite{Park03} in the
heavy-baryon (HB) formulation of covariant perturbation theory.
The one-body axial charge and current operators at leading order in
Eqs.~(\ref{eq:rr-2}) and~(\ref{eq:r-3}) are the same as those listed
in Eqs.~(B1) and~(A3) of Ref.~\cite{Park03}, except for the
pion-pole contribution to ${\bf j}_{5,a}^{(-3)}$, which, while nominally
of the same order ($Q^{-3}$) as the non-pole contribution, is nevertheless
suppressed at low momentum transfer $q$ and is therefore ignored in
Ref.~\cite{Park03} (we note incidentally that in that work ${\bf k}_1=-{\bf q}$,
i.e., the opposite convention adopted here).  Of course, this pion-pole
contribution is crucial for current conservation in the chiral limit.
We have neglected the $1/m^2$ relativistic corrections to the leading order axial
charge.  They are retained in Eq.~(17) of Ref.~\cite{Park03}.  However,
the $1/m^2$ corrections to the leading order axial current in Eq.~(\ref{eq:ed0})
are in agreement with those given in Eq.~(A3) of Ref.~\cite{Park03},
except for the last term proportional to ${\bf q}\,({\bm \sigma}_1\cdot{\bf q})$,
which is again ignored in that work.
\begin{figure}[bth]
\includegraphics[width=1.5in]{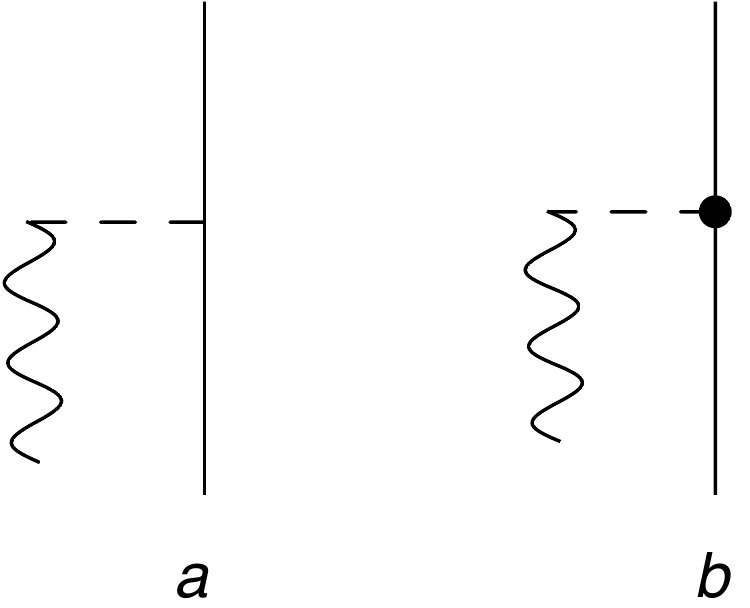}\\
\caption{Feynman amplitudes contributing to the one-body axial
charge at leading order.}
\label{fig:fx}
\end{figure}

Before moving on to the two-body contributions, it is worthwhile
discussing how the one-body axial charge operator emerges in covariant
perturbation theory. The relevant interaction Hamiltonian densities are
\begin{eqnarray}
{\cal H}_{\pi A}(x)&=&f_\pi\, {\bf A}^0(x)\cdot{\bf \Pi}(x)\ ,\\
{\cal H}^{({\rm a})}_{\pi NN}(x)&=&\frac{g_A}{2f_\pi}\, \overline{N}(x){\bm\tau}\cdot{\bm \Pi}(x)\gamma^0\gamma^5 N(x)\ ,\\
{\cal H}^{({\rm b})}_{\pi NN}(x)&=&\frac{g_A}{2f_\pi}\, \overline{N}(x){\bm\tau}\cdot\partial_i{\bm \pi}(x)\gamma^i\gamma^5 N(x)\ ,
\end{eqnarray}
where all fields are in interaction picture.
The $S$-matrix elements associated with the Feynman amplitudes in Fig.~\ref{fig:fx}
are given by
\begin{equation}
S^{(\gamma)}_{fi} =-\frac{1}{2} \int {\rm d}^4 x \,{\rm d}^4 y\,
  \langle {\bf p}^\prime,\lambda^\prime|\, T \left[{\cal H}_{\pi A}(x)\, {\cal H}^{(\gamma)}_{\pi NN}(y)+
{\cal H}^{(\gamma)}_{\pi NN}(x)\, {\cal H}_{\pi A}(y)\right]|{\bf p},\lambda\rangle \ ,
\end{equation}
where $\gamma={\rm a}$ or ${\rm b}$, $T$ denotes the usual chronological
product, and $|{\bf p},\lambda\rangle$ and
$|{\bf p}^\prime,\lambda^\prime\rangle$ are the initial and final nucleon states
with momenta ${\bf p}$ and ${\bf p}^\prime$  in spin-isospin states $\chi_\lambda$ and
$\chi_{\lambda^\prime}$, respectively.
Then for $\gamma={\rm a}$ we obtain
\begin{eqnarray}
S^{({\rm a})}_{fi}&=&-\frac{g_A}{8\,m}\chi^\dagger_{\lambda^\prime}\,
{\bm \sigma}\cdot \left({\bf p}^\prime+{\bf p}\right)\, A^0_c\,  \tau_d\, \chi_{\lambda}
\int {\rm d}^4 x\, {\rm d}^4y
\Big[{\rm e}^{i ({p^\prime}-p)\cdot y-i q\cdot x}\langle 0|T\left[\Pi_c(x)\, \Pi_d(y)\right]|0\rangle \nonumber\\
&&+\,{\rm e}^{i \left(p^\prime-p\right)\cdot x-i q\cdot y}\langle 0|T\left[
\Pi_d(x)\,\Pi_c(y)\right] |0\rangle\Big]\ , 
\end{eqnarray}
where we have considered the leading order in the non-relativistic
expansion of the nucleon matrix element.  Since in the interaction picture
the conjugate field momentum $\Pi_c(x)=\partial^0\pi_c(x)$, it is easily seen that
(see also Ref.~\cite{Gerstein71})
\begin{eqnarray}
\langle 0 |T\left[ \Pi_c(x)\, \Pi_d(y)\right]|0\rangle &=&
\partial^0_x\,\partial^0_y\,\langle 0 |T\left[ \pi_c(x)\, \pi_d(y)\right]|0\rangle -i\,\delta_{cd}\,\delta(x^0-y^0)\,\delta({\bf x}-{\bf y})\nonumber \\
&=&-i\, \delta_{cd} \int \frac{d^4k}{(2\pi)^4}e^{-ik\cdot (x-y)}\left( 1+\frac{ k_0^2}{m_\pi^2-k^2-i\epsilon}\right) \ ,
\label{eq:noncov}
\end{eqnarray}
with the Feynman propagator defined by
\begin{equation}
\langle 0|T\left[ \pi_c(x)\, \pi_d(y)\right] |0\rangle =\int \frac{d^4k}{(2\pi)^4}\frac{-i\, \delta_{cd}}{m_\pi^2-k^2-i\epsilon}e^{-ik\cdot(x-y)} \ .
\end{equation}
The  $T$-matrix element $T_{fi}$ obtained from $S_{fi}=-i\, (2\pi)^4\, \delta(p^\prime-p-q) \, T_{fi}$ reads
\begin{eqnarray}
T^{({\rm a})}_{fi} &=&-\frac{g_A}{4m}\, A^0_c\,
\chi^\dagger_{\lambda^\prime}\,
{\bm \sigma}\cdot \left({\bf p}^\prime+{\bf p}\right)\,  \tau_c\, \chi_{\lambda}
\left(1+ \frac{ q_0^2}{m_\pi^2+{\bf q}^2-q_0^2-i\epsilon}\right) \ ,
\end{eqnarray}
where the term proportional to $q_0=p^\prime_0-p_0$ is suppressed by $Q^2$ in the
power counting.  The leading order term leads to the axial charge operator in Eq.~(\ref{eq:rr-2}).
A similar analysis shows that the leading-order contribution to $S^{({\rm b})}_{fi}$ vanishes.

As already noted, the interaction Hamiltonian in Eq.~(\ref{eq:hhii}) contains
no direct coupling of $A^0_a$ to the nucleon.  However, diagrams of the type
illustrated in Fig.~\ref{fig:fx} are not considered in Refs.~\cite{Park93,Park03}.
It would appear that their contribution is accounted for by retaining the
term $-i\, \delta_{cd}\, \delta(x-y)$ in Eq.~(\ref{eq:noncov}), which effectively
leads to a direct coupling between $A^0_a$ and the nucleon.

Turning to the OPE contributions at tree level, we find that the order
$Q^{-1}$ contribution to the axial charge, $\rho_{5,a}^{(-1)}$, in
Eq.~(\ref{eq:6.66}) reproduces the corresponding contribution,
given by Eqs.~(B2), (B3), and (B5) of Ref.~\cite{Park03} with
$F_1^V(t)=1$, while the order $Q^0$ contribution to the axial
current, ${\bf j}_{5,a}^{(0)}$, in Eq.~(\ref{eq:opejfin}) is the
same as in Eq.~(A5) of Ref.~\cite{Park03}.  We stress again that,
while diagram a2 in Fig.~\ref{fig:f5}
is not explicitly considered in Refs.~\cite{Park93,Park03}, the OPE
axial charge operator derived there has the correct strength.   The
contact terms contributing to the $Q^0$ axial current in Eq.~(A6) of
Ref.~\cite{Park03} can be reduced through Fierz identities to the
form given in Eq.~(\ref{eq:jctct}).
\begin{figure}[bth]
\includegraphics[width=3in]{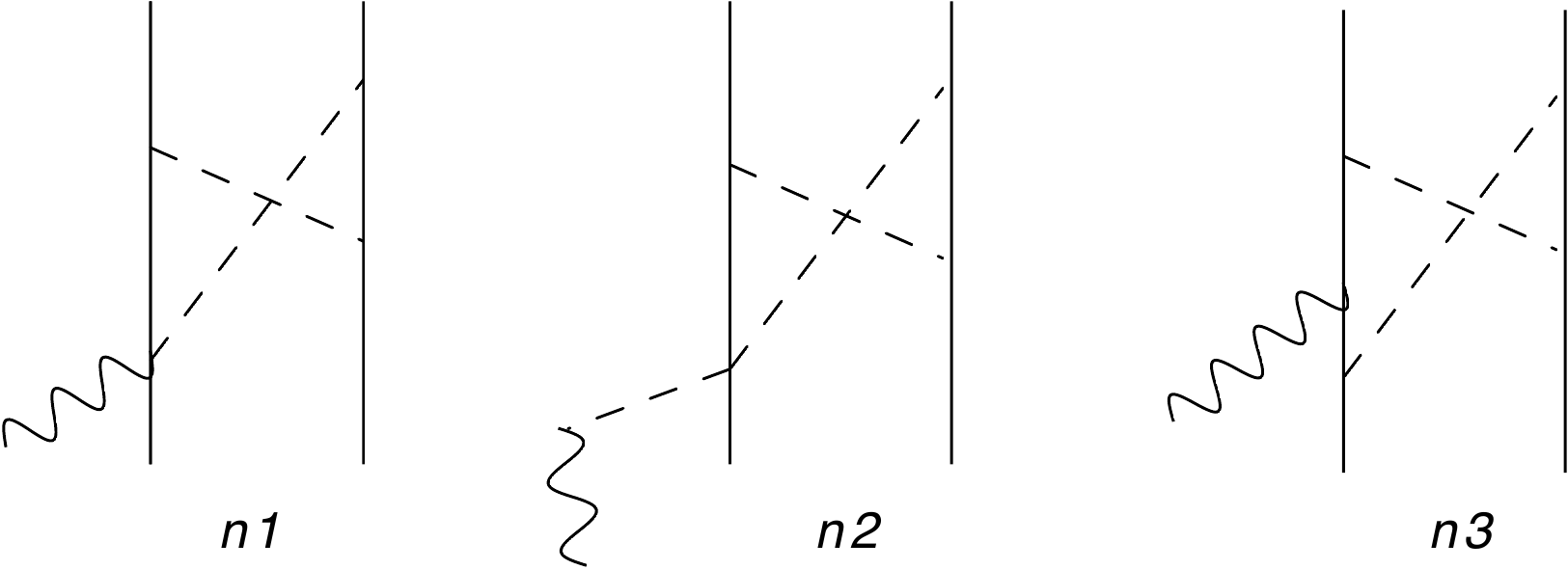}\\
\caption{Diagrams contributing to the axial charge (n1-n2) and current (n3) 
at order $Q$ considered in Ref.~\cite{Park93}.
Nucleons, pions, and axial fields are denoted by solid, dashed, and wavy lines,
respectively.  Only a single time ordering is shown for each of the possible
12 (n1) and 60 (n2 and n3) cross-box topologies. }
\label{fig:f9}
\end{figure}

Next we consider loop corrections to the axial charge.
The contributions of c3-c4, c7-c8, and c9-c12 in Fig.~\ref{fig:f5}
are found to vanish in both approaches, here and in Refs.~\cite{Park93,Park03}.
The contributions of diagrams c1 and c2
are the same as those for $A^{(0)}(a+b)$ in Eq.~(93) of  Ref.~\cite{Park93}.  The contributions
of diagrams c5 and c6 are different from those for $A^{(0)}(c+d)$ reported in Eq.~(94) of
Ref.~\cite{Park93} because
of the different treatment of reducible topologies for these types of terms.
Indeed, if only the (irreducible) cross-box topologies are retained
for diagrams c5 and c6, as illustrated in Fig.~\ref{fig:f9}, then the resulting
operator is the same as in Eq.~(94).
The OPE axial charge operator in Eqs.~(74) and~(90)
of Ref.~\cite{Park93} reads in our notation
\begin{eqnarray}
&&\rho_{5,a}^{\mbox{OPE}}({\rm Park}\,\, et\,\, al.)=
i\frac{g_A}{4\,f_\pi^{2}}\left({\bm \tau}_1\times{\bm \tau}_2\right)_a
{\bm \sigma}_2\cdot{\bf k}_2\frac{1}{\omega_2^2}
\Bigg[1-\frac{k_2^2}{f_\pi^2}\left(\frac{17\,g_A^2+4}
{144\,\pi^2}+ c_3^r\right)-\frac{m_\pi^2\, g_A^2}{12\, \pi^2 f_\pi^2}\nonumber\\
&&+\frac{g_A^2}{96\,\pi^2f_\pi^2}\frac{s_2}{k_2}\ln\left(\frac{s_2+k_2}{s_2-k_2}\right)\left(5\,k_2^2+8\,m_\pi^2\right)+\frac{1}{96\,\pi^2f_\pi^2}\bigg[\frac{s_2^3}{k_2}\ln\left(\frac{s_2+k_2}{s_2-k_2}\right)-8\,m_\pi^2\bigg]\Bigg] \ .
\label{eq:ope-park}
\end{eqnarray}
Provided we define
\[
\widetilde{d}_1^{\,r}+\widetilde{d}_2^{\,r}-\widetilde{d}_4^{\,r}-\frac{(5+13\,g_A^2)}
{288}= -(17\,g_A^2+4)/(144\pi^2\, f_\pi^2)-c_3^{\,r} \ ,
\]
the expression above is in agreement with our Eq.~(\ref{eq:6.66}) in the
limit ${\bf q}=0$ (or ${\bf k}_1=-{\bf k}_2$) which is assumed in Refs.~\cite{Park93,Park03},
except for the term proportional to $m_\pi^2$ in the first line. 

Lastly, the term proportional to the LEC $c_3$
in Ref.~\cite{Park93} (in the HB formulation) is given by
\[
i\,\frac{c_3}{f_\pi^2} \, \overline{N}\, v^\alpha\, \left[\, D^\beta\, ,\, \left[ D_\alpha\, , \, D_\beta\, \right]\, \right]N \, ,
\] 
which can be re-expressed as
\[
i\,\frac{c_3}{2\,f_\pi^2} \, \overline{N} \left[\, D^\beta\, , \, F^+_{0\beta}\, \right]N +\dots  \ ,
\]
and matches the term proportional to $d_6$ in the HB limit of ${\cal L}_{\pi N}^{(3)}$~\cite{Fettes00}---in
the relation above $v^\alpha$ is the velocity, $v^\alpha=(1,{\bf 0})$.

Moving on to the loop corrections to the axial current,  
the sum of the contributions due to diagram m1 of Fig.~\ref{fig:i}
and diagram e15 of Fig.~\ref{fig:f7} gives the same expression as in Eq.~(A7) of Ref.~\cite{Park03},
provided the parameter $\alpha$ in the $3\pi\, A$ vertex of diagram e15 is set to 1/6---the
authors of Refs.~\cite{Park93,Park03} use the exponential parametrization
for the pion field.
The irreducible contributions of diagrams e1 and e4 in Fig.~\ref{fig:f7} are the same
as reported for, respectively, ${\bf A}_{12}^a$($2\pi$:b) and ${\bf A}_{12}^a$($2\pi$:a)
of Eq.~(A13) of Ref.~\cite{Park03},
while the contributions associated with the cross-box topologies of diagram e8 in Fig.~\ref{fig:f7}
and illustrated in panel n3 of Fig.~\ref{fig:f9}, lead to the expression for ${\bf A}_{12}^a$($2\pi$:c) in Eq.~(A13).
Non-vanishing pion-pole diagrams e2, e5, e9, e10, e16, and e17
as well as diagrams e20-e21 (e22 and e23 vanish) in Fig.~\ref{fig:f7} have not been
considered in Refs.~\cite{Park93,Park03}.  In particular, because of this incomplete treatment,
loop corrections to the axial current are $\alpha$-dependent in Refs.~\cite{Park93,Park03}.
Furthermore, the current is not conserved in the chiral limit.

Finally, the OPE axial current at tree-level listed in the recent Ref.~\cite{Hofe15}
(and including pion-pole contributions) is different from that obtained in the present work
in Eqs.~(\ref{eq:ai})--(\ref{eq:aj}).  Moreover, it is not conserved in the chiral limit.

\section{Conclusions}
\label{sec:concl}
In the present work we have carried out an analysis of the weak
axial charge and current operators in a two-nucleon system up to
one loop (i.e., including corrections up to order $Q$ in the power
counting) in $\chi$EFT.  The formalism used in the derivation is
based on standard TOPT, but accounts for cancellations between
the contributions of irreducible diagrams and the contributions due
to non-static corrections from energy denominators of reducible diagrams.
A detailed comparison between the results of this work and those of the early
studies of Park {\it et al.}~\cite{Park93,Park03} in the HB formulation
of $\chi$EFT indicates that there are differences in some of the loop
corrections and in the renormalization of the OPE axial charge, the
former due to a different prescription adopted by the authors of those
papers, one in which only a subset of the irreducible contributions
are retained in the perturbative expansion---for example,  in the
case of box diagrams, only cross-box ones are considered.
Furthermore, while the contribution illustrated by panel e15 in
Fig.~\ref{fig:f7} is accounted for in Refs.~\cite{Park93,Park03},
additional ones involving three- and four-pion vertices, such as
those in panels e5, e16, and  e17, have been ignored.  As a
consequence, the one-loop axial current derived there depends on
the parametrization of the pion field---it is $\alpha$-dependent---and,
furthermore, is not conserved in the chiral limit.

The order $Q$ loop corrections in the axial current are finite, consistently
with the fact that there are no contact terms at this order.  There
is a single LEC (denoted as $z_0$ here and as $d_R$ in Ref.~\cite{Park03})
which enters at lower order $Q^0$.   On the other hand, four
independent LECs (denoted as $z_i$, with $i=1,\dots,4$) multiply contact
terms in the axial charge at order $Q$, two of which are needed to reabsorb the
divergencies from loop corrections in the TPE axial charge.  The loop
corrections to the OPE axial charge instead lead to renormalization of
$\widetilde{d}_1$ which is expressed as linear combinations of the
LECs $d_i$ in the ${\cal L}_{\pi N}^{(3)}$ Lagrangian---some of these $d_i$
having been determined in fits to $\pi N$ scattering data~\cite{Fettes01}.
The LEC $z_0$ has been recently fixed by reproducing the empirical
value of the Gamow-Teller matrix element in $^3$H $\beta$-decay~\cite{Marcucci12}.
However, that calculation ignored MPE loop corrections in ${\bf j}_{5a}$,
and therefore a refitting of $z_0$ will be necessary.  Most calculations of
nuclear axial current matrix elements, such as those reported in Refs.~\cite{Park03,Marcucci13}
for the $pp$ and $p\, ^3$He weak fusions and in Ref.~\cite{Marcucci12} for muon
capture on $^2$H and $^3$He, have used axial current operators
up to order $Q^0$ (one exception is Ref.~\cite{Klos13}, which
included effective one-body reductions, for use in a shell-model study,
of the TPE corrections to the axial current derived in Ref.~\cite{Park03}).
Lastly, there remains the problem of determining
the $z_i$'s in the contact axial charge.  It should be possible to fix at least some
of these LECs by studying muon capture in the few-nucleon systems,
for example, by reproducing data on angular correlation parameters for
the process $^3$He($\mu^-,\nu_\mu)^3$H~\cite{Souder98}, or cross sections
for transitions from the bound state to breakup channels, such as the $^2$H-$n$
two-body breakup, for which data are available~\cite{Kuhn94}.

On a longer time scale, it should be possible to use the weak axial operators constructed
here in quantum Monte Carlo (QMC)~\cite{Carlson15} calculations of $\beta$-decays and
electron- and muon-captures in heavier nuclei with mass number $A>4$ (see
Ref.~\cite{Schiavilla02} for an earlier study of these processes in $^6$He and $^7$Be in the
conventional meson-exchange framework) and of neutrino inclusive cross
sections off light nuclei at low energy and momentum transfers~\cite{Lovato15}.
As a matter of fact, the very recent development
of ``realistic'' and mildly non-local chiral potentials in configuration space~\cite{Piarulli15}, in
which QMC methods are presently formulated, makes it possible to carry out these calculations
in a consistent $\chi$EFT framework (i.e., chiral potentials {\it and} currents), and hence offers
the opportunity to provide first-principles (and numerically exact) predictions, rooted in QCD,
for the rates and cross sections of these weak processes.
\section*{Acknowledgments}
We would like to thank J.\ Goity for a number of interesting discussions during the course
of this work, and
for a critical reading of the manuscript.
A conversation with S.\ Scherer and M.R.\ Schindler is also gratefully acknowledged.
This research is supported by the U.S.~Department of Energy, Office of
Nuclear Physics, under contracts DE-AC05-06OR23177 (A.B.~and R.S.)
and DE-AC52-06NA25396 (S.P.), and by the National Science Foundation
under grant No. PHY-1068305 (S.P.).
\appendix
\section{Chiral Lagrangians}
\label{app:ls}
We adopt the notation and conventions of Ref.~\cite{Fettes00} for the
various fields and covariant derivatives, which we summarize below:
\begin{eqnarray}
\!\!\!\!\!\!U&=&1+\frac{i}{f_\pi}{\bm \tau}\cdot{\bm \pi}-\frac{1}{2\, f_\pi^2}\, 
{\bm \pi}^2-\frac{i\, \alpha}{f_\pi^3}\, {\bm \pi}^2\, {\bm \tau}\cdot{\bm \pi}
+\frac{8\, \alpha-1}{8\,f_\pi^4}\, {\bm \pi}^4
 +\dots  \ , \\
\!\!\!\!\!\!u&=&\sqrt{U}=1+\frac{i}{2\, f_\pi}{\bm \tau}\cdot{\bm \pi}-\frac{1}{8\, f_\pi^2}\, 
{\bm \pi}^2-\frac{i\,(8\, \alpha-1)}{16\, f_\pi^3}\, {\bm \pi}^2\, {\bm \tau}\cdot{\bm \pi}
+\frac{(32\, \alpha-5)}{128\, f_\pi^4}\, {\bm \pi}^4+\dots \ ,\\
\!\!\!\!\!\!u_\mu&=&i\left[ u^\dagger (\partial_\mu -i \, r_\mu )\, u-u\,(\partial_\mu -i \, l_\mu )\, u^\dagger \right] \ ,\\
\!\!\!\!\!\!D_\mu U&=& \partial_\mu U-i \, r_\mu\, U+i\, U\, l_\mu \ ,\\
\!\!\!\!\!\!D_\mu N&=& (\partial_\mu +\Gamma_\mu )\, N
= \partial_\mu N+\frac{1}{2}\left[u^\dagger (\partial_\mu -i \, r_\mu )\, u
+u\,(\partial_\mu -i \,l_\mu )\, u^\dagger \right] N\ , \\
\!\!\!\!\!\!F_{\mu\nu}^{\pm}&=&u^{\dagger}\, F_{\mu\nu}^R\, u\pm u\, F_{\mu\nu}^{L}\, u^{\dagger}\ ,\\
\!\!\!\!\!\!F_{\mu\nu}^R&=&\partial_{\mu}r_\nu-\partial_\nu r_\mu-i\, [\, r_\mu\, ,\, r_\nu\, ]\ ,\qquad r_\mu = v_\mu +a_\mu\ ,\\
\!\!\!\!\!\!F_{\mu\nu}^L&=&\partial_{\mu}l_\nu-\partial_\nu l_\mu-i\, [\, l_\mu\, ,\, l_\nu\, ] \ ,\qquad l_\mu = v_\mu -a_\mu\ , \\
\!\!\!\!\!\!\chi_\pm&=&u^\dagger\, \chi\, u\pm u\, \chi^\dagger\, u=m_\pi^2\left( U^\dagger \pm U \right)\ .
\end{eqnarray}
The parameter $\alpha$ is arbitrary because of the freedom in the choice
of pion field---the only constraint is that $U$ be unitary with ${\rm det}\,U=1$.
Common choices are $\alpha=0$ and $\alpha=1/6$ corresponding, respectively,
to the non-linear sigma model $U=(\sigma+i\, {\bm \tau}\cdot {\bm \pi})/f_\pi$
with $\sigma=\sqrt{f_\pi^2-{\bm \pi}^2}$ and to the exponential parametrization
$U={\rm exp}( i\, {\bm \tau}\cdot{\bm \pi}/f_\pi)$.
In the following we consider only the coupling to the
axial-vector field; further, we ignore isospin-symmetry-breaking effects as well as
the coupling to the isoscalar component
of the axial-vector field, and hence
\begin{eqnarray}
r_\mu&=&-l_\mu= \frac{1}{2} \, {\bm \tau} \cdot {\bf A}_\mu \ ,\\
F_{\mu\nu}^R&=&\frac{1}{2} {\bm \tau}\cdot\left 
( \partial_\mu {\bf A}_\nu- \partial_\nu {\bf A}_\mu + {\bf A}_\mu \times {\bf A}_\nu\right) \ , \\
F_{\mu\nu}^L&=&-\frac{1}{2} {\bm \tau}\cdot\left 
( \partial_\mu {\bf A}_\nu- \partial_\nu {\bf A}_\mu - {\bf A}_\mu \times {\bf A}_\nu\right)\ .
\end{eqnarray}
Inserting the expansions for $U$ and $u$ and keeping terms linear in
the axial-vector field, we find:
\begin{eqnarray}
u_\mu&=& -\frac{1}{f_\pi} \left( 1-\frac{\alpha}{f_\pi^2}\, {\bm \pi}^2\right)  {\bm \tau}\cdot \partial_\mu{\bm \pi}  
+  \frac{4\,\alpha-1}{2\, f_\pi^3}\,{\bm \tau}\cdot {\bm \pi}\,{\bm \pi}\cdot\partial_\mu{\bm \pi} \nonumber \\
&&+\,\, {\bm \tau}\cdot {\bf A}_\mu+\frac{1}{2\, f_\pi^2} 
\left[({\bm \tau}\times{\bm \pi})\times {\bm \pi}\right] \cdot {\bf A}_\mu
+ \dots  \ ,\\
D_\mu U &=&i\, {\bm \tau}\cdot\left[ \frac{1}{f_\pi} \,\partial_\mu{\bm \pi}
-\left(1-\frac{{\bm \pi}^2}{2\, f_\pi^2} \right) {\bf A}_\mu \right] 
- \frac{1}{f_\pi^2} \, {\bm \pi}\cdot\partial_\mu{\bm \pi}
+ \frac{1}{f_\pi}\, {\bm \pi}\cdot{\bf A}_\mu + \dots \ ,\\
D_\mu N&=&\bigg[ \partial_\mu+\frac{i}{4\, f_\pi^2}
\left( {\bm \tau}\times {\bm \pi}\right) \cdot \partial_\mu{\bm \pi} 
-\frac{i}{2\, f_\pi} \left(1-\alpha \frac{{\bm \pi}^2}{f_\pi^2} \right) 
\left( {\bm \tau}\times{\bm \pi} \right) \cdot {\bf A}_\mu \nonumber\\
& &+i\frac{(8\, \alpha-1)}{16\, f_\pi^4}{\bm \pi}^2\,\partial_\mu{\bm \pi}\cdot({\bm \pi}\times{\bm \tau})+ \dots \bigg]N \ , \\
F^+_{\mu\nu} &=& \frac{1}{f_\pi}\left( {\bm \tau} \times{\bm \pi} \right)\cdot  {\bf F}_{\mu\nu}+\dots  \ ,\\
F^-_{\mu\nu} &=&
 \left[  {\bm \tau} + \frac{1}{2\, f_\pi^2} \left({\bm \tau} \times{\bm \pi}\right)\times {\bm \pi}  \right]\cdot {\bf F}_{\mu\nu} +\dots  \ ,\\
\chi_+&=& m_\pi^2\left( 2 -\frac{{\bm\pi}^2}{f_\pi^2} \right) +\dots\ , \\
\chi_-&=& -\frac{2\, i}{f_\pi}\, m_\pi^2\, {\bm \tau} \cdot {\bm \pi} +\dots\ ,
\end{eqnarray}
where ${\bf F}_{\mu\nu}\equiv \partial_\mu {\bf A}_\nu-  \partial_\nu {\bf A}_\mu$ and the $\dots$ denote higher powers of the pion field than shown.
\subsection{$\pi N$ sector}
The $\pi N$ Lagrangians up to order $Q^3$ read:
\begin{eqnarray}
\label{eq:ll1} 
{\cal L}^{(1)}_{\pi N} &=& \overline{N} \left ( i\, \slashed{D}-m +\frac{g_A}{2}\,  \slashed{u} \, \gamma_5 \right) N \ ,\\
\label{eq:l2pi}
{\cal L}^{(2)}_{\pi N}&=&\sum_{i=1}^7c_i\, \overline{N}\, O^{(2)}_i\, N \ ,\\
\label{eq:lp} 
{\cal L}^{(3)}_{\pi N}&=&\sum_{i=1}^{23}d_i\, \overline{N}\, O_i^{(3)}\, N \ ,
\end{eqnarray}
with the operators $O^{(2)}_i$ and $O^{(3)}_i$ defined as in Ref.~\cite{Fettes00}.
Here $g_A$ is the nucleon axial coupling constant, and the $c_i$ and $d_i$ are
LECs.  Below, the $\gamma^\mu$, $\gamma_5$, and
$\sigma^{\mu\nu}$ are $\gamma$ matrices and combinations of $\gamma$ matrices
in standard notation~\cite{Gross93}, and $\epsilon^{\mu\nu\rho\sigma}$ is the Levi-Civita
tensor with $\epsilon^{0123}=+1$.

In terms of the expansions above, ${\cal L}^{(1)}_{\pi N}$ is given by
\begin{eqnarray}
{\cal L}^{(1)}_{\pi N} &=& \overline{N } \bigg [ i\, \slashed{\partial}
 -m-\frac{1}{4\, f_\pi^2} ({\bm \tau}\times{\bm \pi})
\cdot \slashed{\partial}\, {\bm \pi}  -\frac{g_A}{2\, f_\pi}\left(1-\frac{\alpha}{f_\pi^2} {\bm \pi}^2\right)
{\bm \tau}\cdot  \slashed{\partial}\,{\bm \pi} \, \gamma_5\nonumber \\
&&+\frac{g_A}{4\, f_\pi^3}(4\, \alpha -1)\,  {\bm \tau} \cdot {\bm \pi}  \, {\bm \pi}
\cdot \slashed{\partial}\,{\bm \pi} \, \gamma_5 
+\frac{(1-8\, \alpha)}{16\, f_\pi^4}{\bm \pi}^2\, \slashed{\partial}\,{\bm \pi}\cdot({\bm \pi}\times{\bm \tau})\nonumber\\
&&+\frac{1}{2\,f_\pi} \left(1-\frac{\alpha}{f_\pi^2} {\bm \pi}^2\right) 
({\bm \tau} \times{\bm \pi })\cdot \slashed{\bf A} +\frac{g_A}{2}\, {\bm \tau}\cdot \slashed{\bf A}\, \gamma_5
+\frac{g_A}{4\, f_\pi^2} \left[({\bm \tau}\times{\bm \pi})\times
 {\bm \pi}\right] \cdot \slashed{\bf A}\, \gamma_5 \bigg] N \nonumber \ ,\\ 
 \end{eqnarray}
 where $\slashed{\partial}=\gamma^\mu\partial_\mu$ and $\slashed{\bf A}=\gamma^\mu {\bf A}_\mu$.
 The operators $O^{(2)}_i$ in the ${\cal L}^{(2)}_{\pi N}$ Lagrangian are expressed as (below the
 notation $\widetilde{\chi}_+ =\chi_+-\langle \chi_+\rangle/2$ is used, where $\langle \dots \rangle$
 implies a trace in isospin space)
\begin{eqnarray}
O^{(2)}_1&=& \langle\chi_+\rangle  \longrightarrow  4\, m_\pi^2\left(1-\frac{{\bm \pi}^2}{2\, f_\pi^2}\right) \ ,\\
O^{(2)}_2&=&-\frac{1}{8\, m^2} \langle u_\mu u_\nu \rangle \, D^{\mu\nu}+{\rm h.c.} \longrightarrow 
\frac{1}{f_\pi^2} \partial_0 {\bm \pi}  \cdot  \partial_0{\bm \pi} -\frac{2}{f_\pi} \partial_0{\bm \pi} \cdot {\bf A}_0 
\nonumber \\
&& \qquad\qquad +\frac{1}{m\, f_\pi}
\left(  \frac{1}{f_\pi} \partial_0{\bm \pi} \cdot  \partial_i{\bm \pi} - \partial_0{\bm \pi} \cdot{\bf A}_i- \partial_i{\bm\pi} \cdot {\bf A}_0\right) 
 \gamma^0 \, i\overleftrightarrow{\partial}^i 
 \label{eq:a25}\\
O^{(2)}_3&=& \frac{1}{2} \langle u_\mu u^\mu \rangle \longrightarrow
 \frac{1}{f_\pi^2} \, \partial_\mu{\bm \pi} \cdot  \partial^\mu{\bm \pi}-\frac{2}{f_\pi}\,  \partial_\mu{\bm \pi} \cdot{\bf A}^\mu\ , \\
O^{(2)}_4&=& \frac{i}{4}\left[u_\mu\, ,\, u_\nu\right]\sigma^{\mu\nu} \longrightarrow 
 \frac{1}{2}\, {\bm \tau}\cdot \left(-\frac{1}{f^2_\pi} \,\partial_\mu{\bm\pi}\times\partial_\nu{\bm \pi}
+\frac{2}{f_\pi}\, {\bf A}_\mu\times\partial_\nu{\bm \pi}\right)\sigma^{\mu\nu} \ , \\
O^{(2)}_5&=& \widetilde{\chi}_+ \longrightarrow 0\ , \\
O^{(2)}_6&=& \frac{1}{8\, m} F_{\mu\nu}^+\, \sigma^{\mu\nu} \longrightarrow 
\frac{1}{4\,m\,f_\pi} \left( {\bm \tau} \times {\bm \pi} \right) \cdot  \partial_\mu {\bf A}_\nu \, \sigma^{\mu\nu} \ ,\\
O^{(2)}_7&=& \frac{1}{8\, m}\langle F_{\mu\nu}^+\rangle\,  \sigma^{\mu\nu} \longrightarrow 0 \ ,
\end{eqnarray}
while those in the ${\cal L}^{(3)}_{\pi N}$ Lagrangian reduce to
\begin{eqnarray}
O^{(3)}_1&=&-\frac{1}{2\, m}[u_\mu\, ,\, [D_\nu\, ,\, u^\mu]]D^\nu +\mbox{h.c.}\longrightarrow
 \frac{2}{f_\pi}\, {\bm \tau} \cdot \Big( -\frac{1}{f_\pi}\partial_\mu{\bm \pi}\times\partial_0\partial^\mu{\bm \pi}+
 {\bf A}_\mu \times \partial_0\partial^\mu {\bm \pi} \nonumber\\
&&\qquad\qquad\qquad\qquad\qquad -\partial_0 {\bf A}_\mu \times \partial^\mu{\bm \pi} \Big) \gamma^0\ , \\
O^{(3)}_2&=&-\frac{1}{2\, m} [u_\mu\, ,\, [D^\mu\, , \, u_\nu]]D^\nu +\mbox{h.c.} \longrightarrow \frac{2}{f_\pi}\, {\bm \tau} \cdot 
\Big(-\frac{1}{f_\pi}\partial_\mu{\bm \pi}\times\partial^\mu\partial_0{\bm \pi}+ {\bf A}_\mu \times \partial_0\partial^\mu {\bm \pi} \nonumber \\
&&\qquad\qquad\qquad\qquad\qquad -\partial_\mu {\bf A}_0   \times \partial^\mu{\bm \pi}  \Big) \gamma^0\ , \\
O^{(3)}_3&=& \frac{1}{12\, m^3} [u_\mu\, ,\, [D_\nu\, ,\, u_\rho]] D^{\mu\nu\rho}  +\mbox{h.c.}\longrightarrow \frac{2}{f_\pi} \, {\bm \tau} \cdot 
\Big(-\frac{1}{f_\pi}\partial_0{\bm{\pi}}\times\partial_0^2{\bm{\pi}}+ {\bf A}_0 \times \partial^{\, 2}_0\, {\bm \pi} \nonumber\\
&&\qquad\qquad\qquad\qquad\qquad-\partial_0 {\bf A}_0  \times \partial_0 \, {\bm \pi} \Big) \gamma^0\ , \\
O^{(3)}_4&=& - \frac{1}{2\, m}  \, \epsilon^{\mu\nu\alpha\beta}\, \langle u_\mu u_\nu u_\alpha \rangle D_\beta+\mbox{h.c.}\longrightarrow 0 \ ,\\
O^{(3)}_5&=&  \frac{i}{2\, m} \,[\chi_-\, ,\, u_\mu]D^{\mu}+\mbox{h.c.}\longrightarrow -\frac{4\, m^2_\pi}{f_\pi}{\bm \tau}\cdot \Big[ {\bm \pi}\times
\Big( \frac{1}{f_\pi} \partial_0{\bm \pi}- {\bf A}_0\Big) \Big] \gamma^0\ , \\
O^{(3)}_6&=&  \frac{i}{2\, m} \, [D^\mu\, ,\, \widetilde{F}_{\mu\nu}^+]D^\nu+\mbox{h.c.} \longrightarrow  \partial^i F_{i\, 0}^+ \, \gamma^0 \ ,\\
O^{(3)}_7&=&  \frac{i}{2\, m} \, [D^\mu\,  ,\,  \langle F_{\mu\nu}^+ \rangle]D^\nu+\mbox{h.c.} \longrightarrow 0\ , \\
O^{(3)}_8&=&  \frac{i}{2\, m} \, \epsilon^{\mu\nu\alpha\beta}\, \langle \widetilde{F}_{\mu\nu}^+u_\alpha \rangle D_\beta+\mbox{h.c.}
\longrightarrow 0 \ ,\\
 O^{(3)}_9&=&  \frac{i}{2\, m} \,\epsilon^{\mu\nu\alpha\beta}\, \langle F_{\mu\nu}^+ \rangle \, u_\alpha\, D_\beta+\mbox{h.c.} \longrightarrow 0 \ ,\\
  O^{(3)}_{10}&=&  \frac{1}{2}\, \gamma^\mu\gamma_5 \, \langle u\cdot u  \rangle \, u_\mu \longrightarrow 0\ , \\
  O^{(3)}_{11}&=&  \frac{1}{2}\,  \gamma^\mu\gamma_5 \, \langle u_\mu u_\nu \rangle \, u^\nu \longrightarrow 0 \ ,\\
  O^{(3)}_{12}&=& - \frac{1}{8\, m^2}\,  \gamma^\mu\gamma_5 \, \langle u_\lambda u_\nu \rangle \, u_\mu\, D^{\lambda\nu}+\mbox{h.c.}\longrightarrow 0\ ,\\
 O^{(3)}_{13}&=& - \frac{1}{8\, m^2}\, \gamma^\mu\gamma_5\, \langle u_\mu u_\nu \rangle \, u_\lambda \, D^{\lambda\nu}+\mbox{h.c.}\longrightarrow 0\ ,\\
O^{(3)}_{14}&=&  \frac{i}{4\, m} \,\sigma^{\mu\nu}\langle [D_\lambda\, , \, u_\mu] \, u_\nu \rangle \, D^{\lambda}+\mbox{h.c.} 
\longrightarrow  \frac{1}{f_\pi} \Big( \frac{1}{f_\pi} \partial_0\partial_i{\bm \pi} \cdot \partial_j{\bm \pi}- \partial _0 \partial_i {\bm \pi}
\cdot {\bf A}_j\nonumber \\
&&\qquad\qquad\qquad\qquad\qquad - \partial_0 {\bf A}_i \cdot  \partial_j {\bm \pi} \Big) \, \sigma^{ij} \gamma^0\ , \\
 O^{(3)}_{15}&=&  \frac{i}{4\, m} \, \sigma^{\mu\nu} \langle u_\mu [D_\nu\, , \, u_\lambda]  \rangle D^{\lambda}+\mbox{h.c.}
 \longrightarrow  \frac{1}{f_\pi} \Big( \frac{1}{f_\pi}\partial_i{\bm \pi} \cdot  \partial_0 \partial_j{\bm \pi}- \partial_i {\bm \pi}
\cdot  \partial _j{\bf A}_0\nonumber \\
&&\qquad\qquad\qquad\qquad\qquad - {\bf A}_i \cdot  \partial_0 \partial_j {\bm \pi} \Big) \, \sigma^{ij} \gamma^0 \ ,\\
O^{(3)}_{16}&=&  \frac{1}{2} \,\gamma^\mu\gamma_5\, \langle \chi_+\rangle \, u_\mu\longrightarrow 
2\, m_\pi^2\, {\bm \tau} \cdot \Big( -\frac{1}{f_\pi} \partial_i {\bm \pi} 
+{\bf A}_i \Big)\gamma^i \gamma_5 \ ,\\
O^{(3)}_{17}&=&  \frac{1}{2} \, \gamma^\mu\gamma_5\, \langle \chi_+\, u_\mu\rangle \longrightarrow 0 \ ,\\
O^{(3)}_{18}&=&  \frac{i}{2} \, \gamma^\mu\gamma_5\, [D_\mu\, ,\, \chi_-]\longrightarrow 
 \frac{m_\pi^2}{f_\pi} \,  {\bm \tau}\cdot    \partial_i {\bm \pi} \,  \gamma^i \gamma_5  \ ,\\
O^{(3)}_{19}&=&  \frac{i}{2} \,\gamma^\mu\gamma_5\, [D_\mu\, ,\, \langle \chi_-\rangle ]\longrightarrow 0 \ ,\\
O^{(3)}_{20}&=& - \frac{i}{8\, m^2} \, \gamma^\mu\gamma_5\, [\widetilde{F}_{\mu\nu}^+\, ,\, u_\lambda]\, 
D^{\lambda\nu}+\mbox{h.c.}\longrightarrow 0\ , \\
O^{(3)}_{21}&=&  \frac{i}{2} \,\gamma^\mu\gamma_5 \, [\widetilde{F}_{\mu\nu}^+\, ,\, u^\nu]\longrightarrow 0\ , \\
O^{(3)}_{22}&=&  \frac{1}{2} \,\gamma^\mu\gamma_5\, [D^\nu\, ,\, F_{\mu\nu}^-]\longrightarrow \frac{1}{2}\, 
{\bm \tau} \cdot  \partial^\nu{\bf F}_{i\nu}\, \gamma^i \gamma_5\ , \\
O^{(3)}_{23}&=&  \frac{1}{2} \, \gamma_\mu\gamma_5\, \epsilon^{\mu\nu\alpha\beta}\, \langle u_\nu F^-_{\alpha\beta}\rangle
 \longrightarrow -\frac{1}{f_\pi} \, \epsilon^{i \nu \alpha \beta}
\partial_\nu {\bm \pi} \cdot {\bf F}_{\alpha\beta}\, \gamma_i \gamma_5 \ .
\end{eqnarray}
Several comments are now in order.  First, the expressions above for ${\cal L}^{(1)}_{\pi N}$, ${\cal L}^{(2)}_{\pi N}$, and
${\cal L}^{(3)}_{\pi N}$ retain all terms relevant in the present study.  Typically, these
include at most three pion, two pion, and one pion fields for $n=1,2,3$ in ${\cal L}^{(n)}_{\pi N}$,
respectively.  In some instances, for example in $O^{(3)}_1$, terms with two pion fields
are also considered for reasons having to do with the treatment of tadpole-type contributions (see below).
The Lagrangian $\sum_n {\cal L}^{(n)}_{\pi N}$ can now conveniently be expressed as given in Eq.~(\ref{eq:2.3})
with the quantities $\Gamma^0_a(n)$, $\Lambda^i_a(n)$, and $\Delta(n)$, defined in
Eqs.~(\ref{eq:eg0})--(\ref{eq:edelta}), given at leading order by 
\begin{eqnarray}
\Gamma_a^0(0) &=& -\frac{1}{4\, f_\pi^2} ({\bm \tau}\times{\bm \pi})_a\, \gamma^0\ +\frac{8\,\alpha-1}{16\, f_\pi^4}{\bm \pi}^2\,({\bm \tau}\times{\bm \pi})_a\,  \gamma^0 ,
\label{eq:g0}\\
\Lambda_a^i(0) &=&-\frac{g_A}{2\, f_\pi}  \left(1-\frac{\alpha}{f_\pi^2} {\bm \pi}^2\right)\, \tau_a \, \gamma^i \gamma_5 +\frac{g_A}{4\, f_\pi^3}(4\,\alpha-1)({\bm \tau}\cdot {\bm \pi})\, \pi_a \, \gamma^i\gamma_5 \ , 
\label{eq:l0}\\
\Delta(1)&=&\frac{g_A}{2}\, {\bm \tau}\cdot {\bf A}_i\,\gamma^i \gamma_5 +\frac{1}{2\,f_\pi} \left(1-\frac{\alpha}{f_\pi^2} {\bm \pi}^2\right) ({\bm \tau} \times{\bm \pi })\cdot {\bf A}_0\, \gamma^0 \nonumber\\
&&+\frac{g_A}{4\, f_\pi^2} \left[({\bm \tau}\times{\bm \pi})\times {\bm \pi}\right] \cdot {\bf A}_i\,\gamma^i \gamma_5 \ ;
\label{eq:d1}
\end{eqnarray}
at next-to-leading order by
\begin{eqnarray}
\Gamma_a^0(1)&=& -\frac{g_A}{2\, f_\pi} \left(1-\frac{\alpha}{f_\pi^2} {\bm \pi}^2\right)\, \tau_a \, \gamma^0 \gamma_5 +\frac{g_A}{4\, f_\pi^3}(4\,\alpha-1)({\bm \tau}\cdot {\bm \pi})\, \pi_a \, \gamma^0\gamma_5  -2\,\frac{c_2+c_3}{f_\pi}\, A^0_a \ ,
\label{eq:g1}\\
\Lambda_a^i(1)&=&  -\frac{1}{4\, f_\pi^2} ({\bm \tau}\times{\bm \pi})_a\, \gamma^i+\frac{c_3}{f^2_\pi}\, \partial^i\pi_a-2\,\frac{c_3}{f_\pi}\, A^i_a -\frac{c_4}{f_\pi}\,
\left({\bm \tau} \times{\bf A}_j\right)_a \sigma^{ij}\nonumber\\
&& +\frac{c_4}{2\, f^2_\pi}\left({\bm \tau}\times \partial_j{\bm \pi}\right)_a\sigma^{ij}+\frac{(1-8\alpha)}{8\, f_\pi^4}{\bm \pi}^2\,({\bm \pi}\times{\bm \tau})_a\,  \gamma^i
\label{eq:l1} \ , \\
\Delta(2)&=&\frac{g_A}{2}{\bm \tau}\cdot {\bf A}_0\, \gamma^0\gamma^5+\frac{1}{2\, f_\pi} \left(1-\frac{\alpha}{f^2_\pi} {\bm \pi}^2\right) \left( {\bm \tau} \times{\bm \pi}\right)\cdot{\bf A}_i\, \gamma^i
+\frac{g_A}{4f^2_\pi}\left[\left({\bm \tau}\times{\bm \pi}\right)\times{\bm\pi}\right]\cdot {\bf A}_0\, \gamma^0\gamma_5 \nonumber\\
&&+4\, m_\pi^2\,  c_1\left(1-\frac{{\bm \pi}^2}{2\, f_\pi^2}\right)+\frac{c_6}{4\, m\, f_\pi} \, \left({\bm \tau}\times{\bm \pi}\right)
\cdot \partial_i {\bf A}_j \, \sigma^{ij} \ ;
\label{eq:d2}
\end{eqnarray}
and at next-to-next-to-leading order by
\begin{eqnarray}
\Gamma_a^0(2)&=&\frac{c_2}{m\, f_\pi} \left( \frac{1}{f_\pi} \partial_i\pi_a-A_{i,a}\right)\gamma^0\, i\overleftrightarrow{\partial}^i
+\frac{c_4}{f_\pi}\,\left[\frac{1}{f_\pi}\left({\bm \tau}\times \partial_i{\bm \pi}\right)_a-\left({\bm \tau} \times{\bf A}_i\right)_a\,\right] \sigma^{0 i} \,
\nonumber\\
&&+2\frac{d_1+d_2}{f_\pi^2}\left[\left({\bm \tau}\times\partial^i\partial_i{\bm \pi}\right)+\left({\bm \tau}\times\partial^i{\bm \pi}\right)\widetilde{\overleftrightarrow{\partial}}_i\right]\gamma^0 \nonumber\\
&&+2\,\frac{d_1+d_2+d_3}{f_\pi}\,
\left[ -\frac{m_\pi^2}{f_\pi} \left({\bm \tau} \times {\bm \pi} \right)_a -\frac{1}{f_\pi} \left({\bm \tau} \times\partial^i\partial_i {\bm \pi} \right)_a+\left({\bm \tau} \times\partial_i{\bf A}^i\right)_a\right]\gamma^0\nonumber\\
&&-4\, d_5\frac{m^2_\pi}{f^2_\pi}\left({\bm\tau} \times{\bm \pi}\right)_a \gamma^0
+\frac{d_{14}-d_{15}}{f_\pi}\bigg[\frac{1}{f_\pi}\,\partial_i\pi_a\,\sigma^{ij}\widetilde{\overleftrightarrow{\partial}}_j+\partial_i A_{j,a}\,\sigma^{ij}\nonumber\\
&&+ A_{j,a}\,\sigma^{ij}\widetilde{\overleftrightarrow{\partial}}_i\bigg]\gamma^0 +\frac{d_{23}}{f_\pi}\epsilon^{0ijk} F_{jk,a}\,\gamma_i \gamma_5 \ ,
\label{eq:g2}
\end{eqnarray}
\begin{eqnarray}
\Lambda_a^i(2)&=& -\frac{c_2}{m\, f_\pi}\, A_{0,a}\,  \gamma^0\, i\overleftrightarrow{\partial}^i
 +\frac{c_4}{f_\pi}\left({\bm \tau} \times{\bf A}_0\right)_a \sigma^{0i}
 -2\, \frac{d_1}{f_\pi} \left({\bm \tau} \times \partial_0{\bf A}^i\right)_a\gamma^0\nonumber\\
&&-2\, \frac{d_2}{f_\pi} \left({\bm \tau} \times \partial^i{\bf A}_0\right)_a\gamma^0-
\frac{d_6}{f_\pi}\left({\bm \tau}\times{\bf F}^{i\,0}\right)_a\gamma^0
 \nonumber\\
&&+\frac{d_{14}}{f_\pi}\partial_0A_{j,a}\sigma^{ij}\gamma^0
-\frac{d_{15}}{f_\pi}\partial_j A_{0,a}\sigma^{ij}\gamma^0\nonumber\\
&&-\frac{m_\pi^2}{f_\pi} \left( 2\, d_{16}-d_{18}\right) \tau_a\, \gamma^i \gamma_5 
+2\,\frac{d_{23}}{f_\pi}\,
\epsilon^{ijk0} F_{k0,a}\gamma_j\gamma_5 \ ,
\label{eq:l2}
\end{eqnarray}
\begin{eqnarray}
\Delta(3)\!\!&=&\!\!\frac{c_6}{4\, m\,f_\pi} \left({\bm \tau}\times {\bm \pi}\right) \cdot
\left( \partial_0{\bf A}_i-\partial_i {\bf A}_0 \right)\sigma^{0i} -
2\,\frac{d_1+d_2+d_3}{f_\pi}
\left({\bm \tau} \times{\bf A}_0\right) \cdot\left(\partial^i\partial_i{\bm \pi}+m_\pi^2{\bm \pi}\right) \gamma^0 \nonumber\\
&&
+ 4\, d_5\, \frac{m_\pi^2}{f_\pi}\, {\bm \tau}\cdot\left( {\bm \pi} \times {\bf A}_0\right)  \gamma^0
+\frac{d_6}{f_\pi}\, \left({\bm \tau}\times{\bm \pi}\right)\cdot\partial^i {\bf F}_{i\, 0}\, \gamma^0
+ 2\, m_\pi^2\, d_{16} \,{\bm \tau}\cdot{\bf A}_i \, \gamma^i \gamma_5 \nonumber\\
&&+\frac{d_{22}}{2}\, {\bm \tau}\cdot\partial^\nu {\bf F}_{i\nu}\gamma^i\gamma_5  \ .
\label{eq:d3}
\end{eqnarray}

Second, the various derivatives act only on the field to their immediate right, for example $\partial_0{\bm \pi} \cdot{\bf A}_0$
means $\left( \partial_0{\bm \pi}\right)\cdot{\bf A}_0$.  However, the symbols $\overleftrightarrow{\partial}_i 
=\overrightarrow{\partial}_i -\overleftarrow{\partial}_i$ and $\widetilde{\overleftrightarrow{\partial}}_i
=\overrightarrow{\partial}_i +\overleftarrow{\partial}_i$ in Eqs.~(\ref{eq:a25}) and~(\ref{eq:g2})--(\ref{eq:l2}) denote
derivatives acting {\it only} on the right and left nucleon fields, respectively.

Third, the power counting $Q^n$ of ${\cal L}^{(n)}_{\pi N}$ counts powers of derivatives
of the pion field (or of pion mass factors) and factors of $A_a^\mu$ and its derivatives (note
that $A_a^\mu$ is counted as being of order $Q$).  However, the Lorentz structure of the
terms may lead to additional suppression.  For example, in ${\cal L}^{(1)}_{\pi N}$ a term like
\begin{equation}
-\frac{1}{4\, f_\pi^2} ({\bm \tau}\times{\bm \pi})\cdot \partial_0 {\bm \pi}\, \gamma^0 \ ,\nonumber
\end{equation}
is of order $Q$, but a term like
\begin{equation}
-\frac{g_A}{2\, f_\pi} \left(1-\frac{\alpha}{f_\pi^2} {\bm \pi}^2\right)\, 
{\bm \tau} \cdot \partial_0 {\bm \pi} \, \gamma^0 \gamma_5 \nonumber \ ,
\end{equation}
which is nominally of order $Q$, is in fact of order $Q^2$, since $\overline{N} \gamma^0 \gamma_5 N$
couples the lower to the upper components of the spinors, and therefore involves the three-momenta
of the initial and final nucleons (of order $Q$).  We have taken advantage of this suppression
in some of the terms $O^{(3)}_i$ in ${\cal L}^{(3)}_{\pi N}$ by retaining only
the diagonal piece in their Lorentz structure, for example in term $O^{(3)}_{14}$.

Fourth, time derivatives of the nucleon fields in ${\cal L}^{(2)}_{\pi N}$ and ${\cal L}^{(3)}_{\pi N}$
are removed by making use of the equation of motion (to order $Q$)
\begin{equation}
\partial_0 N=-i\, m\, \gamma^0 N+\left[ -\gamma^0 \gamma^i \partial_i +i\, \gamma^0\, 
 \Gamma^0_a(0)\, \partial_0\pi_a  +i\, \gamma^0\Lambda^i_a(0)\, \partial_i \pi_a +i\, \gamma^0\, \Delta(1)\right] N \ ,
  \end{equation}
implying that
\begin{eqnarray}
\!\!\!\!\partial_0^2N\!\!&=&\!\!-m^2\, N-i\, m\, \gamma^0 \big[ \dots \big] N-i\, m\, \big[ \dots \big]\gamma^0 N \\
\!\!\!\!\!\!&=&\!\!-m^2\, N+\Big[  
-\frac{m\, g_A}{f_\pi}{\bm \tau}\cdot \partial_i{\bm \pi} \, \gamma^i\gamma^5 +m\, g_A\, {\bm \tau}\cdot {\bf A}_i\, \gamma^i\gamma_5
-\frac{m}{f_\pi} {\bm \tau} \cdot ({\bf A}_0 \times{\bm \pi}) \gamma^0 \Big]N \ ,
\end{eqnarray}
where in the second line we have ignored non-linear terms in the pion field, since they do not contribute
to the order of interest here.

Fifth, double time derivatives of the pion fields in ${\cal L}_{\pi N}^{(3)}$ are removed by
making use of the equation of motion, see Eq.~(\ref{eq:eom-pi}) below.  Terms containing
both one time derivative and one space derivative of the pion fields have been rewritten by
integrating by parts.  For example, in ${\cal L}_{\pi N}^{(3)}$ a term like
\[
2\frac{d_1+d_2}{f_\pi^2}\, \overline{N}\left({\bm \tau}\times\partial_0\partial^i{\bm \pi}\right)\cdot\partial_i{\bm \pi}N\ ,
\]
can be re-expressed, modulo a total divergence, as
\[
-2\frac{d_1+d_2}{f_\pi^2}\overline{N}\left[\left({\bm \tau}\times\partial_0{\bm \pi}\right)\cdot\partial^i\partial_i{\bm \pi}+\left({\bm \tau}\times\partial_0{\bm \pi}\right)\cdot\partial^i{\bm \pi}\widetilde{\overleftrightarrow{\partial}}_{\!i}\right]N\ .
\]

\subsection{$\pi\pi$ Sector}

The $\pi\pi$ Lagrangians up to order $Q^4$ read~\cite{Sch12}:
\begin{eqnarray}
{\cal L}^{(2)}_{\pi \pi}&=& \frac{f_\pi^2}{4}\, \langle\, D_\mu U \,  \left(D^\mu U\right)^\dagger
+\chi_+ \, \rangle \\
{\cal L}^{(4)}_{\pi \pi}&=&
\frac{l_1}{4} \langle  D_\mu U\left(D^\mu U\right)^\dagger \rangle \,\langle  D_\nu U\, \left(D^\nu U\right)^\dagger \rangle+
\frac{l_2}{4}\langle D_\mu U  \left(D_\nu U\right)^\dagger\rangle\, \langle D^\mu U \left(D^\nu U\right)^\dagger\rangle
+\frac{l_3}{16}\langle\, \chi_+ \,\rangle^2\nonumber\\
&&+\frac{l_4}{16}\left[ 2\, \langle\,
 D_\mu U \left( D^\mu U\right)^\dagger \, \rangle \langle\, \chi_+ \, \rangle +2\, 
 \langle\, \chi^\dagger U\, \chi^\dagger U+
 \chi\, U^\dagger \chi \,U^\dagger\rangle-\langle\chi_-\rangle^2-4\, \langle\, \chi^\dagger \chi\rangle\, \right]\nonumber\\
 &&+l_5\Big(\langle F_{\mu\nu}^R\, U\, F^{\mu\nu}_L \, U^\dagger\rangle 
 -\frac{1}{2}\langle F_{\mu\nu}^LF^{\mu\nu}_L+F_{\mu\nu}^RF^{\mu\nu}_R\rangle\Big) \nonumber\\
&& +i\, \frac{l_6}{2}\langle \, F_{\mu\nu}^R\, D^\mu U\left(D^\nu U\right)^\dagger
+F_{\mu\nu}^L \left( D^\mu U\right)^\dagger \, D^\nu U \rangle
-\frac{l_7}{16}\langle\chi_-\rangle^2+\frac{ h_1+h_3 }{4}\langle\chi\chi^\dagger\rangle\nonumber\\
&&+\frac{h_1-h_3}{16}\left( \langle\chi_+\rangle^2+\langle\chi_-\rangle^2 
-2\, \langle\chi \,U^\dagger\, \chi \,U^\dagger+U\, \chi^\dagger \, U\, \chi^\dagger\rangle\right)\nonumber\\
&&-2\, h_2\langle F_{\mu\nu}^L\, F^{\mu\nu}_L+F_{\mu\nu}^R\, F^{\mu\nu}_R\rangle \ ,
\end{eqnarray}
where in the absence of isospin symmetry breaking (which is assumed throughout
the present work) $\chi$ is proportional to the identity matrix, namely $\chi=m_\pi^2$, and $\langle \chi_-\rangle$ vanishes.
Furthermore, the terms proportional to the LECs $l_1$, $l_2$, $l_5$, $l_6$, and $h_i$ do not contribute
to the order of interest. 
The symmetric matrices $\widetilde{G}_{ab}$, $G_{ab}$, $H_{ab}$, and $F_{ab}$ in the Lagrangian of Eq.~(\ref{eq:2.3})
are obtained as
\begin{eqnarray}
\label{eq:ggtildepi}
\widetilde{G}_{ab}&=& \left(1-\frac{2\, \alpha}{f_\pi^2}
{\bm \pi}^2+2\, l_4 \frac{m_\pi^2}{f_\pi^2}
\right) \delta_{ab} -\frac{4\, \alpha-1}{f_\pi^2}\, \pi_a \pi_b \ , \\
\label{eq:ggpi}
G_{ab}&=& \widetilde{G}_{ab}+
2\,\frac{c_2+c_3}{f_\pi^2}\, \overline{N} N  \,\delta_{ab}  \ , \\
\label{eq:hhpi}
H_{ab}&=&\left[1-\frac{8\, \alpha-1}{4\, f_\pi^2} {\bm \pi}^2
 +2\, \left( l_3+l_4\right)\frac{m_\pi^2}{f_\pi^2}\right]\delta_{ab}\ , \\
 \label{eq:ffpi}
F_{ab}&=& \left(1-\frac{2\, \alpha+1}{2\, f_\pi^2}
{\bm \pi}^2+2 \, l_4 \frac{m_\pi^2}{f_\pi^2}\right) \delta_{ab} -\frac{2\, \alpha-1}{f_\pi^2}\, \pi_a \pi_b \ .
\end{eqnarray}
By retaining only terms linear in the pion field and external axial field,
the equation of motion implied by ${\cal L}_{\pi\pi}^{(2)}$ is
\begin{eqnarray}
\partial_0^{\,2}{\bm\pi}&=&-\left(\partial^i\partial_i+m_\pi^2\right){\bm\pi}+f_\pi\partial_0{\bf A}^{\!0}+f_\pi\partial_i{\bf A}^i\, .\label{eq:eom-pi}
\end{eqnarray}

\section{Interaction vertices}
\label{app:vert}

In this appendix we report expressions for the vertices corresponding to the interaction terms in the
Hamiltonian of Eq.~(\ref{eq:hhii}), which we write as
\begin{eqnarray}
 H_I&=&\sum_{n=1}^3\Big[\left(H_{\pi NN}^{(n)}+H_{2\pi NN}^{(n)} +H_{3\pi NN}^{(n)}+\cdots \right) 
 +\left( H_{NNA}^{(n)}+H_{\pi NNA}^{(n)}+H_{2\pi NNA}^{(n)}+\cdots\right)\Big] \nonumber \\
&&+\sum_{m=1}^2\Big[ \left( H_{2\pi}^{(2m)}+H_{4\pi}^{(2m)}+\cdots\right)+
\left(H_{\pi A}^{(2m)}+H_{3\pi A}^{(2m)}+\cdots \right)\Big] \ ,
\end{eqnarray}
where the superscript $n$ denotes the power counting $Q^n$ and the subscript
specifies the number of pion, nucleon, and axial fields entering a given interaction
term.  We use the following notation: $\lambda={\bf p}\, \sigma\tau$
($\lambda^\prime={\bf p}^\prime\, \sigma^\prime \tau^\prime$)
are the momentum and spin and isospin projections of the initial (final) nucleon;
${\bf k}_1,{\bf k}_2,\dots $ and $a_1,a_2,\dots$ are the momenta and isospin projections of pions
$1,2,\dots$ with energies $\omega_1,\omega_2,\dots$, where $\omega_i=\sqrt{k_i^2+m_\pi^2}$;
${\bf q}$ and $a$ denote the momentum and isospin projection of the external
axial field with energy $\omega_q$ and its spatial and time derivatives expressed as
${\bm \nabla} A^\mu_a \longrightarrow i\, {\bf q} \,A^\mu_a$ and $\partial_0 A^\mu_a \longrightarrow-i\, \omega_q\, A^\mu_a$. 
We also define ${\bf P}=({\bf p}^\prime+{\bf p})/2$
and the (infinite) constants
\begin{equation}
J_{mn}=\int \frac{{\rm d}{\bf l}}{(2\pi)^3} \frac{l^{\, 2 m}}{\omega_l^{\, n}} \ .
\label{eq:jmn}
\end{equation}
\subsection{$\pi NN$ vertices}
The interaction terms read
\begin{eqnarray}
\label{eq:b3}
H^{(1)}_{\pi NN}&=&\frac{g_A}{2f_\pi}\int{\rm d}{\bf x}\, \overline{N}{\bm\tau}\cdot\partial_i{\bm \pi}\gamma^i\gamma^5 N \ ,\\
H^{(2)}_{\pi NN}&=&\frac{g_A}{2f_\pi}\int{\rm d}{\bf x}\,\overline{N}{\bm\tau}\cdot{\bm \Pi}\gamma^0\gamma^5 N \ , \\
H^{(3)}_{\pi NN}&=&\frac{m^2_\pi}{f_\pi}(2\, d_{16}-d_{18})
\int{\rm d}{\bf x}\,\overline{N}{\bm \tau}\cdot{\partial_i{\bm \pi}}\gamma^i\gamma^5 N \ ,
\end{eqnarray}
from which the following vertices for pion absorption are obtained
\begin{eqnarray}
\langle \lambda^\prime\! \mid H^{(1)}_{\pi NN}\mid\! \lambda;{\bf k},a\rangle&=&i\frac{g_A}{2f_\pi}\tau_a\, {\bm \sigma}\cdot{\bf k} \ ,\\
\langle \lambda^\prime\! \mid H^{(2)}_{\pi NN}\mid\! \lambda;{\bf k},a\rangle&=&-i\frac{g_A}{2\, m\,f_\pi} \tau_a\, \omega\, {\bm \sigma}\cdot{\bf P} \ , \\
\langle \lambda^\prime\! \mid H^{(3)}_{\pi NN}\mid\! \lambda;{\bf k},a\rangle&=&
i \frac{m^2_\pi}{f_\pi}(2\, d_{16}-d_{18})
 \tau_a\, {\bm \sigma}\cdot{\bf k}+
 i\frac{g_A}{8m^2f_\pi}\tau_a\bigg[2\,{\bm\sigma}\cdot{\bf P}\, {\bf k}\cdot{\bf P}\nonumber\\
 &&-{\bm\sigma}\cdot\left({\bf p}^\prime-{\bf p}\right)\frac{\left({\bf p}^\prime-{\bf p}\right)\cdot{\bf k}}{2}-2\,P^2\, {\bm \sigma}\cdot{\bf k}-i{\bf k}\cdot \left({\bf p}^\prime-{\bf p}\right)\times{\bf P}\bigg]\ ,
\end{eqnarray}
where on the r.h.s.~of the above equations the $1/\sqrt{2\,\omega}$ normalization factor from the pion field expansion in
normal modes, the initial and final spin-isospin states of the nucleons, and the
three-momentum conserving $\delta$-function $(2\pi)^3 \delta({\bf p}^\prime-{\bf p}-{\bf k})$
have been dropped for simplicity.  We will continue to do so in the equations to follow.  Vertices
in which the pion is in the final state (pion emission) are obtained from those above by
the replacements $\omega, {\bf k}\longrightarrow -\omega,-{\bf k}$.  Lastly, only the leading order
is retained in the non-relativistic expansion of the Lorentz structures associated with the various interaction
terms (here and to follow) unless otherwise noted.
\subsection{$2\pi NN$ vertices}
The interaction term reads
\begin{eqnarray}
H^{(1)}_{2\pi NN}&=&\frac{1}{4f_\pi^2}\int{\rm d}{\bf x}\, \overline{N}\, {\bf \Pi}\cdot\left({\bm \tau}\times{\bm \pi}\right)\gamma^0 N \ ,\\
H^{(2)}_{2\pi NN}&=& \int{\rm d}{\bf x}\,\overline{N}\,\bigg[ \frac{1}{4 f_\pi^2}\partial_i{\bm \pi}\cdot\left({\bm \tau}\times{\bm \pi}\right)\gamma^i  +c_1 \frac{2\,m_\pi^2}{f_\pi^2}{\bm\pi}^2-\frac{c_3}{f_\pi^2}\partial^i{\bm \pi}\cdot\partial_i{\bm \pi}+\nonumber\\
&&-\frac{c_2+c_3}{f_\pi^2}\, {\bm \Pi} \cdot{\bm \Pi}+\frac{c_4}{2 f_\pi^2}{\bm \tau}
\cdot\left( \partial_i{\bm \pi}\times\partial_j{\bm \pi}\right)\sigma^{ij}\bigg]N\ ,\\
H^{(3)}_{2\pi NN}&=& \int{\rm d}{\bf x}\,\overline{N}\,\bigg[-2\,\frac{d_1+d_2+d_3}{f_\pi^2}({\bm \tau}\times{\bm \Pi})\cdot\left(\partial^i\partial_i{\bm \pi}+m_\pi^2{\bm \pi}\right)\gamma^0-4\,\frac{d_5 m_\pi^2}{f_\pi^2}({\bm \Pi}\times{\bm \pi})\cdot{\bm \tau}\gamma^0\nonumber\\
&&+2\,\frac{d_1+d_2}{f_\pi^2}\,({\bm \tau}\times{\bm \Pi})\cdot\left(\partial^i\partial_i{\bm \pi}+\partial^i{\bm \pi}\widetilde{\overleftrightarrow{\partial}}_{\!i}\right)\gamma^0+\frac{d_{15}-d_{14}}{f_\pi^2}\,{\bm \Pi}\cdot\partial_i{\bm \pi }\,\sigma^{ij}\,\widetilde{\overleftrightarrow{\partial}}_{\!j}\,\gamma^0\bigg] N \ ,\nonumber\\
\end{eqnarray}
from which the vertex follows as
\begin{eqnarray}
\langle\lambda^\prime\! \mid H^{(1)}_{2\pi NN}\mid\! \lambda; {\bf k}_1,a_1;{\bf k}_2,a_2\rangle&=&
\frac{i}{4f_\pi^2}\epsilon_{a_1a_2c}\, \tau_c\, \left(\omega_1-\omega_2\right) \ ,\\
\langle\lambda^\prime\! \mid H^{(2)}_{2\pi NN}\mid\! \lambda; {\bf k}_1, a_1;{\bf k}_2,a_2\rangle&=&
-\frac{i}{4f_\pi^2}\frac{ 2\, {\bf P}
+i {\bm \sigma}\times \left({\bf p}^\prime-{\bf p}\right)}{2m}\cdot({\bf k}_1-{\bf k}_2) \,\epsilon_{a_1a_2 a}\tau_a 
+4\, c_1 \frac{m_\pi^2}{f_\pi^2}\,\delta_{a_1,a_2}\nonumber\\
&&-\frac{2\, c_3}{f_\pi^2}\,
{\bf k}_1 \cdot{\bf k}_2\, \delta_{a_1,a_2}
+\frac{2\, \left(c_2+c_3\right)}{f_\pi^2}
 \omega_1 \omega_2\,  \delta_{a_1,a_2}
\nonumber\\
&&-\frac{c_4}{f_\pi^2}{\bm \sigma}\cdot \left({\bf k}_1 
\times{\bf k}_2\right) \epsilon_{a_1a_2 a}\tau_a \ ,\\
\langle\lambda^\prime\! \mid H^{(3)}_{2\pi NN}\mid\! \lambda; {\bf k}_1,a_1;{\bf k}_2,a_2\rangle&=&i(\omega_1-\omega_2)\bigg[\epsilon_{a_1a_2c}\tau_c\bigg(-2\,\frac{d_1+d_2+d_3}{f_\pi^2}\omega_1\omega_2
+4\,\frac{d_5\, m_\pi^2}{f_\pi^2}\nonumber\\
&&+\,2\,\frac{d_1+d_2}{f_\pi^2}\,{\bf k}_1\cdot{\bf k}_2\bigg)+\frac{d_{15}-d_{14}}{f_\pi^2}({\bf k}_1\times{\bf k}_2)\cdot{\bm \sigma}\,\delta_{a_1,a_2}\bigg]
\label{eq:b14}
\end{eqnarray}
and vertices in which either or both pions are in the final state are obtained from the equation
above by replacing ${\bf k}_i,\omega_i \longrightarrow -{\bf k}_i,-\omega_i$.

\subsection{$3\pi NN$ vertices}
The interaction terms read
\begin{equation}
\label{eq:b15}
H_{3\pi NN}^{(1)}=-\frac{g_A}{2f_\pi^3}\int {\rm d}{\bf x}\,
 \overline{N}\left[\alpha \, {\bm \pi}^2\, {\bm \tau}\cdot \partial_i{\bm \pi}
+\frac{1}{2}(4\, \alpha-1) {\bm \tau}\cdot{\bm \pi}\, {\bm \pi}\cdot\partial_i{\bm \pi}\right] \gamma^i
\gamma^5 N \ ,
\end{equation}
which leads to the following interaction vertex
\begin{eqnarray}
\langle \lambda^\prime\! \mid H^{(1)}_{3\pi NN}\mid\! \lambda;{\bf k}_1,a_1;{\bf k}_2,a_2;{\bf k}_3,a_3\rangle&=&-\frac{i\, g_A}{2\, f_\pi^3}
{\bm \sigma}\cdot\Big[\tau_{a_1}\delta_{a_2,a_3}
\left[ \left(2\,\alpha-1/2\right)\left({\bf k}_2+{\bf k}_3\right)+2\, \alpha {\bf k}_1\right]\nonumber\\
&&+\tau_{a_2}\delta_{a_1,a_3}\left[\left(2\,\alpha-1/2\right)\left({\bf k}_1+{\bf k}_3\right)+2\, \alpha {\bf k}_2\right]\nonumber\\
&&+\tau_{a_3}\delta_{a_1,a_2}\left[ \left(2\,\alpha-1/2\right)\left({\bf k}_1+{\bf k}_2\right)+2\, \alpha {\bf k}_3\right]\Big]\ .
\end{eqnarray}
The corresponding tadpole contribution is
\begin{equation}
\langle \lambda^\prime\! \mid H^{(1)}_{3\pi NN} \mid\! \lambda;{\bf k},a\rangle=
-i\frac{g_A}{8f_\pi^3} \left(10\, \alpha-1\right)\,J_{01}\,\tau_a\, {\bm \sigma}\cdot {\bf k} \ ,
\end{equation}
where $J_{01}$ has been defined in Eq.~(\ref{eq:jmn}).
\subsection{$4\pi NN$ vertices}
The interaction term reads
\begin{equation}
H^{(1)}_{4\pi NN}=\frac{1}{32\, f_\pi^4}\int {\rm d}{\bf x}\,
\overline{N}\,  \left(\Pi_a\, {\bm \pi}^2+{\bm \pi}^2\, \Pi_a\right)\left({\bm \tau}\times{\bm \pi}\right)_a \gamma^0\, N\,
\end{equation}
and the tadpole contribution follows as
\begin{equation}
\langle 0 \!\mid H_{4\pi NN}^{(1)}\mid \! {\bf k}_1,a_1;{\bf k}_2,a_2\rangle=\frac{5\, i}{32\, f_\pi^4}\, J_{01}\, 
\epsilon_{a_1 a_2 c}\, \tau_c (\omega_1-\omega_2) \ .
\end{equation}
\subsection{$A NN$ vertices}
The interaction terms read
\begin{eqnarray}
\label{eq:b20}
H_{ANN}^{(1)}&=&-\frac{g_A}{2}\int {\rm d}{\bf x}\, \overline{N}\, \tau_a\, A^i_a \, \gamma_i\, \gamma^5 N \ , \\
\label{eq:b20n}
H_{ANN}^{(3)}&=&-\int {\rm d}{\bf x}\overline{N}\bigg(2\,m_\pi^{2} d_{16}\, {\bm \tau}\cdot{\bf A}_i\gamma^i\gamma_5
+\frac{d_{22}}{2}{\bm \tau}\cdot\partial^j{\bf F}_{ij}\gamma^i\gamma_5\bigg)N\ ,
\end{eqnarray}
from which the vertices follow as
\begin{eqnarray}
\label{eq:b22n}
\langle \lambda^\prime \! \mid H_{ANN}^{(1)} \mid \! \lambda\rangle
&=&\frac{g_A}{2}\tau_a \bigg[  {\bm \sigma}-\frac{1}{2\, m^2}\,P^2\, {\bm \sigma}
-\frac{i}{4\, m^2} \left({\bf p}^\prime-{\bf p}\right)\times{\bf P} +\frac{1}{2\, m^2} {\bm \sigma}\cdot{\bf P}\,\, {\bf P} \nonumber\\
&&-\frac{1}{8\, m^2}{\bm \sigma}\cdot\left({\bf p}^\prime-{\bf p}\right) \,\,\left({\bf p}^\prime-{\bf p}\right)\bigg] \cdot {\bf A}_a \ ,\\
\langle \lambda^\prime \! \mid H_{ANN}^{(3)} \mid \! \lambda\rangle&=& 2\,m_\pi^{2} \, d_{16}\, \tau_a\, {\bm \sigma} \cdot {\bf A}_a
+\frac{d_{22}}{2}\tau_a\left({\bf q}\, {\bf q}\cdot{\bm \sigma}-q^2{\bm \sigma}\right)\cdot {\bf A}_a\ , 
\end{eqnarray}
where in Eq.~(\ref{eq:b22n}) terms of order $Q^2$ have been retained in the expansion of the
bilinear $\overline{N}{\bm \gamma}\gamma_5 N$, since they have been shown to generate significant
corrections to the single-nucleon axial current~\cite{Park03}. 
\subsection{$ \pi NNA$ vertices}
The interaction terms read
\begin{eqnarray}
\label{eq:b24n}
H_{\pi NNA}^{(1)}&=&-\frac{1}{4f_\pi}\int {\rm d}{\bf x}\, \overline{N}{\bf A}_0\cdot\left({\bm \tau}\times {\bm \pi}\right)\gamma^0N\ , \\
H_{\pi NNA}^{(2)}&=&\int{\rm d}{\bf x}\, \overline{N} \bigg[-\frac{1}{2f_\pi}
({\bm \tau}\times{\bm \pi})\cdot{\bf A}_i\gamma^i
-\frac{c_6}{4mf_\pi}({\bm \tau}\times{\bm \pi})\cdot \partial_i{\bf A}_j\,\sigma^{ij}
+\frac{2\,c_3}{f_\pi}{\bf A}^i\cdot\partial_i{\bm \pi}\nonumber\\
&&+\frac{c_4}{f_\pi}(\partial_i{\bm \pi}\times{\bm \tau})\cdot {\bf A}_j \, \sigma^{ij} \bigg] N \ ,\\
 H_{\pi NNA}^{(3)}&=&\int{\rm d}{\bf x}\, \overline{N} 
 \bigg[\frac{2\,d_2+d_6}{f_\pi}(\partial_i{\bm\pi}\times{\bm \tau})\cdot\partial^i{\bf A}^0\gamma^0
+\frac{d_{15}}{f_\pi}\partial_j{\bf A}^0\cdot\partial_i{\bm \pi}\sigma^{ij}\gamma^0 \nonumber \\
&&+2\,\frac{d_{23}}{f_\pi}
\epsilon^{0ijk}\,\partial_i{\bm \pi}\cdot\partial_k{\bf A}^0\,\gamma_j\gamma^5
-\frac{d_6}{f_\pi}
\left({\bm \tau}\times{\bm \pi}\right)\cdot\partial_i\partial^i{\bf A}^0\gamma^0 \nonumber\\
&&+2\frac{d_1+d_2}{f_\pi}({\bm \tau}\times{\bf A}_0)\cdot(\partial^i\partial_i{\bm \pi}+\partial^i{\bm \pi}\widetilde{\overleftrightarrow{\partial}}_{\!i})\gamma^0+\frac{d_{15}-d_{14}}{f_\pi}\partial_i{\bm \pi}\cdot{\bf A}_0\sigma^{ij}\widetilde{\overleftrightarrow{\partial}}_j+\dots\bigg]\, N \ ,\nonumber\\
\label{eq:b26n}
\end{eqnarray}
where the dots indicate terms which do not contribute in tree-level diagrams of order $Q$, for example
\[
\int{\rm d}{\bf x}\, \overline{N} \bigg[
-2\,\frac{d_{23}}{f_\pi} \epsilon^{0ijk}\,\gamma_i\gamma^5\,{\bm \Pi}\cdot\partial_j{\bf A}_k
-2\frac{d_1+d_2+d_3}{f_\pi}\,{\bm \tau}\cdot\left(\partial_i{\bf A}^i\times{\bm \Pi}\right)\gamma^0\bigg]
N \ ,
\]
or
\[
2\,\frac{d_1+d_2+d_3}{f_\pi} \int{\rm d}{\bf x}\, \overline{N}\,
{\bm \tau}\cdot\left(\partial_0{\bf A}_0\times {\bm \Pi} \right) \gamma^0 \, N \ ,
\]
and $\partial_0 {\bf A}_0 \rightarrow -i\, \omega_q {\bf A}_0$ is of order $Q^3$, since
in our counting the energy of the external field is of order $Q^2$.
The interactions in Eqs.~(\ref{eq:b24n})--(\ref{eq:b26n})  lead to the following vertices
\begin{eqnarray}
\langle \lambda^\prime\!\mid H_{\pi NNA}^{(1)}\mid\! \lambda;{\bf k},a\rangle&=&-\frac{1}{4f_\pi}\, \epsilon_{abc}\, A^0_b\, \tau_c\ ,\\
\langle\lambda^\prime\!\mid H_{\pi NNA}^{(2)}\mid \lambda;{\bf k},a\rangle&=&-\frac{1}{2m f_\pi}\epsilon_{abc}\, 
\tau_b \, {\bf A}_c\cdot \left[ {\bf P}+\frac{i}{2}{\bm \sigma}\times \left({\bf p}^\prime-{\bf p}\right) \right]\nonumber\\
&&-i \frac{c_6}{4mf_\pi} \epsilon_{abc}\, \tau_b \,{\bf A}_c\cdot ({\bm \sigma}\times{\bf q})
+2 i \frac{c_3}{f_\pi}\, {\bf k}\cdot{\bf A}_a\nonumber\\
&& - i\frac{c_4}{f_\pi}\, \epsilon_{abc}\, \tau_b\,  {\bf A}_c\cdot ({\bm \sigma}\times{\bf k})\ ,\\
\langle\lambda^\prime\!\mid H_{\pi NNA}^{(3)}\mid \lambda;{\bf k},a\rangle&=&
\frac{2d_1-d_6}{f_\pi}({\bf A}^0\times{\bm \tau})_a\,{\bf q}\cdot{\bf k}+\frac{d_{14}+2\,d_{23}}{f_\pi}
{\bm \sigma}\cdot({\bf q}\times{\bf k}) A^0_a\nonumber\\
&&-\frac{d_6}{f_\pi}({\bf A}^0\times{\bm \tau})_a \,{\bf q}^2\ .
\label{eq:b24}
\end{eqnarray}
\subsection{$2\pi NNA$ vertices}
The interaction term reads
\begin{eqnarray}
\label{eq:b28}
H_{2\pi NNA}^{(1)}&=&-\frac{g_A}{4f_\pi^2}\int {\rm d}{\bf x}\, \overline{N}
{\bf A}_i\cdot\left[\left({\bm \tau}\times{\bm \pi}\right)\times{\bm \pi}\right]\gamma^i\gamma^5 N \ ,
\end{eqnarray}
which leads to the following vertex and tadpole contributions
\begin{eqnarray}
\langle \lambda^\prime\!\mid H_{2\pi NNA}^{(1)} \mid \!\lambda; {\bf k}_1,a_1;{\bf k}_2,a_2\rangle&=&
\frac{g_A}{4f_\pi^2}  \left(\delta_{a,a_1}\, \tau_{a_2}
+\delta_{a,a_2}\, \tau_{a_1}-2\, \delta_{a_1,a_2}\, \tau_a\right){\bf A}_a\cdot {\bm \sigma}\ , \\
\langle\lambda^\prime\!\mid H_{2\pi NNA}^{(1)} \mid\! \lambda\rangle&=&-\frac{g_A}{4f_\pi^2}\, J_{01}\,\tau_a\,
{\bf A}_a\cdot{\bm \sigma}   \ .
\end{eqnarray}
\subsection{$3\pi NNA$ vertices}
The interaction term reads
\begin{equation}
H^{(1)}_{3\pi NNA}=\frac{4\alpha-1}{16\,f_\pi^3}\int d{\bf x}\,\overline{N}\,
{\bm \pi}^2\,{\bf A}^0\cdot\left({\bm \tau}\times {\bm \pi}\right)\, \gamma^0 N\ ,
\end{equation}
from which the tadpole contribution is obtained as
\begin{eqnarray}
\langle\lambda^\prime \!\mid H^{(1)}_{3\pi NNA} \mid \!\lambda;{\bf k}, a\rangle&=&
-\frac{5\left(4\,\alpha-1\right)}{32\,f_\pi^3}\, J_{01}\left( {\bm \tau}\times{\bf A}^0 \right)_a \ .
\end{eqnarray}
\subsection{$2\pi$ vertices}
The interaction terms read
\begin{equation}
H_{2\pi}^{(4)}=\int{\rm d}{\bf x}\, \left[-\frac{m_\pi^2\, l_4}{f_\pi^2}\left( {\bm \Pi}\cdot{\bm \Pi} + \partial^i{\bm \pi}\cdot\partial_i{\bm \pi} \right)
+\frac{m_\pi^4\left(l_3+l_4\right)}{f_\pi^2} {\bm \pi}\cdot{\bm \pi} \right] \ ,
\end{equation}
from which the vertex is obtained as
\begin{equation}
\label{eq:b22}
\langle  0\!\mid H_{2\pi}^{(4)}\mid\!{\bf k}_1,a_1;{\bf k}_2,a_2\rangle=\delta_{a_1, a_2}
\left[ \frac{2\, m_\pi^2\,l_4}{f_\pi^2} \left( \omega_1\omega_2-{\bf k}_1\cdot{\bf k}_2\right)
+ \frac{2 \,m_\pi^4\, (l_3+ l_4)}{f_\pi^2}  \right] \ ,
\end{equation}
where, as noted earlier, the momentum-conserving $\delta$-function $(2\pi)^3\delta({\bf k}_1+{\bf k}_2)$
and the pion field normalization factor $1/\sqrt{4\, \omega_1\omega_2}$ are understood.  Vertices
in which one or both pions are in the final state follow by replacing
$\omega_i,{\bf k}_i \longrightarrow -\omega_i,-{\bf k}_i$.  Enforcing the $\delta$
function requirement ${\bf k}_1=-{\bf k}_2={\bf k}$ and $\omega_1=\omega_2=\omega$,
the vertex in Eq.~(\ref{eq:b22}) reduces to
\begin{equation}
\langle  0\!\mid H_{2\pi}^{(4)}\mid\!{\bf k} ,a;- {\bf k},a\rangle=
 \frac{4\, m_\pi^2\,l_4}{f_\pi^2}  \omega^2
+ \frac{2 \,m_\pi^4\, l_3}{f_\pi^2}  \ .
\end{equation}
Similarly, we find
\begin{equation}
\langle  {\bf k},a \!\mid H_{2\pi}^{(4)}\mid\!{\bf k} ,a\rangle=\frac{2 \,m_\pi^4\, l_3}{f_\pi^2}  \ ,
\end{equation}
according to the prescription given above.  Apart from the factor $1/(2\, \omega)$, which
is not included in the equations above, these vertices are the same as given in
Appendix F of Ref.~\cite{Viv14}.

\subsection{$4\pi$ vertices}
The interaction terms read
\begin{eqnarray}
H_{4\pi}^{(2)}&=&\int d{\bf x}\bigg[ \frac{4\alpha-1}{2f_\pi^2}
\left( {\bm \pi}\cdot{\bm \Pi}\,\,{\bm \Pi}\cdot{\bm \pi}+\partial_i{\bm \pi}\cdot{\bm \pi}\, \,\,\partial^i{\bm \pi}\cdot{\bm \pi}\right)\nonumber\\
&&+\frac{\alpha}{f_\pi^2}\, \left( \pi_a \,{\bm \Pi}\cdot{\bm \Pi}\, \pi_a
+{\bm \pi}^2\partial_i{\bm \pi}\cdot\partial^i{\bm \pi}\right)-\frac{8\alpha-1}{8f_\pi^2}m_\pi^2{\bm \pi}^4\bigg] \ ,
\end{eqnarray}
which leads to the following vertex
\begin{eqnarray}
\!\!\!\!\!\!\!\!\!\!\!\!\!\!\!\!\!\!\!&&\langle 0 \!\mid H_{4\pi}^{(2)}\mid \! {\bf k}_1,a_1;{\bf k}_2,a_2;{\bf k}_3,a_3;{\bf k}_4,a_4\rangle
=\frac{1}{f_\pi^2}\nonumber\\
\!\!\!\!\!\!\!\!\!\!\!\!\!\!\!\!\!\!\!&&\times
 \Big[\delta_{a_1,a_2}\delta_{a_3,a_4}\left[-2\alpha(\omega_1+\omega_2+\omega_3+\omega_4)^2
 +m_\pi^2+({\bf k}_3+{\bf k}_4)^2+(\omega_1+\omega_2)(\omega_3+\omega_4)\right]\nonumber\\
\!\!\!\!\!\!\!\!\!\!\!\!\!\!\!\!\!\!\! &&+\delta_{a_1,a_3}\delta_{a_2,a_4}\left[-2\alpha(\omega_1+\omega_2+\omega_3+\omega_4)^2
  +m_\pi^2+({\bf k}_1+{\bf k}_3)^2+(\omega_1+\omega_3)(\omega_2+\omega_4)\right]\nonumber\\
\!\!\!\!\!\!\!\!\!\!\!\!\!\!\!\!\!\!\!&&+ \delta_{a_1,a_4}\delta_{a_2,a_3}\left[-2\alpha(\omega_1+\omega_2+\omega_3+\omega_4)^2
   +m_\pi^2+({\bf k}_1+{\bf k}_4)^2+(\omega_1+\omega_4)(\omega_2+\omega_3)\right]\Big] \ ,
\end{eqnarray}
and the corresponding tadpole contribution is
\begin{eqnarray}
\langle 0 \!\mid H_{4\pi}^{(2)}\mid \! {\bf k}_1,a_1;{\bf k}_2,a_2\rangle&=&\delta_{a_1,a_2}\, J_{01}
\bigg[ 
\frac{1-10\,\alpha}{2\, f_\pi^2}\,\left(\omega_1\omega_2-{\bf k}_1\cdot {\bf k}_2\right)
-\frac{20\, \alpha-3}{4\, f_\pi^2}\, m_\pi^2\bigg] \ ,
\end{eqnarray}
and the constant $J_{01}$ has been defined in Eq.~(\ref{eq:jmn}).
\subsection{$\pi A$ vertices}
The interaction terms read
\begin{eqnarray}
\label{eq:b40}
H^{(2)}_{\pi A}&=&f_\pi \int {\rm d}{\bf x} \left(  {\bf A}^i\cdot\partial_i{\bm \pi}+ {\bm A}^0\cdot{\bf \Pi}\right) \ ,\\
H^{(4)}_{\pi A}&=&\frac{2\, m_\pi^2\, l_4}{f_\pi}\int {\rm d}{\bf x} \,
{\bf A}^i\cdot\partial_i{\bm \pi}\ ,
\label{eq:b41}
\end{eqnarray}
from which the vertices are obtained as
\begin{eqnarray}
\langle 0\!\mid H_{\pi A}^{(2)}\mid \!{\bf k},a \rangle&=&
i\,f_\pi \left({\bf k}\cdot {\bf A}_a-\omega\,  A^0_a\right) \ , \\
\langle 0\!\mid H_{\pi A}^{(4)}\mid \!{\bf k},a \rangle&=&2\,i \frac{ m_\pi^2\, l_4}{f_\pi}
\, {\bf k}\cdot {\bf A}_a \ .
\end{eqnarray}
\subsection{$3\pi A$ vertices}
The interaction terms read
\begin{eqnarray}
\label{eq:b44}
H_{3\pi A}^{(2)}&=&\frac{1}{2f_\pi}\int {\rm d}{\bf x}\Big[2\,(1-2\,\alpha) {\bf A}^i\cdot{\bm \pi}
\,\, {\bm \pi}\cdot \partial_i{\bm \pi} -(2\, \alpha+1) {\bf A}^i\cdot\partial_i{\bm \pi}\,\, {\bm \pi}\cdot{\bm \pi}\nonumber\\
&&+ 2\left(\alpha-1/2\right)A^0_a \, \pi_b\, \Pi_a\,\pi_b
 + 2 \,\alpha \, A^0_a\,\left( \pi_a\, {\bm \pi}\cdot {\bm \Pi}+{\bm \Pi}\cdot {\bm \pi}\, \pi_a \, \right) \Big]\ ,
\end{eqnarray} 
which lead to the following vertices
\begin{eqnarray}
\langle0\!\mid H_{3\pi A}^{(2)}\mid\! {\bf k}_1,a_1;{\bf k}_2,a_2;{\bf k}_3,a_3\rangle&=&
\frac{i}{f_\pi} \Big[  \delta_{a_2,a_3}\, {\bf A}_{a_1} \cdot \left[ \left( 2\,\alpha-1\right) {\bf q}-2\,{\bf k}_1 \right] \nonumber\\
&&\,\,\,\,\,\,+ \delta_{a_1,a_3}\, {\bf A}_{a_2} \cdot \left[ \left( 2\,\alpha-1\right) {\bf q}-2\,{\bf k}_2 \right]\nonumber\\
&&\,\,\,\,\,\,+ \delta_{a_1,a_2}\, {\bf A}_{a_3} \cdot \left[ \left( 2\,\alpha-1\right) {\bf q}-2\,{\bf k}_3 \right]  \nonumber\\
&&\,\,\,\,\,\,-
\delta_{a_2,a_3}\, A^0_{a_1}\left[2\,\alpha\left(\omega_{1}+\omega_{2}+\omega_{3}\right)-\omega_{1}\right]\nonumber\\
&&\,\,\,\,\,\,-
\delta_{a_1,a_3}\, A^0_{a_2}\left[2\,\alpha\left(\omega_{1}+\omega_{2}+\omega_{3}\right)-\omega_{2}\right]\nonumber\\
&&\,\,\,\,\,\,
-\delta_{a_1,a_2}A^0_{a_3}\left[2\,\alpha\left(\omega_{1}+\omega_{2}+\omega_{3}\right)-\omega_{3}\right]\Big]\ \ ,
\end{eqnarray}
where in the first three lines use has been made of the $\delta$-function
$(2\pi)^3\delta({\bf k}_1+{\bf k}_2+{\bf k}_3+{\bf q})$.   The tadpole
contribution is found to be
\begin{eqnarray}
\langle 0\! \mid H^{(2)}_{3\pi A}\mid \!{\bf k},a\rangle=-\frac{i}{2f_\pi}\, J_{01} \Big[ \left(5\, \alpha+1/2\right) {\bf A}_a\cdot{\bf k}  
+(5\, \alpha-3/2) A^0_a\, \omega  \Big] \ .
\end{eqnarray}
\section{Contact terms at order $Q$}
\label{app:cta0}
The weak-interaction potential $v_5=A^0_a \,\rho_{5,a}-{\bf A}_a\cdot {\bf j}_{5,a}$
is parity (${\cal P}$) and time-reversal (${\cal T}$) invariant, which implies
that $\rho_{5,a} \stackrel{\cal P}{\longrightarrow} -\rho_{5,a}$ and
${\bf j}_{5,a} \stackrel{\cal P}{\longrightarrow} {\bf j}_{5,a}$, and
$\rho_{5,a} \stackrel{\cal T}{\longrightarrow} (-)^{a+1}\, \rho_{5,a}$
and ${\bf j}_{5,a} \stackrel{\cal T}{\longrightarrow} (-)^a\, {\bf j}_{5,a}$.
At order $Q^0$ there is no momentum dependence, and consequently
there are no contact terms which can be constructed for $\rho_{5,a}$, while
two such terms occur for ${\bf j}_{5,a}$, of which only one is
independent (Fierz identities, see below) and is given in Eq.~(\ref{eq:jctct}).
At order $Q$ the contact terms in $\rho_{5,a}$ and ${\bf j}_{5,a}$
must be linear in either ${\bf k}_i={\bf p}_i^\prime-{\bf p}_i$ or ${\bf K}_i=\left({\bf p}_i^\prime+{\bf p}_i\right)/2$ with
$i=1$ and 2.  None can be constructed for
 ${\bf j}_{5,a}$.  A complete, but non minimal, set of hermitian operators
 for the axial charge $\rho_{5,a}$ is the following:
\begin{eqnarray}
\widetilde{O}_1&=&\left(\tau_{1,a}+\tau_{2,a}\right)\,\left({\bm \sigma}_1+{\bm \sigma}_2\right)\cdot\left({\bf K}_1+{\bf K}_2\right)\, ,\nonumber\\
\widetilde{O}_2&=&\left(\tau_{1,a}+\tau_{2,a}\right)\,\left({\bm \sigma}_1-{\bm \sigma}_2\right)\cdot\left({\bf K}_1-{\bf K}_2\right)\, ,\nonumber\\
\widetilde{O}_3&=&i\left(\tau_{1,a}+\tau_{2,a}\right)\,\left({\bm \sigma}_1\times{\bm \sigma}_2\right)\cdot\left({\bf k}_1-{\bf k}_2\right)\, ,\nonumber\\
\widetilde{O}_4&=&\left(\tau_{1,a}-\tau_{2,a}\right)\,\left({\bm \sigma}_1-{\bm \sigma}_2\right)\cdot\left({\bf K}_1+{\bf K}_2\right)\, ,\nonumber\\
\widetilde{O}_{5}&=&\left(\tau_{1,a}-\tau_{2,a}\right)\,\left({\bm \sigma}_1+{\bm \sigma}_2\right)\cdot\left({\bf K}_1-{\bf K}_2\right)\, ,\nonumber\\
\widetilde{O}_{6}&=&i\left(\tau_{1,a}-\tau_{2,a}\right)\,\left({\bm \sigma}_1\times{\bm \sigma}_2\right)\cdot\left({\bf k}_1+{\bf k}_2\right)\, ,\nonumber\\
\widetilde{O}_{7}&=&i\left({\bm\tau}_1\times{\bm\tau}_2\right)_a\,\left({\bm \sigma}_1-{\bm \sigma}_2\right)\cdot\left({\bf k}_1+{\bf k}_2\right)\ ,\nonumber\\
\widetilde{O}_{8}&=&i\left({\bm\tau}_1\times{\bm\tau}_2\right)_a\,\left({\bm \sigma}_1+{\bm \sigma}_2\right)\cdot\left({\bf k}_1-{\bf k}_2\right)\ ,\nonumber\\
\widetilde{O}_{9}&=&\left({\bm\tau}_1\times{\bm\tau}_2\right)_a\,\left({\bm \sigma}_1\times{\bm \sigma}_2\right)\cdot\left({\bf K}_1+{\bf K}_2\right)\, .\nonumber
\end{eqnarray}
The antisymmetry of initial and final two-nucleon states requires
\begin{equation}
\widetilde{O}_i=-P^{\tau} P^{ \sigma} P^{{\rm space}} \widetilde{O}_i\ ,\label{eq:ctrel}
\end{equation}
where $P^{\rm space}$ is the space exchange operator, and
$P^{\sigma}$ and $P^{\tau}$ are the spin and isospin
exchange operators with $P^{\sigma} = \left(1+{\bm \sigma}_1\cdot{\bm \sigma}_2\right)/2$
and similarly for $P^{\tau}$.  Exchange of the final momenta of the two nucleons
${\bf p}_1^\prime \rightleftharpoons {\bf p}_2^\prime$ leads to
\begin{eqnarray}
P^{\rm space}({\bf k}_1+{\bf k}_2)&=&{\bf k}_1+{\bf k}_2\ , \qquad
P^{\rm space}({\bf k}_1-{\bf k}_2) = 2 \left({\bf K}_2-{\bf K}_1\right)\ , \\
P^{\rm space}({\bf K}_1+{\bf K}_2)&=&{\bf K}_1+{\bf K}_2\ ,\qquad
P^{\rm space}({\bf K}_1-{\bf K}_2)=\left({\bf k}_2-{\bf k}_1\right)/2\ ,
\end{eqnarray}
while spin exchange implies
\begin{equation}
P^\sigma\left({\bm \sigma}_1+{\bm \sigma}_2\right)={\bm \sigma}_1+{\bm \sigma}_2\ ,\,\,\,
P^\sigma\left({\bm \sigma}_1-{\bm \sigma}_2\right)= i\left({\bm \sigma}_1\times{\bm \sigma}_2\right)\ ,
\,\,\, P^\sigma\left({\bm \sigma}_1\times{\bm \sigma}_2\right)= -i\left({\bm \sigma}_1-{\bm \sigma}_2\right)\ ,
\label{eq:ps}
\end{equation}	
and similar relations follow under isospin exchange.  The following (Fierz) identities
are obtained from Eq.~(\ref{eq:ctrel}):
\begin{eqnarray}
\widetilde{O}_2=\widetilde{O}_3/2\ , \qquad  \widetilde{O}_4&=&\widetilde{O}_{9}\ , \qquad \widetilde{O}_{5}=\widetilde{O}_{8}/2\ , \qquad  \widetilde{O}_{6}=-\widetilde{O}_{7} \ , 
\end{eqnarray}
while $\widetilde{O}_1$ is required to vanish.  Hence only 4 of the
above 9 operators are independent, and a convenient set is
\begin{equation}
O_1=\left(\widetilde{O}_{7}-\widetilde{O}_{8}\right)/2\ , \>\>
O_2=\left(\widetilde{O}_{7}+\widetilde{O}_{8}\right)/2\ , \>\>
O_3=\left(\widetilde{O}_{6}-\widetilde{O}_{3}\right)/2\ ,  \>\>
O_4= \widetilde{O}_4 \ .
\end{equation}
We note that $O_1$ and $O_3$ have the same operator structures associated with
the divergent parts of the loop diagrams.
\section{Regularized loop contributions to ${\bf j}^{\rm MPE}_{5,a}$}
\label{app:jforms}
The regularized contributions of diagrams in Fig.~\ref{fig:f7} read:
\begin{eqnarray}
\!\!\!{\bf j}_{5,a}^{(1)}({\rm e1})\!&=&\!\frac{g_A^3}{64\, \pi f_\pi^4}\tau_{2,a}\int_0^1 dz\bigg[{\bm \sigma}_1M(k_2,z)+ {\bf k}_2\,{\bm \sigma}_1\cdot{\bf k}_2\, \frac{z\overline{z}}{M(k_2,z)}\bigg]\ ,\\
{\bf j}_{5,a}^{(1)}({\rm e4})&=&-\frac{g_A^3}{64\, \pi f_\pi^4}\, \tau_{2,a}\, 
{\bm \sigma}_2\int_0^1dz\bigg[\frac{k_1^2 \overline{z}z}{M(k_1,z)}+3\, M(k_1,z)\bigg]\ ,\\
{\bf j}_{5,a}^{(1)}({\rm e5})&=&\frac{g_A^3}{128\pi f_\pi^4}
\frac{{\bf q}}{q^2+m_\pi^2}\int_0^1 dz\Bigg[\tau_{2,a}\, {\bm \sigma}_2\cdot
\left({\bf k}_1-{\bf k}_2\right)\bigg[ \frac{k_1^2\, z\overline{z}}{M(k_1,z)}+3\, M(k_1,z)\bigg]\nonumber\\
&&-\left({\bm \tau}_1\times{\bm \tau}_2\right)_a
\left({\bm \sigma}_1\times{\bm \sigma}_2\right)\cdot{\bf k}_1\, M(k_1,z)\Bigg]\ ,\\
\!\!\!{\bf j}_{5,a}^{(1)}({\rm e8})&=&
-\frac{g_A^5}{64\, \pi f_\pi^4}\int_0^1 dz\Bigg[\tau_{2,a}\bigg[5\, {\bm \sigma}_1M(k_2,z)
+\frac{{\bf k}_2}{2} \,  {\bm \sigma}_1\cdot{\bf k}_2
\bigg[\frac{k_2^2\, (z\overline{z})^2}{M(k_2,z)^3}+\frac{1-7 z\overline{z}}{M(k_2,z)}\bigg]
\nonumber\\
&&+\frac{k_2^2}{2} \, {\bm \sigma}_1
\bigg[ \frac{10\, z\overline{z}-1}{M(k_2,z)}+\frac{1}{4}\frac{z\overline{z}\left(1-8z\overline{z}\right)}{M(k_2,z)^3}\bigg] \bigg]
+2\,\tau_{1,a}\left({\bm \sigma}_2\times{\bf k}_2\right)\times{\bf k}_2\nonumber\\
&&\times
\bigg[\frac{1}{4\, M(k_2,z)}+\frac{1}{48}\frac{k_2^2\, (2z-1)^2}{M(k_2,z)^3}\bigg]\Bigg]\ ,\\
{\bf j}_{5,a}^{(1)}({\rm e10})&=&\frac{g_A^3}{128\, \pi\, f_\pi^4}\frac{{\bf q}}{q^2+m_\pi^2}\int_0^1 dz
\Bigg[\left(2\,\tau_{2,a}-\tau_{1,a}\right)\left[\frac{k_2^2}{M(k_2,z)}+3M(k_2,z)\right]{\bm\sigma}_1\cdot{\bf k}_2\nonumber\\
&&+\left({\bm \tau}_1\times{\bm \tau}_2\right)_aM(k_2,z)\left({\bm \sigma}_1\times{\bm \sigma}_2\right)\cdot{\bf k}_2\bigg]\ ,\\
{\bf j}_{5,a}^{(1)}({\rm e15})&=&\frac{g_A^3}{128\, \pi\, f_\pi^4}\int_0^1 dz
\Bigg[\tau_{2,a}\bigg[ \frac{k_1^2\, z\overline{z}}{M(k_1,z)}+3\, M(k_1,z)\bigg] ({\bf k}_2-3\, {\bf k}_1)\nonumber\\
&&+ 4\left({\bm \tau}_1\times{\bm \tau}_2\right)_a\left({\bm \sigma}_1\times{\bf k}_1\right)M(k_1,z)\Bigg]
\frac{{\bm \sigma}_2\cdot{\bf k}_2}{\omega_2^2}\ ,\\
{\bf j}_{5,a}^{(1)}({\rm e16})&=&\frac{g_A^3}{128\,\pi \,  f_\pi^4}\frac{{\bf q}}{q^2+m_\pi^2}
\int_0^1dz\Bigg[\tau_{2,a}\bigg[-10 \, M(k_1,z)^3 +M(k_1,z) (15\, m_\pi^2+11\,k_1^2\nonumber\\
&&+3\,k_2^2+3\, q^2-20 \, k_1^2z\overline{z})
+\frac{k_1^2\, z\overline{z}}{M(k_1,z)}(5\, m_\pi^2+k_2^2+q^2
+3\,k_1^2-2\,k_1^2z\overline{z})\bigg]\nonumber\\
&&-2\left({\bm \tau}_1\times{\bm \tau}_2\right)_a\left({\bm \sigma}_1\times{\bf k}_1\right)\cdot\left({\bf k}_2+{\bf q}\right)M(k_1,z)\Bigg]\frac{{\bm \sigma}_2\cdot{\bf k}_2}{\omega_2^2}\ ,\\
{\bf j}_{5,a}^{(1)}({\rm e17})&=&\frac{g_A^3}
{32\, \pi } \frac{m_\pi^3}
{f_\pi^4}\,\tau_{2,a}\,\frac{{\bf q}}{q^2+m _\pi^2}\frac{{\bm \sigma}_2\cdot{\bf k}_2}{\omega_2^2} \ ,\\
{\bf j}_{5,a}^{(1)}({\rm e20})&=&-\frac{g_A^3\, m_\pi}{8 \,\pi \,f_\pi^2}\, C_T\, \tau_{1,a} \, {\bm \sigma}_2\ ,
\end{eqnarray}
where $M(k,z)$ and $\overline{z}$ have been defined in Eqs.~(\ref{eq:mmfnt}) and~(\ref{eq:zzzz}).
The contributions corresponding to diagrams e2, e9, and e21 easily follow from those
for e1, e8, and e20.

The loop functions $W_i$ and $Z_i$ introduced in Eqs.~(\ref{eq:mpejfin}) and~(\ref{eq:mpej1fin}) are
defined as
\begin{eqnarray}
W_1(k)&=&\int_0^1 dz\left[ \left(1-5\,g_A^2\right)M(k,z)
-\frac{g_A^2 \,k^2}{2}\left[ \frac{10\, z\,\overline{z}-1}{M(k,z)}
+\frac{z\, \overline{z}\left(1-8\, z\,\overline{z}\right)}{4\,M(k,z)^3}\right]\right]\ , \\
W_2(k)&=&\int_0^1 dz\left[-\frac{g_A^2\, z\,\overline{z}\, k^2}{M(k,z)^3}+\frac{z\, \overline{z}\left(7\, g_A^2+2\right)
-g_A^2}{2M(k,z)}\right]\ , \\
W_3(k)&=&\frac{1}{2}\int_0^1 dz\left[ \frac{k^2\,(z-\overline{z}\,)^2}{12\, M(k,z)^3}+\frac{1}{M(k,z)}\right]\ ,\\
Z_1(k)&=&\int_0^1 dz \left[ \frac{z\, \overline{z}\, k^2}{M(k,z)}+3\, M(k,z)\right] \ , \\
Z_2({\bf k}) &=&\int_0^1 dz  \bigg[4m_\pi^3
-10\, M(k,z)^3+M(k,z)\,(15\, m_\pi^2+14\,k^2-6\,{\bf q}\cdot{\bf k}+6\,q^2 \nonumber\\
&&-20\,z\,\overline{z}\, k^2) +\frac{z\, \overline{z}\, k^2}{M(k,z)}
\left(5\, m_\pi^2+4\, k^2+2\,q^2-2\, {\bf q}\cdot{\bf k}-2\, k^2\, z\,\overline{z}\right)\bigg]\ , \\
Z_3(k)&=& \int_0^1 dz \, M(k,z)\ .
\end{eqnarray}
\section{Counter-terms to order $Q^3$}
\label{app:renora}
Having made the replacements in Eqs.~(\ref{eq:zz})--(\ref{eq:mm}), the bare Lagrangian
${\cal L}$ can be rewritten in terms of the renormalized fields and
physical masses as
\begin{eqnarray}
{\cal L}&=&{\cal L}^{r}+\delta {\cal L}^r\, ,
\end{eqnarray}
where ${\cal L}^r$ is the same as in Eq.~(\ref{eq:2.3}) but now in terms of
renormalized fields and masses, and $\delta{\cal L}^r$ includes the set
of counter-terms
\begin{eqnarray}
\delta{\cal L}^r&\!\!=\!&\delta m\, \overline{N}^{\,r}N^r+\delta Z_N\overline{N}^{\,r}\left(i\gamma^\mu\partial_\mu-m^r\right)N^r
+\delta Z_N\overline{N}^{\,r}\left[\Gamma^{0,r}_a(0)\partial_0\pi_a^r+\Lambda^{i,r}_a(0)\partial_i\pi_a^r+\Delta^{r}(1)\right]N^r\nonumber\\
&&+\delta Z_\pi\overline{N}^r\bigg[\left[\Gamma^{0,r}(0)+\delta\Gamma^{0,r}_a(0)\right]\partial_0\pi_a^r+\left[\Lambda^{i,r}_a(0)/2+\delta\Lambda_a^{i,r}(0)\right]\partial_i\pi_a^r+\delta\Delta^r(1)\bigg]N^r\nonumber\\
&&+\frac{\delta m_\pi^2}{2} \pi_a^r \pi_a^r
+\frac{\delta Z_\pi}{2}\bigg[\partial_0\pi_a^r\left(\widetilde{G}^r_{ab}+\delta\widetilde{G}_{ab}^r\right)\partial^0\pi^r_b+\partial_i\pi_a^r\left(\widetilde{G}^r_{ab}+\delta\widetilde{G}^r_{ab}\right)\partial^i\pi^r_b\nonumber\\
&&-m_\pi^{r\,2}\, \pi_a^r\left(H_{ab}^r+\delta H_{ab}^r\right)\pi_b^r\bigg]-\delta Z_\pi \,f_\pi\,A^\mu_a(F^r_{ab}/2+\delta F^r_{ab})\partial_\mu\pi^r_b\ ,
\end{eqnarray}
where $\Gamma^{0,r}_a(0)$, $\Lambda^{i,r}_a(0)$ and $\Delta^r(1)$ are the field combinations defined in
Eqs.~(\ref{eq:g0}), (\ref{eq:l0}) and~(\ref{eq:d1}) expressed in terms of renormalized fields
and physical masses.  The remaining quantities are given by
\begin{eqnarray}
\delta\Gamma^{0r}_a(0)&=&\frac{8\alpha-1}{8\,f_\pi^4}{\bm \pi}^r\cdot{\bm \pi}^r\left({\bm \tau}\times{\bm \pi}^r\right)_a\gamma^0\ ,\\
\delta\Lambda^{i,r}_a(0)&=&\frac{g_A}{4f_\pi^3}\big[ 2\,\alpha\, {\bm \pi}^r\cdot{\bm \pi}^r\,\tau_a+\left(4\,\alpha-1\right){\bm \tau}\cdot{\bm \pi}^r\pi_a^r\big]\gamma^i\gamma_5\ ,\\
\delta\Delta^r(1)&=&\frac{1}{4f_\pi}\left(1-\frac{3\,\alpha}{f_\pi^2}\, {\bm \pi}^r\cdot{\bm \pi}^r \right)
\left({\bm \tau}\times{\bm \pi}^r\right)\cdot{\bf A}_0\gamma^0 \nonumber\\
&&+\frac{g_A}{4f_\pi^2}\big[\left({\bm \tau}\times{\bm\pi}^r\right)\times{\bm \pi}^r\big]\cdot{\bf A}_i\, \gamma^i\gamma_5\ ,\\
\delta\widetilde{G}^r_{ab}&=&-\frac{2\, \alpha}{f_\pi^2}
{\bm \pi}^r\cdot{\bm \pi}^r \,\delta_{ab}+\frac{1-4\,\alpha}{f_\pi^2}\pi_a^r\pi_b^r\ ,\\
\delta H^r_{ab}&=&\frac{1-8\, \alpha}{4 f_\pi^2}\,{\bm \pi}^r\cdot{\bm \pi}^r\,\delta_{ab}\ ,\\
\delta F_{ab}^r&=&-\frac{2\,\alpha+1}{2f_\pi^2}\, {\bm \pi}^r\cdot{\bm \pi}^r\delta_{ab}+\frac{1-2\,\alpha}
{f_\pi^2}\pi_a^r\,\pi_b^r\, .
\end{eqnarray}
It is convenient to define
\begin{eqnarray}
\label{eq:counter1}
\widetilde{G}_{ab}^{\,\prime}&=&\widetilde{G}_{ab}^r+\delta Z_\pi\big
(\widetilde{G}_{ab}^r+\delta\widetilde{G}_{ab}^r\big)\ ,\\
G_{ab}^{\,\prime}&=&\widetilde{G}_{ab}^{\,\prime}+2\,\frac{c_2+c_3}{f_\pi^2}\,\overline{N}^rN^r\delta_{ab}\ ,\\
F_{ab}^{\,\prime}&=&F_{ab}^r+\delta Z_\pi\left(F_{ab}^r/2+\delta F_{ab}^r\right)\ ,\\
H_{ab}^{\,\prime}&=&H_{ab}^r+\delta Z_\pi\left(H_{ab}^r+\delta H_{ab}^r\right)\ ,\\
\Gamma_a^{0\,\prime}&=&\Gamma_a^{0,r}+\delta Z_N\,
\Gamma^{0,r}_a(0)+\delta Z_\pi\, \big[ \Gamma_a^{0,r}(0)+\delta\Gamma_a^{0,r}(0)\big]\ ,\\
\Lambda^{i\,\prime}_a&=&\Lambda^{i,r}_a+\delta Z_N\Lambda^{i,r}_a(0)+\delta Z_\pi
\big[ \Lambda^{i,r}_a(0)/2+\delta\Lambda^{i,r}_a(0)\big]\ ,\\
\Delta^{\,\prime}&=&\Delta^r+\delta Z_N\, \Delta^r(1)+\delta Z_\pi\, \delta\Delta^r(1)\ ,
\label{eq:counter6}
\end{eqnarray}
which then leads to the Lagrangian as given in Eq.~(\ref{eq:renL}).
\end{document}